\documentclass[12pt]{article}
\usepackage{epsfig}
\usepackage{a4}
\usepackage{latexsym}

\usepackage{pslatex}
\usepackage[latin1]{inputenc}
\usepackage[T1]{fontenc}

\usepackage{color,colordvi}
\def\colour4colour#1{\Blue{#1}}

\newcommand{\gsim}{\raisebox{-0.07cm}{$\:\:\stackrel{>}{{\scriptstyle
 \sim}}\:\: $} }
\newcommand{\lsim}{\raisebox{-0.07cm}{$\:\:\stackrel{<}{{\scriptstyle
 \sim}}\:\: $} }
\newcommand{\beq}{\begin{equation}}
\newcommand{\eeq}{\end{equation}}
\newcommand{\bea}{\begin{eqnarray}}
\newcommand{\eea}{\end{eqnarray}}
\newcommand{\nn}{\nonumber}
\newcommand{\MSb}{$\overline{\mbox{MS}}$}
\newcommand{\as}{\alpha_{\rm s}}
\newcommand{\ra}{\rightarrow}
\newcommand{\DD}{{\cal D}}

\begin{document}
\setlength{\parskip}{0.3cm}
\setlength{\baselineskip}{0.55cm}

\def\plus{{\!+\!}}
\def\minus{{\!-\!}}
\def\z#1{{\zeta_{#1}}}
\def\ca{{C^{}_A}}
\def\cf{{C^{}_F}}
\def\nf{{n^{}_{\! f}}}
\def\n2f{{n^{\,2}_{\! f}}}

\def\dabc2n{{{d^{abc}d_{abc}}\over{n_c}}}
\def\S(#1){{{S}_{#1}}}
\def\Ss(#1,#2){{{S}_{#1,#2}}}
\def\Sss(#1,#2,#3){{{S}_{#1,#2,#3}}}
\def\Ssss(#1,#2,#3,#4){{{S}_{#1,#2,#3,#4}}}
\def\Sssss(#1,#2,#3,#4,#5){{{S}_{#1,#2,#3,#4,#5}}}
\def\Npm{{{\bf N_{\pm}}}}
\def\Npmi{{{\bf N_{\pm i}}}}
\def\Nminus{{{\bf N_{-}}}}
\def\Nplus{{{\bf N_{+}}}}
\def\Nminustwo{{{\bf N_{-2}}}}
\def\Nplustwo{{{\bf N_{+2}}}}
\def\Nminusthree{{{\bf N_{-3}}}}
\def\Nplusthree{{{\bf N_{+3}}}}

\def\pqq(#1){p_{\rm{qq}}(#1)}
\def\pqg(#1){p_{\rm{qg}}(#1)}
\def\pgq(#1){p_{\rm{gq}}(#1)}
\def\pgg(#1){p_{\rm{gg}}(#1)}
\def\H(#1){{\rm{H}}_{#1}}
\def\Hh(#1,#2){{\rm{H}}_{#1,#2}}
\def\Hhh(#1,#2,#3){{\rm{H}}_{#1,#2,#3}}
\def\Hhhh(#1,#2,#3,#4){{\rm{H}}_{#1,#2,#3,#4}}

\begin{titlepage}
\noindent
NIKHEF 04-004 \hfill {\tt hep-ph/0404111}\\
DESY 04--060 \\ 
SFB/CPP-04-12 \\
April 2004 \\
\vspace{1.3cm}
\begin{center}
\Large
{\bf The Three-Loop Splitting Functions in QCD:} \\
\vspace{0.15cm}
{\bf The Singlet Case} \\
\vspace{1.5cm}
\large
A. Vogt$^{\, a}$, S. Moch$^{\, b}$ and J.A.M. Vermaseren$^{\, a}$\\
\vspace{1.2cm}
\normalsize
{\it $^a$NIKHEF Theory Group \\
\vspace{0.1cm}
Kruislaan 409, 1098 SJ Amsterdam, The Netherlands} \\
\vspace{0.5cm}
{\it $^b$Deutsches Elektronensynchrotron DESY \\
\vspace{0.1cm}
Platanenallee 6, D--15735 Zeuthen, Germany}\\
\vspace{3.0cm}
\large
{\bf Abstract}
\vspace{-0.2cm}
\end{center}
We compute the next-to-next-to-leading order $\:$(NNLO)$\:$ 
contributions to the splitting functions governing the evolution of 
the unpolarized flavour-singlet parton densities in perturbative QCD. 
The exact expressions are presented in both Mellin-$N$ and Bjorken-$x$ 
space. We also provide accurate parametrizations for practical 
applications.
Our results agree with all partial results available in the literature.
As in the non-singlet case, the correct leading logarithmic predictions 
for small momentum fractions $x$ do not provide good estimates of the
respective complete splitting functions. 
We investigate the size of the corrections and the stability of the
NNLO evolution under variation of the renormalization scale. The
perturbative expansion appears to converge rapidly at $x \gsim
10^{\,-3}$. Relatively large third-order corrections are found at
smaller values of $x$.
\vfill
\end{titlepage}
%
%
\section{Introduction}
\label{sec:introduction}
%
%
Parton distributions form indispensable ingredients for the analysis 
of all hard-scattering processes involving initial-state hadrons. The 
scale-dependence (evolution) of these distributions can be derived from 
first principles in terms of an expansion in powers of the strong 
coupling constant~$\as$. The corresponding $n\,$th-order coefficients 
governing the evolution are referred to as the $n$-loop anomalous 
dimensions or splitting functions. Parton distributions evolved by 
including the terms up to order $\as^{\, n+1}$ in this expansion 
constitute, together with the corresponding results for the partonic 
cross sections for the observable under consideration, the 
N$^{\rm n}$LO (leading-order, next-to-leading-order, 
next-to-next-to-leading-order, etc.) approximation of perturbative QCD.

Presently the next-to-leading order is the standard approximation for
most important processes. The corresponding one- and two-loop splitting
functions have been known for a long time 
\cite{Gross:1973rr,Georgi:1974sr,Altarelli:1977zs,Floratos:1977au,%
Floratos:1979ny,Gonzalez-Arroyo:1979df,Gonzalez-Arroyo:1980he,%
Curci:1980uw,Furmanski:1980cm,Floratos:1981hs,Hamberg:1992qt}.
The NNLO corrections need to be included, however, in order to arrive at
quantitatively reliable predictions for hard processes at present and
future high-energy colliders. These corrections are so far known only
for structure functions in deep-inelastic scattering (DIS)
\cite{ vanNeerven:1991nn, Zijlstra:1991qc,Zijlstra:1992kj,%
Zijlstra:1992qd}
and for Drell-Yan lepton-pair and gauge-boson production in 
proton--(anti-)proton collisions 
\cite{Hamberg:1991np,Harlander:2002wh,Anastasiou:2003yy,%
Anastasiou:2003ds}
and the related cross sections for Higgs production in the 
heavy-top-quark approximation 
\cite{Harlander:2002wh,Anastasiou:2002yz,Ravindran:2003um,%
Harlander:2003ai}.
Work on NNLO cross sections for jet production is under way and 
expected to yield results in the near future, see 
Ref.~\cite{Glover:2002gz} and references therein.
 
For the three-loop splitting functions, on the other hand, only partial
results had been computed until very recently, especially the lowest 
six/seven (even or odd) integer-$N$ Mellin moments 
\cite{Larin:1994vu,Larin:1997wd,Retey:2000nq} and the leading 
$(\ln x)/x$ small-$x$ terms of three of the four singlet splitting 
functions \cite{Catani:1994sq,Fadin:1998py}. 
The results of Refs.~\cite{Larin:1994vu,Larin:1997wd,Retey:2000nq} 
have been employed -- directly 
\cite{Santiago:1999pr,Kataev:1999bp,Santiago:2001mh,Kataev:2001kk} 
and indirectly~\cite{Martin:2002dr,Alekhin:2002fv} via Bjorken 
$x$-space approximations constructed in Refs.\ 
\cite{vanNeerven:1999ca,vanNeerven:2000uj,vanNeerven:2000wp} 
from them and the \mbox{small-$x$} constraints 
\cite{Catani:1994sq,Fadin:1998py} -- to improve the analysis of DIS 
data and hadron-collider predictions. This information is however not 
sufficient for quantitative predictions at small values of $x$. 

We have recently published the non-singlet part of the unpolarized 
three-loop splitting functions~\cite{Moch:2004pa}. In the present 
article we compute the corresponding singlet quantities. 
The article is organized as follows: In section 2 we set up our 
notations and very briefly discuss the method of our calculation.
The Mellin-$N$ space results are written down in section~3. 
The $(\ln N)/N$ subleading large-$N$ term of the three-loop gluon-gluon 
splitting function is found to be related to the leading $\ln N$
contribution at second order, in complete analogy to the relation
found for the non-singlet quark-quark case.
In section~4 we present the exact results as well as compact 
parametrizations for the $x$-space splitting functions and study their
behaviour at small~$x$. We demonstrate that neither do the $(\ln x)/x$ 
terms dominate the splitting functions at experimentally relevant 
values of $x$, nor do even all $1/x$ terms dominate the Mellin 
convolutions by which the splitting functions enter the evolution equations.
The numerical implications of our results for the scale dependence of 
the singlet-quark and gluon distributions are illustrated in section 5. 
As in the non-singlet case the perturbation series converges rapidly
for $x \gsim 10^{-3}$, while relatively large corrections occur for
smaller momentum fractions.
Finally we briefly summarize our findings in section~6.
%
%
\section{Notations and method} 
\label{sec:method}
%
%
We start by setting up our notations for the singlet parton densities
and the splitting functions governing their evolution. The singlet 
quark distribution of a hadron is given~by
\beq
  q_{\rm s}^{}(x,\mu_f^{\,2}) \: = \: \sum_{i=1}^{\nf} \left[ 
  q_i^{}(x,\mu_f^{\,2}) + \bar{q}_i^{}(x,\mu_f^{\,2}) \right] \:\: .
\eeq
Here $q_i^{}(x,\mu_f^{\,2})$ and $\bar{q}_i^{}(x,\mu_f^{\,2})$ 
represent the respective number distributions of quarks and antiquarks
in the fractional hadron momentum $x$.  The corresponding gluon 
distribution is denoted by $g(x,\mu_f^{\,2})$.  The subscript $i$ 
indicates the flavour of the (anti-) quarks, and $\nf$ stands for the 
number of effectively massless flavours.  Finally $\mu_f$ represents 
the factorization scale. For the time being we do not need to introduce
a renormalization scale $\mu_{\,r}$ different from $\mu_f$.

Suppressing the functional dependences, the evolution equations
for the singlet parton distributions read
\beq
\label{eq:evol}
  \frac{d}{d \ln\mu_f^{\,2}}
  \left( \begin{array}{c} \! q_{\rm s}^{} \! \\ g  \end{array} \right) 
  \: = \: \left( \begin{array}{cc} P_{\rm qq} & P_{\rm qg} \\ 
  P_{\rm gq} & P_{\rm gg} \end{array} \right) \otimes 
  \left( \begin{array}{c} \!q_{\rm s}^{}\! \\ g  \end{array} \right) 
  \:\: ,
\eeq
where $\otimes$ stands for the Mellin convolution in the momentum
variable,
\beq
  [ a \otimes b ](x) \:\equiv\: \int_x^1 \! \frac{dy}{y} \: a(y)\,
  b\bigg(\frac{x}{y}\bigg) \:\: .
\eeq
The quark-quark splitting function $P_{\rm qq}$ in Eq.~(\ref{eq:evol})
can be expressed as 
\beq
\label{eq:Pqq}
  P_{\rm qq} \: =\: P_{\rm ns}^{\,+} + \nf (P_{\rm qq}^{\:\rm s}
  + P_{\rm {\bar{q}q}}^{\:\rm s})
  \:\equiv\:  P_{\rm ns}^{\,+} + P_{\rm ps}^{} \:\: .
\eeq
Here $P_{\rm ns}^{\,+}$ is the non-singlet splitting function which we
have recently computed up to the third order in Ref.~\cite{Moch:2004pa}.
The ${\cal O}(\alpha_{\rm s}^2)$ quantities $P_{\rm qq}^{\,\rm s}$ and 
$P_{\rm {\bar{q}q}}^{\,\rm s}$ are the flavour independent (`sea') 
contributions to the quark-\-quark and quark-antiquark splitting 
functions $P_{{\rm q}_{i}{\rm q}_{k}}$ and $P_{{\bar{\rm q}_{i} 
{\rm q}_{k}}}$, respectively. The non-singlet contribution dominates 
Eq.~(\ref{eq:Pqq}) at large $x$, where the `pure singlet' term 
$P_{\rm ps} ^{}$ is very small. At small $x$, on the other hand, the 
latter contribution takes over as $xP_{\rm ps}^{}$ does not vanish for 
$x \ra 0$, unlike $xP_{\rm ns}^{\,+}$. The gluon-quark and quark-gluon 
entries in Eq.~(\ref{eq:evol}) are given by
\beq
\label{eq:Poffd}
  P_{\rm qg} \: =\: \nf\, P_{{\rm q}_{i}\rm g} \:\: , \quad
  P_{\rm gq} \: =\: P_{{\rm gq}_{i}}
\eeq
in terms of the flavour-independent splitting functions $P_{{\rm q}_{i}
\rm g} = P_{\bar{\rm q}_{i}\rm g}$ and $P_{{\rm gq}_{i}} = P_{{\rm g}
\bar{\rm q}_{i}}$. With the exception of the $\alpha_{\rm s}^1$ part of 
$P_{\rm qg}$, neither of the quantities $xP_{\rm qg}$, $xP_{\rm gq}$ 
and $xP_{\rm gg}$ vanishes for $x \ra 0$.  

Our calculation is performed in Mellin-$N$ space, i.e., we compute the
singlet anomalous dimensions $\gamma_{\,\rm ab}(N,\as)$ which are 
related to the splitting functions by a Mellin transformation,
\beq
\label{eq:Pdef}
  \gamma_{\,\rm ab}(N,\as) \: = \:
  - \int_0^1 \!dx\:\, x^{\,N-1}\, P_{\,\rm ab}(x,\as) \:\: .
\eeq
The additional relative sign is the standard convention. Note that in 
the older literature an additional factor of two is often included in 
Eq.~(\ref{eq:Pdef}).

The calculation follows the approach of 
Refs.~\cite{Larin:1997wd,Retey:2000nq,Kazakov:1988jk,Moch:1999eb}.
The optical theorem and the operator product expansion are employed to 
compute the Mellin moments of (partly fictitious, see below) 
deep-inelastic structure functions. Since the moment variable $N$ is 
now an analytical parameter, we cannot apply the techniques of 
Refs.~\cite{Larin:1997wd,Retey:2000nq}, where the integrals were solved 
using the {\sc Mincer} program \cite{Gorishnii:1989gt,Larin:1991fz}.  
The introduction of new techniques was therefore necessary, and various 
aspects of those have already been discussed in 
Refs.~\cite{Moch:1999eb,Vermaseren:2000we,Moch:2002sn,%
Vermaseren:2002rn,Moch:2004pa}.
A salient feature of our method is, however, that we can check our
extensive manipulations at almost any stage by falling back on a
{\sc Mincer} evaluation of fixed low-integer moments. Note also that 
we will obtain the three-loop coefficient functions in DIS as well, 
once the present calculation is supplemented by a second Lorentz 
projection required to disentangle the structure functions $F_2^{}$ and 
$F_L^{}$ \cite{MVV5}.

The complete set of NNLO singlet anomalous dimensions can be extracted
from the third-order amplitudes of the forward Compton processes
\beq
\label{eq:ampl}
 \mbox{parton}\,(P)+\mbox{probe}\,(Q) \:\longrightarrow\: 
 \mbox{parton}\,(P)+\mbox{probe}\,(Q) \:\: ,
\end{equation}
where the probes are the photon ($\gamma$) and a fictitious classical
scalar $\phi $ coupling directly only to the gluon field via $\phi\, 
G_{\mu\nu}^{\,a}G_a^{\,\mu\nu}$. The inclusion of the latter, required 
for obtaining also the anomalous dimensions $\gamma_{\,\rm gq}$ and 
$\gamma_{\,\rm gg}$ to the desired accuracy, leads to a substantial 
increase of the number of diagrams as shown in Table 1. Among the
partons in Eq.~(\ref{eq:ampl}) we also include an external ghost~$h$.
This is done in order to allow us to take the sum over external gluon 
spins by contracting with $-g_{\mu\nu}$ instead of the full physical 
expression which would, due to the presence of extra powers of $P$,
lead to a complication of our task. For similar reasons we do not
keep the gauge dependence in our all-$N$ computations, but check its
cancellation only for a few fixed values of $N$.

\begin{table}[htp]
\label{table1}
\begin{center}
\begin{tabular}{c r r r r}\\
\hline & & & & \\[-3mm]
{process} &{tree} &1-loop &2-loop &3-loop \\[1mm]
\hline & & & & \\[-3mm]
$q\,\gamma\:\ra\: q\,\gamma$ & 1 &  3 &  25 &  359 \\
$g\,\gamma\:\ra\: g\,\gamma$ &   &  2 &  17 &  345 \\
$h\,\gamma\:\ra\: g\,\gamma$ &   &    &   2 &   56 \\
$q\,\phi \:\ra\: q\,\phi $   &   &  1 &  23 &  696 \\
$g\,\phi \:\ra\: g\,\phi $   & 1 &  8 & 218 & 6378 \\
$h\,\phi \:\ra\: h\,\phi $   &   &  1 &  33 & 1184 \\[1mm]
\hline & & & & \\[-3mm]
 sum & 2 & 15 & 318 & 9018 \\[1mm] \hline
\end{tabular} 
\end{center}
\vspace{-1mm}
\caption{The number of diagrams for the amplitudes employed for the 
 calculation of the three-loop singlet anomalous dimensions. The roles 
 of the ghost $h$ and the scalar $\phi$ are discussed in the text.} 
\end{table}
 
The diagrams are generated automatically with the diagram generator 
{\sc Qgraf}~\cite{Nogueira:1991ex}. For all symbolic manipulations we 
use the latest version of 
{\sc Form}~\cite{Vermaseren:2000nd,Vermaseren:2002rp}.
The calculation is performed in dimensional regularization~\cite
{'tHooft:1972fi,Bollini:1972ui,Ashmore:1972uj,Cicuta:1972jf}.
The renormalization is carried out in the \MSb-scheme 
\cite{'tHooft:1973mm,Bardeen:1978yd} as described in detail in
Ref.~\cite{Larin:1997wd}, using the result of Refs.~\cite{ZG2a,ZG2b}
for the renormalization of the operator 
$G_{\mu\nu}^{\,a}G_a^{\,\mu\nu}$ entering the scalar case.
%
%
\setcounter{equation}{0}
\section{Results in Mellin space}
\label{sec:results}
%
%
Here we present the anomalous dimensions $\gamma_{\,\rm ab}(N,\as)$ in 
the \MSb-scheme up to the third order in the running coupling constant 
$\as$. The N$^{\rm n}$LO expansion coefficients 
$\gamma_{\,\rm ab}^{\,(n)}(N)$ are normalized as
\beq
 \gamma_{\,\rm ab}\left(\as,N\right) \: = \: \sum_{n=0}\,
  \left(\frac{\as}{4\pi}\right)^{n+1} \gamma^{\,(n)}_{\,\rm ab}(N)
  \:\: .
\eeq
The anomalous dimensions can be expressed in terms of harmonic 
sums~\cite{Gonzalez-Arroyo:1979df,Gonzalez-Arroyo:1980he,%
Vermaseren:1998uu,Blumlein:1998if,Moch:2001zr}. Recall that, following 
the notation of Ref.~\cite{Vermaseren:1998uu}, these sums are 
recursively defined by
\beq
\label{eq:Hsum1}
  S_{\pm m}(M) \: = \: \sum_{i=1}^{M}\: \frac{(\pm 1)^i}{i^{\, m}}
\eeq
and 
\beq
\label{eq:Hsum2}
  S_{\pm m_1,m_2,\ldots,m_k}(M) \: = \: \sum_{i=1}^{M}\: 
  \frac{(\pm 1)^{i}}{i^{\, m_1}}\: S_{m_2,\ldots,m_k}(i) \:\: .
\eeq
The sum of the absolute values of the indices $m_k$ defines the weight
of the harmonic sum. Sums up to weight $2l-1$ occur in the $l$-loop 
results written down below.
 
In order to arrive at a reasonably compact representation of our 
results, we employ the abbreviation $S_{\vec{m}}\,\equiv\, 
S_{\vec{m}}(N)$ in what follows, together with the notation
\beq
\label{eq:shiftN}
  \Npm \, S_{\vec{m}} \: = \: S_{\vec{m}}(N \pm 1) \:\: , \quad\quad
  \Npmi\, S_{\vec{m}} \: = \: S_{\vec{m}}(N \pm i) 
\eeq
for arguments shifted by $\pm 1$ or a larger integer $i$. In this
notation the well-known one-loop (LO) singlet anomalous dimensions 
\cite{Gross:1973rr,Georgi:1974sr} read
\bea
  \gamma^{\,(0)}_{\,\rm ps}(N) & \! = \! & 0 
\nn \\[1mm]
  \gamma^{\,(0)}_{\,\rm qg}(N) & \! = \! &  
         2\, \* \colour4colour{\nf} \, \*  \big(
            \Nminus
          + 4 \* \Nplus
          - 2 \* \Nplustwo
          - 3
          \big) \, \* \S(1)
 \nn \\[1mm]
  \gamma^{\,(0)}_{\,\rm gq}(N) & \! = \! & 
       2\, \* \colour4colour{\cf}  \, \*  \big(
            2 \* \Nminustwo
          - 4 \* \Nminus
          - \Nplus
          + 3
          \big) \, \* \S(1)
  \nn \\[1mm]
  \gamma^{\,(0)}_{\,\rm gg}(N) & \! = \! &
         \colour4colour{\ca}  \, \*  \bigg(
            4 \* ( \Nminustwo
          - 2 \* \Nminus
          - 2 \* \Nplus
          + \Nplustwo
          + 3 ) \, \* \S(1)
          - {11 \over 3}
          \bigg)
       + {2 \over 3} \, \* \colour4colour{\nf}
%
\:\ .\label{eq:gij0}
\eea
The corresponding second-order (NLO) quantities \cite{Floratos:1979ny,%
Gonzalez-Arroyo:1979df,Floratos:1981hs,Hamberg:1992qt} are given by
\bea
 \label{eq:gps1}
  &&\gamma^{\,(1)}_{\,\rm ps}(N) \:\: = \:\:
         4\,  \*  \colour4colour{\cf \* \nf}  \*  \bigg(
            {20 \over 9} \* (\Nminustwo-\Nminus) \* \S(1)
          - (\Nplus-\Nplustwo) \* \bigg[
            {56 \over 9} \* \S(1)
          + {8 \over 3} \* \S(2)
       \bigg]
          + (1-\Nplus) \* \bigg[
            8 \* \S(1)
          - 4 \* \S(2)
       \bigg]
  \nonumber\\&& \mbox{}
          - (\Nminus-\Nplus) \* \bigg[
            2 \* \S(1)
          + \S(2)
          + 2 \* \S(3)
       \bigg]
          \bigg)
 \\[4mm]
 \label{eq:gqg1}
  &&\gamma^{\,(1)}_{\,\rm qg}(N) \:\: = \:\: 
       4\,  \*  \colour4colour{\ca \* \nf} \*  \bigg(
            {20 \over 9} \* (\Nminustwo-\Nminus) \* \S(1) 
          - (\Nminus-\Nplus) \* \bigg[ 
            2 \* \S(1) 
          + \S(2) 
          + 2 \* \S(3)
  	  \bigg]
          - (\Nplus-\Nplustwo) \* \bigg[
            {218 \over 9} \* \S(1) 
  \nonumber\\&& \mbox{}
          + 4 \* \Ss(1,1) 
          + {44 \over 3} \* \S(2) 
  	  \bigg]
          + (1-\Nplus) \* \bigg[ 
            27 \* \S(1)
          + 4 \* \Ss(1,1)
          - 7 \* \S(2)
          - 2 \* \S(3)
  	  \bigg]
          - 2 \* (\Nminus+4\*\Nplus-2\*\Nplustwo-3) \* \bigg[ 
            \Ss(1,-2) 
  \nonumber\\&& \mbox{}
          + \Sss(1,1,1)
  	  \bigg]
          \bigg)
       + 4\,  \*  \colour4colour{\cf \* \nf}  \*  \bigg(
            2 \* (\Nplus-\Nplustwo) \* \bigg[ 
            5 \* \S(1)
          + 2 \* \Ss(1,1) 
          - 2 \* \S(2) 
	  + \S(3) 
  	  \bigg]
          - (1-\Nplus) \* \bigg[ 
            {43 \over 2} \* \S(1)
          + 4 \* \Ss(1,1)
          - {7 \over 2} \* \S(2)
  	  \bigg]
  \nonumber\\&& \mbox{}
          + (\Nminus-\Nplus) \* \bigg[
            7 \* \S(1) 
          - {3 \over 2} \* \S(2) 
  	  \bigg]
          + 2 \* (\Nminus+4\*\Nplus-2\*\Nplustwo-3) \* \bigg[ 
            \Sss(1,1,1)
          - \Ss(1,2) 
          - \Ss(2,1)
          + {1 \over 2} \* \S(3)
  	  \bigg]
          \bigg)
 \\[4mm]
 \label{eq:ggq1}
  &&\gamma^{\,(1)}_{\,\rm gq}(N) \:\: = \:\: 
       4\,  \*  \colour4colour{\ca \* \cf}  \*  \bigg(
            2\*(2\*\Nminustwo-4\*\Nminus-\Nplus+3) \* \bigg[
            \Sss(1,1,1) 
          - \Ss(1,-2)
          - \Ss(1,2) 
          - \Ss(2,1) 
  	  \bigg]
          + (1-\Nplus) \* \bigg[ 
            2 \* \S(1)
  \nonumber\\&& \mbox{}
          - 13 \* \Ss(1,1)
          - 7 \* \S(2)
          - 2 \* \S(3)
  	  \bigg]
          + (\Nminustwo-2\*\Nminus+\Nplus) \* \bigg[
            \S(1) 
          - {22 \over 3} \* \Ss(1,1)
  	  \bigg]
          + 4 \* (\Nminus-\Nplus) \* \bigg[
            {7 \over 9} \* \S(1) 
          + 3 \* \S(2) 
          + \S(3) 
  	  \bigg]
  \nonumber\\&& \mbox{}
          + (\Nplus-\Nplustwo) \* \bigg[
            {44 \over 9} \* \S(1)
          + {8 \over 3} \* \S(2) 
  	  \bigg]
          \bigg)
       + 4\,  \*  \colour4colour{\cf \* \nf}  \*  \bigg(
            (\Nminustwo-2\*\Nminus+\Nplus) \* \bigg[
            {4 \over 3} \* \Ss(1,1)
          - {20 \over 9} \* \S(1) 
  	  \bigg]
          - (1-\Nplus) \* \bigg[
            4 \* \S(1)
  \nonumber\\&& \mbox{}
          - 2 \* \Ss(1,1)
  	  \bigg]
          \bigg)
       + 4\,  \*  \colour4colour{\cf^2}  \*  \bigg(
            (2\*\Nminustwo-4\*\Nminus-\Nplus+3)  \* \bigg[
            3 \* \Ss(1,1) 
          - 2 \* \Sss(1,1,1) 
  	  \bigg]
          - (1-\Nplus) \* \bigg[
            \S(1)
          - 2 \* \Ss(1,1)
          + {3 \over 2} \* \S(2)
  \nonumber\\&& \mbox{}
          - 3 \* \S(3) 
  	  \bigg]
          - (\Nminus-\Nplus) \* \bigg[
            {5 \over 2} \* \S(1) 
          + 2 \* \S(2) 
          + 2 \* \S(3) 
  	  \bigg]
          \bigg)
 \\[4mm]
 \label{eq:ggg1}
  &&\gamma^{\,(1)}_{\,\rm gg}(N) \:\: = \:\: 
       4\,  \*  \colour4colour{\ca \* \nf}  \*  \bigg(
            {2 \over 3}
          - {16 \over 3} \* \S(1)
          - {23 \over 9} \* (\Nminustwo+\Nplustwo) \* \S(1) 
          + {14 \over 3} \* (\Nminus+\Nplus) \* \S(1) 
          + {2 \over 3} \* (\Nminus-\Nplus) \* \S(2) 
          \bigg)
  \nonumber\\&& \mbox{}
       + 4\,  \*  \colour4colour{\ca^2}  \*  \bigg(
            2 \* \S(-3)
          - {8 \over 3}
          - {14 \over 3} \* \S(1)
          + 2 \* \S(3)
          - (\Nminustwo-2\*\Nminus-2\*\Nplus+\Nplustwo+3) \* \bigg[
            4 \* \Ss(1,-2) 
          + 4 \* \Ss(1,2) 
          + 4 \* \Ss(2,1) 
  	  \bigg]
  \nonumber\\&& \mbox{}
          +  {8 \over 3} \* (\Nplus-\Nplustwo) \* \S(2) 
          - 4 \* (\Nminus-3\*\Nplus+\Nplustwo+1) \* \bigg[
            3 \* \S(2)
          - \S(3)
  	  \bigg]
          + {109 \over 18} \* (\Nminus+\Nplus) \* \S(1) 
          + {61 \over 3} \* (\Nminus
  \nonumber\\&& \mbox{}
	  -\Nplus) \* \S(2) 
          \bigg)
       + 4\,  \*  \colour4colour{\cf \* \nf}  \*  \bigg(
            {1 \over 2}
          + {2 \over 3} \* (\Nminustwo-13\*\Nminus-\Nplus-5\*\Nplustwo+18) \* \S(1) 
          + (3\*\Nminus-5\*\Nplus+2) \* \S(2)
  \nonumber\\&& \mbox{}
          - 2 \* (\Nminus-\Nplus) \* \S(3)
          \bigg)
\:\: . 
\eea

The pure-singlet contribution (\ref{eq:Pqq}) to the three-loop (NNLO) 
anomalous dimension $\gamma_{\,\rm qq}^{\,(2)}(N)$ is
\bea
  &&\gamma^{\,(2)}_{\,\rm ps}(N) \:\: = \:\: 
         16\,  \*  \colour4colour{\ca \* \cf \* \nf}  \*  \bigg(
	    {1 \over 3} \* (4\*\Nminustwo-\Nminus-\Nplus+4\*\Nplustwo-6) \* \bigg[
            3 \* \S(1) \* \z3 
          + \Sss(1,-2,1)
          - \Sss(1,1,-2) 
          + \Ssss(1,1,1,1) 
  \nonumber\\&& \mbox{}
          - \Sss(1,1,2) 
  	  \bigg]
          + (\Nminustwo-\Nminus)  \* \bigg[
            {571 \over 108} \* \Ss(1,1) 
          - {6761 \over 324} \* \S(1) 
          - {3 \over 2} \* \Ss(1,2) 
          - {52 \over 9} \* \Ss(1,-2) 
          + {56 \over 27} \* \S(2) 
          - {20 \over 9} \* \Ss(2,1) 
  	  \bigg]
  \nonumber\\&& \mbox{}
          - (\Nminustwo-\Nminus-\Nplus+\Nplustwo)  \* \bigg[
            {8 \over 3} \* \Ss(1,-3) 
          + 2 \* \S(1,3) 
          + {1 \over 9} \* \Sss(1,1,1) 
          + {2 \over 3} \* \Sss(2,1,1) 
  	  \bigg]
          + (\Nplus-\Nplustwo) \* \bigg[
            {10279 \over 162} \* \S(1) 
  \nonumber\\&& \mbox{}
          + {106 \over 9} \* \Ss(1,-2) 
          + {151 \over 54} \* \Ss(1,1) 
          + {9 \over 2} \* \Ss(1,2) 
          + 4 \* \Ss(2,-2) 
          + {2299 \over 54} \* \S(2) 
          + {28 \over 9} \* \Ss(2,1) 
          + {2 \over 3} \* \Ss(2,2) 
          + {83 \over 6} \* \S(3) 
          + {2 \over 3} \* \Ss(3,1) 
  	  \bigg]
  \nonumber\\&& \mbox{}
          + (1-\Nplus) \* \bigg[
            {4 \over 3} \* \Ss(1,2)
          - {251 \over 4} \* \S(1)
          - {50 \over 3} \* \Ss(1,-2)
          - {29 \over 12} \* \S(2)
          - {1165 \over 36} \* \Ss(1,1)
          + 5 \* \Ss(2,-2)
          + {33 \over 4} \* \Ss(2,1)
          + \Sss(2,1,1)
          + {3 \over 2} \* \Ss(2,2)
  \nonumber\\&& \mbox{}
          - {37 \over 2} \* \S(3) 
          - 4 \* \Ss(3,-2)
          + \Ss(3,1)
          - 10 \* \S(4)
          - 7 \* \S(5)
  	  \bigg]
	  - (\Nminus+\Nplus-2) \* \bigg[
            {1 \over 2} \* \Ss(1,-3)
          + 3 \* \Sss(1,-2,1)
          + {3 \over 4} \* \Sss(1,1,1)
          + {9 \over 4} \* \Ss(1,3)
  	  \bigg]
  \nonumber\\&& \mbox{}
          + (\Nminus-\Nplus) \* \bigg[
            {121 \over 12} \* \S(1) 
          + {16 \over 3} \* \Ss(1,-2) 
          + {437 \over 36} \* \Ss(1,1) 
          - {13 \over 6} \* \Ss(1,2) 
          + {3565 \over 108} \* \S(2) 
          - 6 \* \S(2) \* \z3
          + 3 \* \Ss(2,-3) 
          + {3 \over 2} \* \Ss(2,-2) 
  \nonumber\\&& \mbox{}
          - {479 \over 36} \* \Ss(2,1) 
          + 2 \* \Sss(2,1,-2) 
          + {11 \over 6} \* \Sss(2,1,1) 
          - 2 \* \Ssss(2,1,1,1) 
          + 2 \* \Sss(2,1,2) 
          + \Ss(2,2) 
          + {7 \over 2} \* \Ss(2,3)
          + {269 \over 36}  \* \S(3) 
          + 5 \* \Ss(3,-2)
          + {29 \over 6} \* \S(4)  
  \nonumber\\&& \mbox{}
          + {59 \over 12} \* \Ss(3,1)
          + \Sss(3,1,1) 
          + {1 \over 2} \* \Ss(4,1) 
          + 4 \* \S(5) 
  	  \bigg]
          \bigg)
       + 16\,  \*  \colour4colour{\cf \* \nf^2}  \*  \bigg(
	    {2 \over 9} \* (\Nminustwo-\Nminus-\Nplus+\Nplustwo) \* \bigg[
            \Sss(1,1,1) 
          + {5 \over 3} \* \Ss(1,1) 
  \nonumber\\&& \mbox{}
          + {2 \over 3} \* \S(1) 
  	  \bigg] 
	  + (\Nplus-\Nplustwo) \* \bigg[
            2\* \S(1) 
          - \Ss(1,1) 
          + {19 \over 9} \* \S(2)
          - {2 \over 3} \* \Ss(2,1) 
  	  \bigg] 
	  - (1-\Nplus) \* \bigg[
	    2 \* \S(1) 
          - \Ss(1,1)
          + \Ss(2,1)
          + {2 \over 27} \* \S(2)
  	  \bigg] 
  \nonumber\\&& \mbox{}
	  + (\Nminus+\Nplus-2) \* \bigg[
            {77 \over 54} \* \S(1) 
          - {63 \over 54} \* \Ss(1,1)
          - {37 \over 27} \* \S(2) 
          + {1 \over 6} \* \Sss(1,1,1)
  	  \bigg] 
	  + {1 \over 3} \* (\Nminus-\Nplus) \* \bigg[
            {5 \over 3} \* \Ss(2,1) 
          - \Sss(2,1,1)
          + {29 \over 6} \* \S(3) 
  \nonumber\\&& \mbox{} 
          - 2 \* \Ss(3,1) 
          - \S(4) 
  	  \bigg]
          \bigg)
       + 16\,  \*  \colour4colour{\cf^2 \* \nf}  \*  \bigg(
	    (\Nplus-\Nplustwo) \* \bigg[
            {16 \over 3} \* \Ss(3,1)
          + {13 \over 54} \* \S(2) 
          - {163 \over 12} \* \S(1) 
          - {85 \over 12} \* \Ss(1,1) 
          + {28 \over 9} \* \Ss(2,1) 
          - {22 \over 3} \* \S(3) 
  \nonumber\\&& \mbox{}
          + 4 \* \Ss(2,2) 
          - {4 \over 3} \* \Sss(2,1,1) 
          + 3 \* \Ss(1,2) 
          - {22 \over 3} \* \S(4)
  	  \bigg]
	  - {1 \over 3} \* (4\*\Nminustwo-\Nminus-\Nplus+4\*\Nplustwo-6) \* \bigg[
            3 \* \S(1) \* \z3
          - \Sss(1,1,1) 
          - \Sss(1,1,2) 
  \nonumber\\&& \mbox{}
          + \Ssss(1,1,1,1) 
          - {1 \over 2} \* \Ss(1,3) 
  	  \bigg]
	  + (\Nminustwo+\Nplustwo) \* \bigg[
            {55 \over 12} \* \S(1) 
          - {523 \over 108} \* \Ss(1,1) 
          - {23 \over 9} \* \Ss(1,2) 
  	  \bigg]
          - {55 \over 6} \* \S(1) 
          + {46 \over 9} \* \Ss(1,2) 
          + {523 \over 54} \* \Ss(1,1)
  \nonumber\\&& \mbox{}
	  + (1-\Nplus) \* \bigg[
            {298 \over 27} \* \S(1) 
          - {121 \over 9} \* \Ss(1,2) 
          + {2707 \over 108} \* \Ss(1,1) 
          - {497 \over 18} \* \S(2)
          - {63 \over 4} \* \Ss(2,1)
          + {5 \over 6} \* \Sss(1,1,1) 
          + 5 \* \Ss(2,2)
          + {181 \over 12} \* \S(3)
          - \S(4)
  \nonumber\\&& \mbox{}
          - \Sss(2,1,1)  
          + 5 \* \Ss(3,1)
  	  \bigg]
	  + (\Nminus-\Nplus) \* \bigg[
            {47 \over 9} \* \Ss(1,2) 
          - {971 \over 108} \* \Ss(1,1) 
          + {275 \over 216} \* \S(1) 
          - {755 \over 72} \* \S(2) 
          - {5 \over 12} \* \Sss(1,1,1) 
          + 6 \* \S(2) \* \z3
          - \Ss(2,3) 
  \nonumber\\&& \mbox{}
          + 17 \* \Ss(2,1) 
          + 2 \* \Ssss(2,1,1,1) 
          - 2 \* \Sss(2,1,2) 
          - 3 \* \Sss(2,1,1) 
          + 2 \* \Ss(2,2) 
          - {32 \over 3} \* \S(3) 
          - 2 \* \Ss(3,1) 
          - \Sss(3,1,1) 
          + 4 \* \Ss(3,2) 
          - {3 \over 2} \* \S(4) 
          + 6 \* \Ss(4,1)
  \nonumber\\&& \mbox{}
          - 4 \* \S(5) 
  	  \bigg]
          \bigg)
\:\: .\label{eq:gps2}
\eea
The non-singlet part of $\gamma_{\,\rm qq}^{\,(2)}(N)$ can be found in
Eq.~(3.7) of Ref.~\cite{Moch:2004pa}.
The third-order results for the off-diagonal anomalous dimensions 
$\gamma_{\,\rm qg}(N)$ and $\gamma_{\,\rm gq}(N)$ in 
Eq.~(\ref{eq:evol}) are given by
\bea
  &&\gamma^{\,(2)}_{\,\rm qg}(N) \:\: = \:\:  
16\,  \*   \colour4colour{\ca \* \cf \* \nf}  \*  \bigg(
	  (\Nminus+4\*\Nplus-2\*\Nplustwo-3) \* \bigg[
            {31 \over 2} \* \S(1) \* \z3
          - {3997 \over 96} \* \S(1)
          - {11 \over 2} \* \Ss(1,-4)
          + 6 \* \Sss(1,-3,1)
  \nonumber\\&& \mbox{}
          - {3 \over 2} \* \Ss(1,-3)
          - {9 \over 2} \* \Ss(1,-2)
          - 3 \* \Sss(1,-2,-2)
          - {5 \over 2} \* \Sss(1,-2,1)
          - 2 \* \Ssss(1,-2,1,1)
          + 2 \* \Sss(1,-2,2)
          - {2405 \over 216} \* \Ss(1,1)
          + 6 \* \Sss(1,1,-3)
  \nonumber\\&& \mbox{}
          + 3 \* \Ss(1,1) \* \z3
          + {5 \over 2} \* \Sss(1,1,-2)
          - 6 \* \Ssss(1,1,-2,1)
          - {128 \over 9} \* \Sss(1,1,1)
          - 6 \* \Ssss(1,1,1,-2)
          - {13 \over 3} \* \Ssss(1,1,1,1)
          - 4 \* \Sssss(1,1,1,1,1)
          - 3 \* \Ssss(1,1,1,2)
  \nonumber\\&& \mbox{}
          - {35 \over 12} \* \Sss(1,1,2)
          + 3 \* \Ssss(1,1,2,1)
          + \Sss(1,1,3)
          + {53 \over 8} \* \Ss(1,2)
          + 3 \* \Sss(1,2,-2)
          + {15 \over 4} \* \Sss(1,2,1)
          + 6 \* \Ssss(1,2,1,1)
          - 6 \* \Sss(1,3,1)
          - {2833 \over 216} \* \S(2)
  \nonumber\\&& \mbox{}
          + {3 \over 2} \* \Ss(1,4)
          + 3 \* \S(2) \* \z3
          - 6 \* \Ss(2,-3)
          - {5 \over 2} \* \Ss(2,-2)
          + 6 \* \Sss(2,-2,1)
          + {49 \over 4} \* \Ss(2,1)
          + 6 \* \Sss(2,1,-2)
          - 6 \* \Sss(2,1,1)
          + 3 \* \Sss(2,1,2)
          - \Sss(2,2,1)
  \nonumber\\&& \mbox{}
          + 2 \* \Ssss(2,1,1,1)
          + {49 \over 4} \* \Ss(2,2)
          - 3 \* \Ss(2,3)
          - {551 \over 72} \* \S(3)
          + {173 \over 12} \* \Ss(3,1)
          - 2 \* \Sss(3,1,1)
          - {79 \over 6} \* \S(4)
          + 2 \* \Ss(4,1)
	\bigg]
	 + (\Nminustwo-1) \* \bigg[
            {55 \over 12} \* \S(1)
  \nonumber\\&& \mbox{}
          - 4 \* \S(1) \* \z3
          - {371 \over 108} \* \Ss(1,1)
          + {23 \over 9} \* \Sss(1,1,1)
          - {2 \over 3} \* \Ssss(1,1,1,1)
          + {4 \over 3} \* \Sss(1,1,2)
          - {23 \over 9} \* \Ss(1,2)
          + {2 \over 3} \* \Ss(1,3)
	\bigg]
	 + (\Nminus-\Nplus) \* \bigg[
            {8543 \over 192}  \* \S(1)
  \nonumber\\&& \mbox{}
          - {71 \over 2}  \* \S(1) \* \z3
          - \Ss(1,-3)
          + 23  \* \Ss(1,-2)
          - {9 \over 2}  \* \Sss(1,-2,1)
          + {1301 \over 216}  \* \Ss(1,1)
          + {13 \over 2}  \* \Sss(1,1,-2)
          + {109 \over 18}  \* \Sss(1,1,1)
          - {5 \over 2}  \* \Sss(1,2,1)
          + 4  \* \Ss(3,2)
  \nonumber\\&& \mbox{}
          + {55 \over 6}  \* \Ss(1,3)
          + {23 \over 6}  \* \Ssss(1,1,1,1)
          + {4 \over 3}  \* \Sss(1,1,2)
          - {235 \over 72}  \* \Ss(1,2)
          + {55 \over 8}  \* \S(2)
          + 9  \* \S(2) \* \z3
          - {21 \over 2}  \* \Ss(2,-2)
          - {269 \over 36}  \* \Ss(2,1)
          - 4  \* \Sss(2,1,-2)
  \nonumber\\&& \mbox{}
          + 2  \* \Ss(2,-3)
          + {83 \over 12}  \* \Sss(2,1,1)
          + {3 \over 2}  \* \Ssss(2,1,1,1)
          - 3  \* \Sss(2,1,2)
          - {41 \over 4}  \* \Ss(2,2)
          + \Sss(2,2,1)
          - {5 \over 2}  \* \Ss(2,3)
          - {55 \over 48}  \* \S(3)
          + 3  \* \Ss(3,-2)
          - {143 \over 12}  \* \Ss(3,1)
  \nonumber\\&& \mbox{}
          - 2  \* \Sss(3,1,1)
          + {49 \over 4}  \* \S(4)
          + 4  \* \Ss(4,1)
          - 2  \* \S(5)
	\bigg]
	 + (1-\Nplus) \* \bigg[
            {145 \over 2} \* \S(1) \* \z3
          - {3571 \over 64} \* \S(1)
          + 2 \* \Ss(1,-3)
          - {58 \over 3} \* \Ss(1,3)
          - {25 \over 9} \* \Sss(1,1,1)
  \nonumber\\&& \mbox{}
          + {23 \over 2} \* \Sss(1,-2,1)
          + {335 \over 216} \* \Ss(1,1)
          - {31 \over 2} \* \Sss(1,1,-2)
          - {11 \over 3} \* \Ssss(1,1,1,1)
          - {5 \over 3} \* \Sss(1,1,2)
          + {245 \over 72} \* \Ss(1,2)
          + {3 \over 2} \* \Ssss(2,1,1,1)
          + 8 \* \Ss(4,1)
          - 2 \* \S(5)
  \nonumber\\&& \mbox{}
          + {1 \over 2} \* \Sss(1,2,1)
          - {83 \over 2} \* \Ss(1,-2)
          + 27 \* \S(2) \* \z3
          - 8 \* \Ss(2,-3)
          + {3 \over 2} \* \Ss(2,-2)
          + 8 \* \Sss(2,-2,1)
          - {183 \over 4} \* \S(4)
          + 8 \* \Sss(2,1,-2)
          - {117 \over 4} \* \Sss(2,1,1)
  \nonumber\\&& \mbox{}
          - 3 \* \Sss(2,1,2)
          + {157 \over 4} \* \Ss(2,2)
          - 3 \* \Sss(2,2,1)
          - {9 \over 2} \* \Ss(2,3)
          - {581 \over 16} \* \S(3)
          - \Ss(3,-2)
          + {237 \over 4} \* \Ss(3,1)
          - 8 \* \Sss(3,1,1)
          + 8 \* \Ss(3,2)
          + {73 \over 3} \* \Ss(2,1)
  \nonumber\\&& \mbox{}
          - {4319 \over 48} \* \S(2)
	\bigg]
          \bigg)
       +16\,  \*   \colour4colour{\ca \* \nf^2}  \*  \bigg(
	  {1 \over 6} \* (\Nminus+4\*\Nplus-2\*\Nplustwo-3) \* \bigg[
            {175 \over 27} \* \S(1)
          - 2 \* \Ss(1,-3)
          + {7 \over 3} \* \Ss(1,-2)
          - {7 \over 9} \* \Ss(1,1)
          + {4 \over 3} \* \S(3)
  \nonumber\\&& \mbox{}
          + {7 \over 3} \* \Sss(1,1,1)
          - \Ssss(1,1,1,1)
          + \Sss(1,1,2)
          - \Sss(1,2,1)
          - \Ss(1,3)
          + {229 \over 18} \* \S(2)
	\bigg]
	+ {1 \over 6} \* (\Nminus-1) \* \bigg[
            \Ss(1,-2)
          - {4 \over 3} \* \Ss(1,1)
          + \Sss(1,1,1)
	\bigg]
  \nonumber\\&& \mbox{}
          - {53 \over 162} \* (\Nminustwo-1) \* \S(1)
	- (\Nminus-\Nplus) \* \bigg[
            {149 \over 648} \* \S(1)
          + {7 \over 4} \* \S(2)
          - {2 \over 9} \* \S(3)
          - {1 \over 3} \* \S(4)
	\bigg]
	- (1-\Nplus) \* \bigg[
            {473 \over 648} \* \S(1)
          - {169 \over 36} \* \S(2)
  \nonumber\\&& \mbox{}
          + {1 \over 6} \* \Ss(2,1)
          - {43 \over 18} \* \S(3)
          + {5 \over 3} \* \S(4)
	\bigg]
          \bigg)
       + 16\,  \*   \colour4colour{\ca^2 \* \nf}  \*   \bigg(
	  (\Nminus+4\*\Nplus-2\*\Nplustwo-3) \* \bigg[
            {3220 \over 27} \* \S(1)
          - {3 \over 2} \* \Ss(1,-4)
          + {277 \over 12} \* \Ss(1,-2)
  \nonumber\\&& \mbox{}
          - {31 \over 2} \* \S(1) \* \z3
          + {61 \over 6} \* \Ss(1,-3)
          + 2 \* \Sss(1,-3,1)
          + 3 \* \Sss(1,-2,-2)
          - {8 \over 3} \* \Sss(1,-2,1)
          + 2 \* \Ssss(1,-2,1,1)
          - 2 \* \Ssss(1,1,-2,1)
          + 6 \* \Ssss(1,1,1,-2)
  \nonumber\\&& \mbox{}
          - {95 \over 54} \* \Ss(1,1)
          - 3 \* \Ss(1,1) \* \z3
          + 2 \* \Sss(1,1,-3)
          + {20 \over 3} \* \Sss(1,1,-2)
          + {47 \over 8} \* \Sss(1,1,1)
          + {4 \over 3} \* \Ssss(1,1,1,1)
          + 2 \* \Sssss(1,1,1,1,1)
          - \Sss(1,1,3)
          + {37 \over 6} \* \Ss(1,3)
  \nonumber\\&& \mbox{}
          + 4 \* \Ssss(1,1,1,2)
          + {21 \over 4} \* \Sss(1,1,2)
          + 2 \* \Ssss(1,1,2,1)
          + {69 \over 8} \* \Ss(1,2)
          - \Sss(1,2,-2)
          + {23 \over 12} \* \Sss(1,2,1)
          - 3 \* \Ss(4,1)
          + 2 \* \Ss(2,3)
          - {5 \over 2} \* \Ss(1,4)
          + 95 \* \S(2)
  \nonumber\\&& \mbox{}
          - 3 \* \S(2) \* \z3
          - \Ss(2,-3)
          + {25 \over 2} \* \Ss(2,-2)
          + 2 \* \Sss(2,-2,1)
          - {155 \over 72} \* \Ss(2,1)
          + {53 \over 6} \* \Sss(2,1,1)
          + 3 \* \Sss(1,3,1)
          - {5 \over 12} \* \Ss(2,2)
          + {31 \over 12} \* \Ss(3,1)
          - 3 \* \S(4)
  \nonumber\\&& \mbox{}
          + {2561 \over 72} \* \S(3)
          - 2 \* \Sss(1,2,2)
	\bigg]
	+ (\Nminustwo-1) \* \bigg[
            4 \* \S(1) \* \z3
          - {2351 \over 108} \* \S(1)
          - {8 \over 3} \* \Ss(1,-3)
          - {4 \over 3} \* \Sss(1,1,2)
          - {52 \over 9} \* \Ss(1,-2)
          + {4 \over 3} \* \Sss(1,-2,1)
  \nonumber\\&& \mbox{}
          + {161 \over 36} \* \Ss(1,1)
          - {4 \over 3} \* \Sss(1,1,-2)
          - {10 \over 9} \* \Sss(1,1,1)
          + {2 \over 3} \* \Ssss(1,1,1,1)
          - {3 \over 2} \* \Ss(1,2)
          + {56 \over 27} \* \S(2)
          - {20 \over 9} \* \Ss(2,1)
          - 2 \* \Ss(1,3)
          - {2 \over 3} \* \Sss(2,1,1)
	\bigg]
  \nonumber\\&& \mbox{}
          - (\Nminus-1) \* \Sss(1,2,1)
	+ (\Nminus-\Nplus) \* \bigg[
            22 \* \S(1) \* \z3
          - {1759 \over 24} \* \S(1)
          - {13 \over 6} \* \Ss(1,-3)
          - {799 \over 36} \* \Ss(1,-2)
          - {8 \over 3} \* \Sss(1,-2,1)
          - {21 \over 2} \* \Ss(1,3)
  \nonumber\\&& \mbox{}
          - {37 \over 3} \* \Sss(1,1,-2)
          - {425 \over 72} \* \Sss(1,1,1)
          - {7 \over 12} \* \Ssss(1,1,1,1)
          - {35 \over 6} \* \Sss(1,1,2)
          - {217 \over 24} \* \Ss(1,2)
          - {1385 \over 18} \* \S(2)
          + {593 \over 36} \* \Ss(1,1)
          - {49 \over 6} \* \Sss(2,1,1)
  \nonumber\\&& \mbox{}
          + {5 \over 2} \* \Ss(2,-3)
          - 8 \* \Ss(2,-2)
          - {209 \over 24} \* \Ss(2,1)
          + 3 \* \Sss(2,1,-2)
          - \Ssss(2,1,1,1)
          + 2 \* \Sss(2,1,2)
          + {17 \over 12} \* \Ss(2,2)
          - 6 \* \S(2) \* \z3
          + {13 \over 4} \* \Ss(2,3)
          + {9 \over 4} \* \Ss(4,1)
  \nonumber\\&& \mbox{}
          - {1363 \over 72} \* \S(3)
          + {9 \over 2} \* \Ss(3,-2)
          + {1 \over 6} \* \Ss(3,1)
          + 3 \* \Sss(3,1,1)
          + {25 \over 6} \* \S(4)
          + 4 \* \S(5)
	\bigg]
	+ (1-\Nplus) \* \bigg[
            {15 \over 4} \* \Ss(2,2)
          + {1783 \over 24} \* \S(1)
          - 41 \* \S(1) \* \z3
  \nonumber\\&& \mbox{}
          + {4 \over 3} \* \Ss(1,-3)
          + {995 \over 36} \* \Ss(1,-2)
          + {16 \over 3} \* \Sss(1,-2,1)
          - {2731 \over 72} \* \Ss(1,1)
          + {62 \over 3} \* \Sss(1,1,-2)
          + {319 \over 72} \* \Sss(1,1,1)
          - {7 \over 12} \* \Ssss(1,1,1,1)
          + {49 \over 6} \* \Sss(1,1,2)
  \nonumber\\&& \mbox{}
          + {287 \over 24} \* \Ss(1,2)
          + {79 \over 4} \* \Ss(1,3)
          + {73141 \over 216} \* \S(2)
          - 24 \* \S(2) \* \z3
          + {17 \over 2} \* \Ss(2,-3)
          + {93 \over 2} \* \Ss(2,-2)
          - {1567 \over 72} \* \Ss(2,1)
          - {34 \over 3} \* \S(4)
          - {15 \over 4} \* \Ss(4,1)
  \nonumber\\&& \mbox{}
          + 7 \* \Sss(2,1,-2)
          + {167 \over 6} \* \Sss(2,1,1)
          - 3 \* \Ssss(2,1,1,1)
          + 6 \* \Sss(2,1,2)
          + {53 \over 4} \* \Ss(2,3)
          + {7385 \over 72} \* \S(3)
          - {7 \over 2} \* \Ss(3,-2)
          + {47 \over 4} \* \Ss(3,1)
          + 5 \* \Sss(3,1,1)
  \nonumber\\&& \mbox{}
          - 19 \* \S(5)
	\bigg]
          \bigg)
       +16\,  \*   \colour4colour{\cf \* \nf^2}  \*  \bigg(
	  (\Nminus+4\*\Nplus-2\*\Nplustwo-3) \* \bigg[
            {2303 \over 324} \* \S(1)
          + {7 \over 54} \* \Ss(1,1)
          - {7 \over 18} \* \Sss(1,1,1)
          - {1 \over 6} \* \Sss(2,1,1)
          - \S(4)
  \nonumber\\&& \mbox{}
          + {4 \over 9} \* \Ss(1,2)
          + {1 \over 6} \* \Ssss(1,1,1,1)
          - {1 \over 3} \* \Ss(1,3)
          + {35 \over 18} \* \S(2)
          + {7 \over 18} \* \Ss(2,1)
          - {11 \over 9} \* \S(3)
	\bigg]
	- {1 \over 6} \* (\Nminus-1) \* \bigg[
            \Sss(1,1,1)
          + \Ss(1,2)
          - \Ss(2,1)
	\bigg]
  \nonumber\\&& \mbox{}
	- (\Nminus-\Nplus) \* \bigg[
            {59963 \over 2592} \* \S(1)
          - {7 \over 18} \* \Ss(1,1)
          - {251 \over 27} \* \S(2)
          + {199 \over 24} \* \S(3)
          - {25 \over 6} \* \S(4)
          + 2 \* \S(5)
	\bigg]
	+ (1-\Nplus) \* \bigg[
            {163 \over 24} \* \S(2)
          + 6 \* \S(5)
  \nonumber\\&& \mbox{}
          + {96277 \over 2592} \* \S(1)
          - {17 \over 36} \* \Ss(1,1)
          - {7 \over 24} \* \S(3)
          - {19 \over 2} \* \S(4)
	\bigg]
          + {77 \over 81} \* (\Nminustwo-1) \* \S(1)
          \bigg)
       +16\,  \*   \colour4colour{\cf^2 \* \nf}  \*  \bigg(
	  (\Nminus-1) \* \bigg[
            4 \* \Sss(2,1,-2)
  \nonumber\\&& \mbox{}
          + {1 \over 2} \* \Ss(2,2)
	\bigg]
	 + (\Nminus+4\*\Nplus-2\*\Nplustwo-3) \* \bigg[
            {81 \over 32} \* \S(1)
          - \Ss(1,-4)
          + 5 \* \Ss(1,-3)
          - {5 \over 2} \* \Ss(1,-2)
          + 2 \* \Sss(1,-2,-2)
          + 4 \* \Sss(1,1,-3)
  \nonumber\\&& \mbox{}
          + {87 \over 8} \* \Ss(1,1)
          - 4 \* \Sss(1,1,-2)
          + {61 \over 8} \* \Sss(1,1,1)
          + 3 \* \Ssss(1,1,1,1)
          + 2 \* \Sssss(1,1,1,1,1)
          - \Ssss(1,1,1,2)
          - {5 \over 2} \* \Sss(1,1,2)
          + 7 \* \Sss(1,3,1)
          - 3 \* \Ss(1,4)
  \nonumber\\&& \mbox{}
          - 5 \* \Ssss(1,1,2,1)
          + 4 \* \Sss(1,1,3)
          - {17 \over 2} \* \Ss(1,2)
          + 2 \* \Sss(1,2,-2)
          - {11 \over 2} \* \Sss(1,2,1)
          - 6 \* \Ssss(1,2,1,1)
          + 6 \* \Sss(1,2,2)
          + {5 \over 2} \* \Ss(1,3)
          - {87 \over 8} \* \S(2)
          + 4 \* \S(5)
  \nonumber\\&& \mbox{}
          - 4 \* \Ss(2,-3)
          + 4 \* \Ss(2,-2)
          - {61 \over 8} \* \Ss(2,1)
          - 3 \* \Sss(2,1,1)
          - 2 \* \Ssss(2,1,1,1)
          + \Sss(2,1,2)
          + {5 \over 2} \* \Ss(2,2)
          + 5 \* \Sss(2,2,1)
          - 4 \* \Ss(2,3)
          + 6 \* \Sss(3,1,1)
  \nonumber\\&& \mbox{}
          + 11 \* \S(3)
          - 4 \* \Ss(3,-2)
          + {11 \over 2} \* \Ss(3,1)
          - 6 \* \Ss(3,2)
          - {15 \over 2} \* \S(4)
          - 7 \* \Ss(4,1)
	\bigg]
	+ (\Nminus-\Nplus) \* \bigg[
            {801 \over 64} \* \S(1)
          + {27 \over 2} \* \S(1) \* \z3
          - {3 \over 2} \* \Ss(1,2)
  \nonumber\\&& \mbox{}
          + 3 \* \Ss(1,-3)
          - {35 \over 2} \* \Ss(1,-2)
          - {103 \over 8} \* \Ss(1,1)
          - 4 \* \Sss(1,1,-2)
          - {7 \over 8} \* \Sss(1,1,1)
          - {13 \over 4} \* \Ssss(1,1,1,1)
          + {9 \over 2} \* \Sss(1,1,2)
          + {7 \over 2} \* \Sss(1,2,1)
          - {1 \over 2} \* \Ssss(2,1,1,1)
  \nonumber\\&& \mbox{}
          - {9 \over 2} \* \Ss(1,3)
          + {1 \over 4} \* \S(2)
          - 3 \* \S(2) \* \z3
          + 7 \* \Ss(2,-2)
          + {27 \over 8} \* \Ss(2,1)
          + {3 \over 4} \* \Sss(2,1,1)
          + \Sss(2,1,2)
          - \Sss(2,2,1)
          + 3 \* \Ss(2,3)
          - {87 \over 16} \* \S(3)
          - \Sss(3,1,1)
  \nonumber\\&& \mbox{}
          - {13 \over 4} \* \Ss(3,1)
          + 2 \* \Ss(3,2)
          + {27 \over 4} \* \S(4)
          + {7 \over 2} \* \Ss(4,1)
          - 3 \* \S(5)
	\bigg]
	+ (1-\Nplus) \* \bigg[
            17 \* \Ss(1,1)
          - {1759 \over 64} \* \S(1)
          - {63 \over 2} \* \S(1) \* \z3
          + {17 \over 4} \* \Ssss(1,1,1,1)
  \nonumber\\&& \mbox{}
          - 11 \* \Ss(1,-3)
          + {71 \over 2} \* \Ss(1,-2)
          + 12 \* \Sss(1,1,-2)
          - {19 \over 8} \* \Sss(1,1,1)
          - {13 \over 2} \* \Sss(1,1,2)
          + {13 \over 2} \* \Ss(1,2)
          - {3 \over 2} \* \Sss(1,2,1)
          + {13 \over 2} \* \Ss(1,3)
          - 3 \* \S(2) \* \z3
  \nonumber\\&& \mbox{}
          - {409 \over 16} \* \S(2)
          - 4 \* \Ss(2,-3)
          - \Ss(2,-2)
          + {59 \over 8} \* \Ss(2,1)
          - {1 \over 4} \* \Sss(2,1,1)
          + {3 \over 2} \* \Ssss(2,1,1,1)
          - 3 \* \Sss(2,1,2)
          + 3 \* \Sss(2,2,1)
          - 5 \* \Ss(2,3)
          + {565 \over 16} \* \S(3)
  \nonumber\\&& \mbox{}
          - 8 \* \Ss(3,-2)
          + {17 \over 4} \* \Ss(3,1)
          + 3 \* \Sss(3,1,1)
          - 6 \* \Ss(3,2)
          - {103 \over 4} \* \S(4)
          - {21 \over 2} \* \Ss(4,1)
          + 11 \* \S(5)
	\bigg]
          \bigg)
\label{eq:gqg2}
\eea
and
\bea
  &&\gamma^{\,(2)}_{\,\rm gq}(N) \:\: = \:\:
        16\,  \*   \colour4colour{\ca \* \cf \* \nf}  \*  \bigg(
	  (2\*\Nminustwo-4\*\Nminus-\Nplus+3) \* \bigg[
            {967 \over 144} \* \S(1)
          - 2 \* \S(1) \* \z3
          + {2 \over 3} \* \Ss(1,-3)
          + {41 \over 18} \* \Ss(1,-2)
          - {1 \over 3} \* \Ss(1,3)
  \nonumber\\&& \mbox{}
          - {2 \over 3} \* \Sss(1,-2,1)
          + {251 \over 108} \* \Ss(1,1)
          - {4 \over 3} \* \Sss(1,1,-2)
          - {13 \over 4} \* \Sss(1,1,1)
          + {5 \over 6} \* \Ssss(1,1,1,1)
          - {5 \over 6} \* \Sss(1,1,2)
          + {10 \over 9} \* \Ss(1,2)
          - {5 \over 6} \* \Sss(1,2,1)
          - {151 \over 108} \* \S(2)
  \nonumber\\&& \mbox{}
          - {1 \over 3} \* \Ss(2,-2)
          + {10 \over 9} \* \Ss(2,1)
          - {5 \over 6} \* \Sss(2,1,1)
          + {1 \over 3} \* \Ss(2,2)
	\bigg]
	+ (\Nminus-\Nplus) \* \bigg[
            {331 \over 72} \* \S(1)
          - 4 \* \Ss(2,-2)
          + {28 \over 9} \* \Ss(1,-2)
          - {11 \over 18} \* \Sss(1,1,1)
  \nonumber\\&& \mbox{}
          + {4 \over 3} \* \Ss(3,1)
          - {2 \over 9} \* \Ss(2,1)
          + {53 \over 54} \* \Ss(1,1)
          - {733 \over 54} \* \S(2)
          + {4 \over 3} \* \Sss(2,1,1)
          - {22 \over 3} \* \S(3)
	\bigg]
	+ (1-\Nplus) \* \bigg[
            {10 \over 3} \* \Ss(2,-2)
          + {1 \over 12} \* \Ss(2,1)
          - {1 \over 4} \* \Ss(1,1)
  \nonumber\\&& \mbox{}
          - {17 \over 3} \* \Ss(1,-2)
          - {137 \over 144} \* \S(1)
          + {5 \over 6} \* \Ss(1,2)
          + {1 \over 4} \* \Sss(1,1,1)
          + {565 \over 36} \* \S(2)
          - \Sss(2,1,1)
          + {35 \over 12} \* \S(3)
          - {2 \over 3} \* \Ss(3,1)
	\bigg]
	- {2 \over 9} \* (\Nminus-\Nplustwo) \* \bigg[
            \S(3)
  \nonumber\\&& \mbox{}
          - 3 \*\Ss(2,1)
          + {131 \over 4} \* \S(1)
          + \Ss(1,-2)
          - {25 \over 6} \* \Ss(1,1)
          - \Sss(1,1,1)
          + {125 \over 6} \* \S(2)
	\bigg]
          - {2 \over 3} \* (\Nminus-1) \* \S(4)
          \bigg)
       +16\,  \*   \colour4colour{\ca \* \cf^2}  \*  \bigg(
	  (2\*\Nminustwo
  \nonumber\\&& \mbox{}
-4\*\Nminus-\Nplus+3) \* \bigg[
            {163 \over 32} \* \S(1)
          - {3 \over 2} \* \Ss(1,-4)
          - {3 \over 2} \* \Ss(1,-3)
          + {6503 \over 432} \* \Ss(1,1)
          - 5 \* \Sss(1,-2,-2)
          - 3 \* \Sss(1,-2,1)
          - 4 \* \Sssss(1,1,1,1,1)
  \nonumber\\&& \mbox{}
          + \Ss(1,-2)
          + 2 \* \Ssss(1,-2,1,1)
          - 9 \* \Ss(1,1) \* \z3
          - 4 \* \Sss(1,1,-3)
          + 3 \* \Sss(1,1,-2)
          + 2 \* \Ssss(1,1,-2,1)
          + 5 \* \Sss(1,1,3)
          + 6 \* \Ssss(1,1,1,-2)
          + \Ssss(1,1,2,1)
  \nonumber\\&& \mbox{}
          + 3 \* \Ssss(1,1,1,2)
          + {35 \over 3} \* \Ssss(1,1,1,1)
          + {2 \over 9} \* \Sss(1,1,1)
          - {1 \over 12} \* \Sss(1,1,2)
          - {191 \over 24} \* \Ss(1,2)
          - 3 \* \Sss(1,2,-2)
          - {41 \over 12} \* \Sss(1,2,1)
          + 4 \* \Ss(1,3)
          - 4 \* \Ss(2,1)
  \nonumber\\&& \mbox{}
          + 2 \* \Ssss(1,2,1,1)
          - {5 \over 2} \* \Ss(1,4)
          - {9 \over 2} \* \Sss(2,1,1)
          + 2 \* \Ssss(2,1,1,1)
          + \Sss(2,1,2)
          + 3 \* \Ss(2,2)
          + \Sss(2,2,1)
          - 2 \* \Ss(2,3)
	\bigg]
	  + (\Nminus-\Nplustwo) \* \bigg[
            6 \* \Ss(2,1)
  \nonumber\\&& \mbox{}
          + {173 \over 54} \* \Ss(1,1)
          - {26 \over 9} \* \Sss(1,1,1)
          - {2 \over 3} \* \Ssss(1,1,1,1)
          - {335 \over 54} \* \S(2)
          + {7 \over 2} \* \S(1)
          - 2 \* \Sss(2,1,1)
          - {28 \over 9} \* \S(3)
          + {8 \over 3} \* \S(4)
	\bigg]
	  - 6 \* (\Nminus-1) \* \bigg[
            \Ss(2,-3)
  \nonumber\\&& \mbox{}
          - 2 \* \Sss(2,1,-2)
          + 3 \* \S(2) \* \z3
	\bigg]
	+ (\Nminus-\Nplus) \* \bigg[
            36 \* \S(1) \* \z3
          - {9703 \over 288} \* \S(1)
          + 12 \* \Ss(1,-3)
          - 36 \* \Ss(1,-2)
          - {2263 \over 216} \* \Ss(1,1)
          + 4 \* \Ss(3,2)
  \nonumber\\&& \mbox{}
          - 16 \* \Ss(1,3)
          - 24 \* \Sss(1,1,-2)
          - {101 \over 36} \* \Sss(1,1,1)
          + {5 \over 6} \* \Ssss(1,1,1,1)
          - {23 \over 12} \* \Ss(1,2)
          + 2 \* \Sss(1,2,1)
          + {12605 \over 432} \* \S(2)
          + 36 \* \Ss(2,-2)
          + {79 \over 6} \* \S(4)
  \nonumber\\&& \mbox{}
          + {55 \over 18} \* \Ss(2,1)
          - {10 \over 3} \* \Sss(2,1,1)
          - 3 \* \Ssss(2,1,1,1)
          + {17 \over 3} \* \Ss(2,2)
          - 2 \* \Sss(2,2,1)
          - {119 \over 8} \* \S(3)
          - 14 \* \Ss(3,-2)
          + {47 \over 3} \* \Ss(3,1)
          - 7 \* \Sss(3,1,1)
          + 4 \* \S(5)
  \nonumber\\&& \mbox{}
          + 10 \* \Ss(2,3)
	\bigg]
	+ (1-\Nplus) \* \bigg[
            {2005 \over 64} \* \S(1)
          - {117 \over 2} \* \S(1) \* \z3
          - {39 \over 2} \* \Ss(1,-3)
          + {315 \over 4} \* \Ss(1,-2)
          - \Sss(1,-2,1)
          + 3 \* \Ssss(1,1,1,1)
          - 2 \* \S(4,1)
  \nonumber\\&& \mbox{}
          + {2525 \over 144} \* \Ss(1,1)
          + 40 \* \Sss(1,1,-2)
          - {55 \over 12} \* \Sss(1,1,1)
          - 3 \* \Sss(1,1,2)
          + {197 \over 24} \* \Ss(1,2)
          - {11 \over 2} \* \Sss(1,2,1)
          + {53 \over 2} \* \Ss(1,3)
          + {13 \over 2} \* \Sss(3,1,1)
          - 4 \* \Ss(2,2)
  \nonumber\\&& \mbox{}
          - {2831 \over 72} \* \S(2)
          - 37 \* \Ss(2,-2)
          + 13 \* \Ss(3,-2)
          + {1 \over 2} \* \Sss(2,1,1)
          + {3 \over 2} \* \Ssss(2,1,1,1)
          - {15 \over 2} \* \Ss(3,1)
          + 3 \* \Ss(2,2,1)
          - 12 \* \Ss(2,3)
          + {2407 \over 48} \* \S(3)
  \nonumber\\&& \mbox{}
          + {3 \over 2} \* \Ss(2,1)
          - 6 \* \Ss(3,2)
          - {57 \over 2} \* \S(4)
	\bigg]
          \bigg)
       +16\,  \*  \colour4colour{\ca^2 \* \cf}  \*  \bigg(
	  (2\*\Nminustwo-4\*\Nminus-\Nplus+3) \* \bigg[
            {138305 \over 2592} \* \S(1)
          - 2 \* \Ssss(1,-2,1,1)
  \nonumber\\&& \mbox{}
          - {11 \over 2} \* \Ss(1,-4)
          + {49 \over 6} \* \Ss(1,-3)
          + \Sss(1,-2,-2)
          - 10 \* \Ssss(1,1,-2,1)
          + {109 \over 12} \* \Ss(1,-2)
          - {3 \over 2} \* \Sss(1,-2,1)
          + 2 \* \Sss(1,-2,2)
          - {3379 \over 216} \* \Ss(1,1)
  \nonumber\\&& \mbox{}
          + 8 \* \Sss(1,-3,1)
          + 3 \* \Ss(1,1) \* \z3
          + 12 \* \Sss(1,1,-3)
          + {19 \over 2} \* \Sss(1,1,-2)
          + 2 \* \Sssss(1,1,1,1,1)
          + {65 \over 24} \* \Sss(1,1,1)
          - 6 \* \Ssss(1,1,1,-2)
          - {43 \over 6} \* \Ssss(1,1,1,1)
  \nonumber\\&& \mbox{}
          - 4 \* \Ssss(1,1,1,2)
          + {55 \over 12} \* \Sss(1,1,2)
          - 4 \* \Ssss(1,1,2,1)
          + 2 \* \Sss(1,1,3)
          + {71 \over 24} \* \Ss(1,2)
          + 5 \* \Sss(1,2,-2)
          + {55 \over 12} \* \Sss(1,2,1)
          - 4 \* \Ssss(1,2,1,1)
          + 6 \* \Sss(1,2,2)
  \nonumber\\&& \mbox{}
          + {11 \over 2} \* \Ss(1,3)
          + 4 \* \Sss(1,3,1)
          - {3 \over 2} \* \Ss(1,4)
          - {395 \over 54} \* \S(2)
          - 7 \* \Ss(2,-3)
          - {11 \over 6} \* \Ss(2,-2)
          + 4 \* \Sss(2,-2,1)
          + 2 \* \Sss(2,1,-2)
          - 2 \* \Ssss(2,1,1,1)
  \nonumber\\&& \mbox{}
          + {17 \over 3} \* \Sss(2,1,1)
          + 3 \* \Sss(2,1,2)
          - {1 \over 3} \* \Ss(2,2)
          + 3 \* \Sss(2,2,1)
          - 3 \* \Ss(2,3)
          + 4 \* \Sss(3,1,1)
          - 4 \* \Ss(3,2)
	\bigg]
	+ (\Nminus-1) \* \bigg[
            6 \* \S(2) \* \z3
          - 8 \* \Sss(2,-2,1)
	\bigg]
  \nonumber\\&& \mbox{}
	+ (\Nminus-\Nplus) \* \bigg[
            {57595 \over 1296} \* \S(1)
          - 12 \* \S(1) \* \z3
          - {31 \over 6} \* \Ss(1,-3)
          - {143 \over 6} \* \Ss(2,-2)
          + {25 \over 3} \* \Sss(1,-2,1)
          - {689 \over 54} \* \Ss(1,1)
          + {50 \over 3} \* \Sss(1,1,-2)
  \nonumber\\&& \mbox{}
          + {11 \over 18} \* \Sss(1,1,1)
          - {11 \over 6} \* \Ssss(1,1,1,1)
          + {229 \over 36} \* \Ss(1,2)
          + {113 \over 12} \* \Ss(1,3)
          - {2200 \over 27} \* \S(2)
          - 3 \* \Ss(2,-3)
          - 12 \* \Ss(3,2)
          + 9 \* \Ss(1,-2)
          + {31 \over 2} \* \Ss(2,1)
  \nonumber\\&& \mbox{}
          - 18 \* \Sss(2,1,-2)
          + {13 \over 6} \* \Sss(2,1,1)
          + 4 \* \Ssss(2,1,1,1)
          - {37 \over 3} \* \Ss(2,2)
          - {25 \over 2} \* \Ss(2,3)
          - 31 \* \S(3)
          - 9 \* \Ss(3,-2)
          - {463 \over 12} \* \Ss(3,1)
          + 4 \* \Sss(3,1,1)
          + \S(4)
  \nonumber\\&& \mbox{}
          - {13 \over 2} \* \Ss(4,1)
          - 8 \* \S(5)
	\bigg]
	+ (\Nminus-\Nplustwo) \* \bigg[
            {4 \over 3} \* \Sss(1,-2,1)
          - {2105 \over 81} \* \S(1)
          - {8 \over 3} \* \Ss(1,-3)
          - 10 \* \Ss(1,-2)
          - {109 \over 27} \* \Ss(1,1)
          - {4 \over 3} \* \Sss(1,1,-2)
  \nonumber\\&& \mbox{}
          + {37 \over 9} \* \Sss(1,1,1)
          + {2 \over 3} \* \Ssss(1,1,1,1)
          - {145 \over 18} \* \Ss(1,2)
          - {4 \over 3} \* \Ss(1,3)
          - {584 \over 27} \* \S(2)
          - 4 \* \Ss(2,-2)
          - {104 \over 9} \* \Ss(2,1)
          + {8 \over 3} \* \Sss(2,1,1)
          - {14 \over 3} \* \Ss(2,2)
  \nonumber\\&& \mbox{}
          - {77 \over 18} \* \S(3)
          - 6 \* \Ss(3,1)
          + {14 \over 3} \* \S(4)
	\bigg]
	+ (1-\Nplus) \* \bigg[
            {39 \over 2} \* \S(1) \* \z3
          - {29843 \over 864} \* \S(1)
          + {17 \over 2} \* \Ss(3,-2)
          + {145 \over 6} \* \Ss(3,1)
          - {29 \over 2} \* \Sss(1,-2,1)
  \nonumber\\&& \mbox{}
          - {25 \over 2} \* \Ss(1,-2)
          - {57 \over 2} \* \Sss(1,1,-2)
          - {13 \over 12} \* \Sss(1,1,1)
          + {5 \over 4} \* \Ssss(1,1,1,1)
          + 4 \* \Sss(1,1,2)
          - {97 \over 24} \* \Ss(1,2)
          + 4 \* \Sss(1,2,1)
          - {41 \over 2} \* \Ss(1,3)
          + {7417 \over 72} \* \S(2)
  \nonumber\\&& \mbox{}
          + {1 \over 2} \* \Ss(2,-3)
          + {92 \over 3} \* \Ss(2,-2)
          - {53 \over 12} \* \Ss(2,1)
          + 15 \* \Sss(2,1,-2)
          - {9 \over 4} \* \Sss(2,1,1)
          - 3 \* \Ssss(2,1,1,1)
          + 5 \* \Ss(2,2)
          + {1 \over 4} \* \Ss(4,1)
          + 38 \* \S(3)
          + 8 \* \Ss(3,2)
  \nonumber\\&& \mbox{}
          + {41 \over 4} \* \Ss(2,3)
          + {9 \over 2} \* \Ss(1,-3)
          + {92 \over 3} \* \Ss(1,1)
          - 2 \* \Sss(3,1,1)
          + {25 \over 3} \* \S(4)
          + {31 \over 2} \* \S(5)
	\bigg]
          \bigg)
       +16\,  \*  \colour4colour{\cf \* \nf^2}  \*  \bigg(
	 {1 \over 6} \* (1-\Nplus) \* \bigg[
            {5 \over 3} \* \S(1)
          - \Ss(1,1)
	\bigg]
  \nonumber\\&& \mbox{}
	- {1 \over 6} \* (2\*\Nminustwo-4\*\Nminus-\Nplus+3) \* \bigg[
            {1 \over 3} \* \S(1)
          + {5 \over 3} \* \Ss(1,1)
          - \Sss(1,1,1)
	\bigg]
          \bigg)
       +16\,  \*  \colour4colour{\cf^2 \* \nf}  \*  \bigg(
	  (\Nminus-\Nplus) \* \bigg[
            {2 \over 3} \* \Ss(1,2)
          - {371 \over 432} \* \S(1)
  \nonumber\\&& \mbox{}
          - {35 \over 9} \* \Ss(1,-2)
          - {1 \over 9} \* \Ss(1,1)
          - {1 \over 3} \* \Sss(1,1,1)
          + {1057 \over 72} \* \S(2)
          + {16 \over 3} \* \Ss(2,-2)
          - {8 \over 9} \* \Ss(2,1)
          + {1 \over 3} \* \Sss(2,1,1)
          - {2 \over 3} \* \Ss(2,2)
          + {181 \over 12} \* \S(3)
          - {2 \over 3} \* \Ss(3,1)
  \nonumber\\&& \mbox{}
          - {1 \over 3} \* \S(4)
          + 4 \* \S(5)
	\bigg]
	+ (2\*\Nminustwo-4\*\Nminus-\Nplus+3) \* \bigg[
            2 \* \S(1) \* \z3
          - {1 \over 3} \* \Sss(1,2,1)
          - {31 \over 18} \* \Ss(1,-2)
          + {95 \over 54} \* \S(2)
          + {1 \over 2} \* \Ss(1,3)
          + {1 \over 3} \* \Ss(1,2)
  \nonumber\\&& \mbox{}
          - {1625 \over 144} \* \S(1)
          - {5 \over 6} \* \Ssss(1,1,1,1)
          - {2 \over 3} \* \Sss(1,1,2)
          - {7 \over 108} \* \Ss(1,1)
          + {83 \over 36} \* \Sss(1,1,1)
          + {2 \over 3} \* \Ss(2,-2)
	\bigg]
	- {4 \over 9} \* (\Nminus-\Nplustwo) \* \bigg[
            {7 \over 2} \* \S(1)
          - {11 \over 6} \* \S(2)
  \nonumber\\&& \mbox{}
          - \S(1,-2)
          - \S(3)
	\bigg]
	+ (1-\Nplus) \* \bigg[
            {15137 \over 864} \* \S(1)
          + {49 \over 6} \* \Ss(1,-2)
          - {107 \over 36} \* \Ss(1,1)
          + {19 \over 12} \* \Sss(1,1,1)
          - {5 \over 6} \* \Ss(1,2)
          - 10 \* \S(2)
          - 4 \* \Ss(2,-2)
  \nonumber\\&& \mbox{}
          - {1 \over 2} \* \Sss(2,1,1)
          + \Ss(2,2)
          - {155 \over 24} \* \S(3)
          + \Ss(3,1)
          + \S(4)
          - 6 \* \S(5)
	\bigg]
          \bigg)
       +16\,  \*  \colour4colour{\cf^3}  \*  \bigg(
	    (2\*\Nminustwo-4\*\Nminus-\Nplus+3) \* \bigg[
            6 \* \Sss(1,-2,-2)
  \nonumber\\&& \mbox{}
          - {47 \over 16} \* \S(1)
          - \Ss(1,-4)
          - {7 \over 2} \* \Ss(1,-2)
          + 6 \* \Ss(1,-3)
          - {47 \over 16} \* \Ss(1,1)
          + 6 \* \Ss(1,1) \* \z3
          + 4 \* \Sss(1,1,-3)
          - 6 \* \Sss(1,1,-2)
          - 3 \* \Sss(1,1,2)
          - 3 \* \Sss(1,1,3)
  \nonumber\\&& \mbox{}
          - {23 \over 8} \* \Sss(1,1,1)
          - {9 \over 2} \* \Ssss(1,1,1,1)
          + 2 \* \Sssss(1,1,1,1,1)
          + \Ssss(1,1,1,2)
          + 3 \* \Ssss(1,1,2,1)
          + {7 \over 4} \* \Ss(1,2)
          + 2 \* \Sss(1,2,-2)
          + 2 \* \Ssss(1,2,1,1)
          - 2 \* \Sss(1,2,2)
  \nonumber\\&& \mbox{}
          - {3 \over 2} \* \Ss(1,3)
	\bigg]
	 + 2 \* (\Nminus-1) \* \bigg[
            6 \* \S(2) \* \z3
          - 4 \* \Sss(2,1,-2)
          + 8 \* \Ss(3,-2)
	\bigg]
	+ (\Nminus-\Nplus) \* \bigg[
            {287 \over 32} \* \S(1)
          - 24 \* \S(1) \* \z3
          + \Ssss(1,1,1,1)
  \nonumber\\&& \mbox{}
          - 12 \* \Ss(1,-3)
          + 36 \* \Ss(1,-2)
          + {111 \over 8} \* \Ss(1,1)
          + 16 \* \Sss(1,1,-2)
          + {1 \over 4} \* \Sss(1,1,1)
          + {9 \over 2} \* \Ss(1,2)
          - 2 \* \Sss(1,2,1)
          + 9 \* \Ss(1,3)
          - 4 \* \Ss(3,1)
          - 5 \* \Ss(2,3)
  \nonumber\\&& \mbox{}
          + 3 \* \Sss(3,1,1)
          - {91 \over 16} \* \S(2)
          + 8 \* \Ss(2,-3)
          - 30 \* \Ss(2,-2)
          - {41 \over 4} \* \Ss(2,1)
          + \Sss(2,1,1)
          - \Ssss(2,1,1,1)
          + 2 \* \Sss(2,2,1)
          - {35 \over  8} \* \S(3)
          - \S(4)
          + 3 \* \Ss(4,1)
  \nonumber\\&& \mbox{}
          - 2 \* \S(5)
	\bigg]
	+ (1-\Nplus) \* \bigg[
            39 \* \S(1) \* \z3
          - {749 \over 64} \* \S(1)
          + 20 \* \Ss(1,-3)
          - {141 \over 2} \* \Ss(1,-2)
          - {433 \over 16} \* \Ss(1,1)
          + 6 \* \Sss(1,1,1)
          - {17 \over 4} \* \Ssss(1,1,1,1)
  \nonumber\\&& \mbox{}
          - 30 \* \Sss(1,1,-2)
          - \Sss(1,1,2)
          - {19 \over 4} \* \Ss(1,2)
          + {3 \over 2} \* \Sss(1,2,1)
          - {57 \over 4} \* \Ss(1,3)
          + 21 \* \S(2)
          - 10 \* \Ss(2,-3)
          + 35 \* \Ss(2,-2)
          - {9 \over 2} \* \Sss(3,1,1)
          + {37 \over 4} \* \S(4)
  \nonumber\\&& \mbox{}
          + {19 \over 4} \* \Ss(2,1)
          + {9 \over 4} \* \Sss(2,1,1)
          + {3 \over 2} \* \Ssss(2,1,1,1)
          + 3 \* \Ss(2,2)
          - 3 \* \Sss(2,2,1)
          + {11 \over 2} \* \Ss(2,3)
          - {485 \over 16} \* \S(3)
          + {27 \over 4} \* \Ss(3,1)
          - {9 \over 2} \* \Ss(4,1)
	\bigg]
          \bigg)
 \:\: . \label{eq:ggq2}
\eea
Finally the three-loop gluon-gluon anomalous dimension reads
\bea
  &&\gamma^{\,(2)}_{\,\rm gg}(N) \:\: = \:\:
16\,  \*  \colour4colour{\ca \* \cf \* \nf}  \*  \bigg(
            {241 \over 288}
	+ (\Nminustwo-2\*\Nminus-2\*\Nplus+\Nplustwo+3) \* \bigg[
            4 \* \S(1) \* \z3
          - {15331 \over 648} \* \S(1)
          - {44 \over 9} \* \Ss(1,-2)
  \nonumber\\&& \mbox{}
          - {2 \over 3} \* \Ss(1,-3)
          + {4 \over 3} \* \Sss(1,-2,1)
          - {521 \over 108} \* \Ss(1,1)
          - {16 \over 3} \* \Sss(1,1,-2)
          + {1 \over 9} \* \Sss(1,1,1)
          - {4 \over 3} \* \Ssss(1,1,1,1)
          + {4 \over 3} \* \Sss(1,1,2)
          - {17 \over 18} \* \Ss(1,2)
          - {8 \over 3} \* \Ss(1,3)
  \nonumber\\&& \mbox{}
          + {86 \over 27} \* \S(2)
          + {4 \over 3} \* \Ss(2,-2)
          - {2 \over 3} \* \Ss(2,1)
          + {2 \over 3} \* \Sss(2,1,1)
          - {4 \over 3} \* \Ss(2,2)
  	  \bigg]
	+ (\Nminus+\Nplus-2) \* \bigg[
            17 \* \S(1) \* \z3
          + {25 \over 3} \* \Ss(1,-3)
          - {8 \over 3} \* \Sss(1,-2,1)
  \nonumber\\&& \mbox{}
          - {70 \over 3} \* \Sss(1,1,-2)
          + {31 \over 36} \* \Sss(1,1,1)
          - {7 \over 3} \* \Ssss(1,1,1,1)
          + {7 \over 3} \* \Sss(1,1,2)
          - {55 \over 6} \* \Ss(1,3)
  	  \bigg]
	+ (\Nminus-\Nplus) \* \bigg[
            {133 \over 18} \* \Ss(1,2)
          - {221 \over 9} \* \Ss(1,-2)
  \nonumber\\&& \mbox{}
          - {673 \over 54} \* \Ss(1,1)
          + {4948 \over 81} \* \S(1)
          - {49 \over 108} \* \S(2)
          - 12 \* \S(2) \* \z3
          - 4 \* \Ss(2,-3)
          + 17 \* \Ss(2,-2)
          + {119 \over 6} \* \Ss(2,1)
          + 16 \* \Sss(2,1,-2)
          + 6 \* \Ss(4,1)
  \nonumber\\&& \mbox{}
          - {7 \over 6} \* \Sss(2,1,1)
          + 2 \* \Ssss(2,1,1,1)
          - 2 \* \Sss(2,1,2)
          - \Ss(2,2)
          + 7 \* \Ss(2,3)
          + {251 \over 12} \* \S(3)
          - {10 \over 3} \* \Ss(3,1)
          - \Sss(3,1,1)
          + 4 \* \Ss(3,2)
          - {29 \over 6} \* \S(4)
          + 8 \* \S(5)
	\bigg]
  \nonumber\\&& \mbox{}
          - 8 \* (\Nminus-1) \* \S(3,-2)
	+ (\Nminus-\Nplustwo) \* \bigg[
            {127 \over 18} \* \S(3)
          - {511 \over 12} \* \S(1)
          - 6 \* \Ss(1,-2)
          - {97 \over 12} \* \Ss(1,1)
          - 3 \* \Ss(1,2)
          + 2 \* \Ss(3,1)
          - {103 \over 27} \* \S(2)
  \nonumber\\&& \mbox{}
          - {8 \over 3} \* \Ss(2,-2)
          - {16 \over 9} \* \Ss(2,1)
          - {2 \over 3} \* \Ss(2,2)
	\bigg]
	+ (1-\Nplus) \* \bigg[
            {1807 \over 324} \* \S(1)
          + {604 \over 9} \* \Ss(1,-2)
          + {5311 \over 108} \* \Ss(1,1)
          - {52 \over 9} \* \Ss(1,2)
          - {1667 \over 54} \* \S(2)
  \nonumber\\&& \mbox{}
          - {68 \over 3} \* \Ss(2,-2)
          - {53 \over 4} \* \Ss(2,1)
          - {7 \over 3} \* \Sss(2,1,1)
          + {19 \over 6} \* \Ss(2,2)
          + {67 \over 12} \* \S(3)
          + {9 \over 2} \* \Ss(3,1)
          + {33 \over 2} \* \S(4)
          - 20 \* \S(5)
	\bigg]
          + {6923 \over 324} \* \S(1)
          - 2 \* \S(1) \* \z3
  \nonumber\\&& \mbox{}
          + {2 \over 3} \* \Ss(1,-3)
          + {44 \over 9} \* \Ss(1,-2)
          - {4 \over 3} \* \Sss(1,-2,1)
          + {521 \over 108} \* \Ss(1,1)
          + {16 \over 3} \* \Sss(1,1,-2)
          - {1 \over 9} \* \Sss(1,1,1)
          + {4 \over 3} \* \Ssss(1,1,1,1)
          - {4 \over 3} \* \Sss(1,1,2)
          + {8 \over 3} \* \Ss(1,3)
  \nonumber\\&& \mbox{}
          + {17 \over 18} \* \Ss(1,2)
          - {86 \over 27} \* \S(2)
          - {4 \over 3} \* \Ss(2,-2)
          + {2 \over 3} \* \Ss(2,1)
          - {2 \over 3} \* \Sss(2,1,1)
          + {4 \over 3} \* \Ss(2,2)
          \bigg)
       +16\,  \*   \colour4colour{\ca \* \nf^2}  \*  \bigg(
           {11 \over 72} \* (1-\Nplus) \* \S(2)
          - {65 \over 162} \* \S(1)
  \nonumber\\&& \mbox{}
          + {13 \over 54} \* \Ss(1,1)
          - {29 \over 288}
	 + (\Nminustwo-2\*\Nminus-2\*\Nplus+\Nplustwo+3) \* \bigg[
            {59 \over 162} \* \S(1)
          - {13 \over 54} \* \Ss(1,1)
  	  \bigg]
	-  {1 \over 9} \* (\Nminus-\Nplus) \* \bigg[
            \S(2)
  \nonumber\\&& \mbox{}
          - 2 \* \Ss(2,1)
          + \S(3)
  	  \bigg]
	+ (\Nminus+\Nplus-2) \* \bigg[
            {47 \over 648} \* \S(1)
          - {19 \over 216} \* \Ss(1,1)
  	  \bigg]
          - {13 \over 54} \* (\Nminus-\Nplustwo) \* \S(2)
          \bigg)
       +16\,  \*   \colour4colour{\ca^2 \* \nf}  \*  \bigg(
            {233 \over 288}
  \nonumber\\&& \mbox{}
	  + (\Nminustwo-2\*\Nminus-2\*\Nplus+\Nplustwo+3) \* \bigg[
            {1204 \over  81} \* \S(1)
          - 4 \* \S(1) \* \z3
          - {2 \over  3} \* \Ss(1,-3)
          + {19 \over  3} \* \Ss(1,-2)
          + 2 \* \Sss(1,1,-2)
          + {11 \over  3} \* \Ss(1,2)
  \nonumber\\&& \mbox{}
          - {2 \over  3} \* \Sss(1,-2,1)
          + {205 \over  108} \* \Ss(1,1)
          - {71 \over  27} \* \S(2)
          - {2 \over  3} \* \Ss(2,-2)
          + {11 \over  3} \* \Ss(2,1)
  	  \bigg]
	+ (\Nminus+\Nplus-2) \* \bigg[
            {305 \over 18} \* \Ss(1,-2)
          - {1405 \over 648} \* \S(1)
  \nonumber\\&& \mbox{}
          - 8 \* \S(1) \* \z3
          - {31 \over 6} \* \Ss(1,-3)
          + {4 \over 3} \* \Sss(1,-2,1)
          + {2441 \over 216} \* \Ss(1,1)
          + 9 \* \Sss(1,1,-2)
          + {4 \over 9} \* \Ss(1,2)
          + {25 \over 12} \* \Ss(1,3)
  	  \bigg]
	+ (\Nminus-\Nplus) \* \bigg[
            {109 \over 108} \* \S(2)
  \nonumber\\&& \mbox{}
          + 6 \* \S(2) \* \z3
          + 3 \* \Ss(2,-3)
          - {59 \over 6} \* \Ss(2,-2)
          - {71 \over 12} \* \Ss(2,1)
          - 6 \* \Sss(2,1,-2)
          - {2 \over 3} \* \Ss(2,2)
          - {3 \over 2} \* \Ss(2,3)
          - {64 \over 9} \* \S(3)
          + 5 \* \Ss(3,-2)
          + {5 \over 12} \* \Ss(3,1)
  \nonumber\\&& \mbox{}
          - 2 \* \S(4)
          - {3 \over 2} \* \Ss(4,1)
	\bigg]
	+ (\Nminus-\Nplustwo) \* \bigg[
            {2 \over 3} \* \Ss(2,-2)
          - {2243 \over 108} \* \S(2)
          + {31 \over 9} \* \S(3)
          - {2 \over 3} \* \Ss(3,1)
	\bigg]
	+ (1-\Nplus) \* \bigg[
            {6815 \over 216} \* \S(2)
          + \S(5)
  \nonumber\\&& \mbox{}
          + {25 \over 3} \* \Ss(2,-2)
          - {8 \over 9} \* \Ss(2,1)
          - {473 \over 36} \* \S(3)
          - 4 \* \Ss(3,-2)
          - {25 \over 6} \* \Ss(3,1)
          + {31 \over 6} \* \S(4)
	\bigg]
          - {10 \over 9} \* \S(-3)
          - {1 \over 3} \* \Ss(1,3)
          - {5443 \over 324} \* \S(1)
          + 2 \* \S(1) \* \z3
  \nonumber\\&& \mbox{}
          + {2 \over 3} \* \Ss(1,-3)
          - {37 \over 9} \* \Ss(1,-2)
          + {2 \over 3} \* \Sss(1,-2,1)
          - {205 \over 108} \* \Ss(1,1)
          - 2 \* \Sss(1,1,-2)
          - {13 \over 9} \* \Ss(1,2)
          + {2 \over 3} \* \Ss(-2,-2)
          + {151 \over 54} \* \S(2)
          + {2 \over 3} \* \Ss(2,-2)
  \nonumber\\&& \mbox{}
          - {13 \over 9} \* \Ss(2,1)
          - {10 \over 9} \* \S(3)
          - {1 \over 3} \* \Ss(3,1)
          \bigg)
       +16\,  \*   \colour4colour{\ca^3}  \*  \bigg(
	  (\Nminustwo-2\*\Nminus-2\*\Nplus+\Nplustwo+3) \* \bigg[
            {73091 \over 648} \* \S(1)
          - 16 \* \Ss(1,-4)
  \nonumber\\&& \mbox{}
          + {88 \over 3} \* \Ss(1,-3)
          + 16 \* \Sss(1,-3,1)
          + {85 \over 6} \* \Ss(1,-2)
          + 4 \* \Sss(1,-2,-2)
          - 11 \* \Sss(1,-2,1)
          + 4 \* \Sss(1,-2,2)
          - {413 \over 108} \* \Ss(1,1)
          + 24 \* \Sss(1,1,-3)
  \nonumber\\&& \mbox{}
          + 11 \* \Sss(1,1,-2)
          - 16 \* \Ssss(1,1,-2,1)
          + 8 \* \Sss(1,1,3)
          - {67 \over 9} \* \Ss(1,2)
          + 8 \* \Sss(1,2,-2)
          + 8 \* \Sss(1,2,2)
          + {55 \over 3} \* \Ss(1,3)
          + 8 \* \Sss(1,3,1)
          - 8 \* \Ss(1,4)
  \nonumber\\&& \mbox{}
          - {395 \over 27} \* \S(2)
          - 14 \* \Ss(2,-3)
          - {11 \over 3} \* \Ss(2,-2)
          + 8 \* \Sss(2,-2,1)
          - {67 \over 9} \* \Ss(2,1)
          + 4 \* \Sss(2,1,-2)
          + 8 \* \Sss(2,1,2)
          + {22 \over 3} \* \Ss(2,2)
          + 8 \* \Sss(2,2,1)
  \nonumber\\&& \mbox{}
          - 10 \* \Ss(2,3)
          + 8 \* \Sss(3,1,1)
          - 8 \* \Ss(3,2)
  	  \bigg]
	+ (\Nminus+\Nplus-2) \* \bigg[
            14 \* \Sss(1,-2,1)
          - {713 \over 324} \* \S(1)
          - {26 \over 3} \* \Ss(1,-3)
          - {61 \over 9} \* \Ss(1,-2)
  \nonumber\\&& \mbox{}
          - {80 \over 27} \* \Ss(1,1)
          + 14 \* \Sss(1,1,-2)
          - {109 \over 18} \* \Ss(1,2)
          + 4 \* \Ss(1,3)
  	  \bigg]
	+ (\Nminus-\Nplus) \* \bigg[
            {473 \over 216} \* \S(2)
          - 12 \* \Ss(2,-3)
          + 5 \* \Ss(2,-2)
          - 2 \* \Ss(2,1)
  \nonumber\\&& \mbox{}
          - 8 \* \Sss(2,1,-2)
          + {23 \over 3} \* \Ss(2,2)
          - 10 \* \Ss(2,3)
          + {665 \over 36} \* \S(3)
          - 20 \* \Ss(3,-2)
          + {34 \over 3} \* \Ss(3,1)
          - 16 \* \Ss(3,2)
          - 21 \* \S(4)
          - 26 \* \Ss(4,1)
	\bigg]
  \nonumber\\&& \mbox{}
	+ (\Nminus-\Nplustwo) \* \bigg[
            8 \* \Ss(2,-3)
          - {9533 \over 108} \* \S(2)
          - {77 \over 3} \* \Ss(2,-2)
          - 8 \* \Sss(2,-2,1)
          - 8 \* \Sss(2,1,-2)
          - {44 \over 3} \* \Ss(2,2)
          - {1517 \over 18} \* \S(3)
          - 8 \* \S(5)
  \nonumber\\&& \mbox{}
          + 8 \* \Ss(3,-2)
          - {121 \over 3} \* \Ss(3,1)
          + 4 \* \Ss(3,2)
          + 44 \* \S(4)
          + 16 \* \Ss(4,1)
	\bigg]
	+ (1-\Nplus) \* \bigg[
            {8533 \over 108} \* \S(2)
          + {103 \over 3} \* \Ss(2,-2)
          + {1579 \over 18} \* \S(3)
  \nonumber\\&& \mbox{}
          - 8 \* \Ss(2,-3)
          + 8 \* \Sss(2,-2,1)
          + {109 \over 9} \* \Ss(2,1)
          + 8 \* \Sss(2,1,-2)
          + {28 \over 3} \* \Ss(2,2)
          - 4 \* \Ss(3,2)
          + 8 \* \Ss(3,-2)
          + {71 \over 3} \* \Ss(3,1)
          - 16 \* \Ss(4,1)
          + 36 \* \S(5)
  \nonumber\\&& \mbox{}
          - {98 \over 3} \* \S(4)
	\bigg]
          - {79 \over 32}
          + 4 \* \S(-5)
          - 8 \* \Ss(-4,1)
          + {67 \over 9} \* \S(-3)
          - 4 \* \Ss(-3,-2)
          - 2 \* \Ss(-3,2)
          - 4 \* \Ss(-2,-3)
          - {67 \over 9} \* \Ss(1,2)
          + {413 \over 108} \* \Ss(1,1)
  \nonumber\\&& \mbox{}
          - {11 \over 3} \* \Ss(-2,-2)
          + 4 \* \Sss(-2,-2,1)
          + 4 \* \Sss(-2,1,-2)
          - {16619 \over 162} \* \S(1)
          - {88 \over 3} \* \Ss(1,-3)
          - {523 \over 18} \* \Ss(1,-2)
          + 11 \* \Sss(1,-2,1)
          - {22 \over 3} \* \Ss(2,2)
  \nonumber\\&& \mbox{}
          - 11 \* \Sss(1,1,-2)
          - {33 \over 2} \* \Ss(1,3)
          + {781 \over 54} \* \S(2)
          - 4 \* \Ss(2,-3)
          + {11 \over 3} \* \Ss(2,-2)
          + 4 \* \Sss(2,-2,1)
          - {67 \over 9} \* \Ss(2,1)
          + 4 \* \Sss(2,1,-2)
          + {11 \over 6} \* \Ss(3,1)
  \nonumber\\&& \mbox{}
          + {67 \over 9} \* \S(3)
          - 4 \* \Ss(3,-2)
          - 2 \* \Ss(3,2)
          - 8 \* \Ss(4,1)
          + 4 \* \S(5)
          \bigg)
       +16\,  \*   \colour4colour{\cf \* \nf^2}  \*  \bigg(
	  (\Nminustwo-2\*\Nminus-2\*\Nplus+\Nplustwo+3) \* \bigg[
            {4 \over 9} \* \Ss(1,2)
  \nonumber\\&& \mbox{}
          - {77 \over 81} \* \S(1)
          + {16 \over 27} \* \Ss(1,1)
          - {2 \over 9} \* \Sss(1,1,1)
  	  \bigg]
	+ {7 \over 9} \* (\Nminus+\Nplus-2) \* \bigg[
            \Ss(1,2)
          - {1 \over 2} \* \Sss(1,1,1)
  	  \bigg]
          - {11 \over 144}
          + {2 \over 9} \* \Sss(1,1,1)
          - {16 \over 27} \* \Ss(1,1)
  \nonumber\\&& \mbox{}
          + {77 \over 81} \* \S(1)
          - {4 \over 9} \* \Ss(1,2)
	+ {1 \over 3} \* (\Nminus-\Nplus) \* \bigg[
            {211 \over 27} \* \S(1)
          - {139 \over 18} \* \Ss(1,1)
          + {11 \over 3} \* \S(2)
          + \Ss(2,1)
          + \Sss(2,1,1)
          - 2 \* \Ss(2,2)
          - 2 \* \Ss(3,1)
          + \S(4)
  \nonumber\\&& \mbox{}
          + {5 \over 2} \* \S(3)
	\bigg]
	- (\Nminus-\Nplustwo) \* \bigg[
            2 \* \S(1)
          - \Ss(1,1)
          + {11 \over 27} \* \S(2)
          + {2 \over 9} \* \Ss(2,1)
          - {4 \over 9} \* \S(3)
	\bigg]
	+ (1-\Nplus) \* \bigg[
            {64 \over 81} \* \S(1)
          + {58 \over 27} \* \Ss(1,1)
          + {1 \over 3} \* \S(3)
  \nonumber\\&& \mbox{}
          - {10 \over 3} \* \S(2)
          + {1 \over 3} \* \Ss(2,1)
	\bigg]
          \bigg)
       +16\,  \*   \colour4colour{\cf^2 \* \nf}  \*  \bigg(
	  {4 \over 3} \* (\Nminustwo-2\*\Nminus-2\*\Nplus+\Nplustwo+3) \* \bigg[
            {5 \over 4} \* \Ss(1,2)
          + {1 \over 2} \* \Ss(1,3)
	  - \Sss(1,1,1)
  \nonumber\\&& \mbox{}
          - \Ss(1,-3)
          + 2 \* \Sss(1,1,-2)
          + {31 \over 16} \* \Ss(1,1)
          + \Ssss(1,1,1,1)
          - {11 \over 16} \* \S(1)
          - \Sss(1,1,2)
	\bigg]
	+ (\Nminus+\Nplus-2) \* \bigg[
            {25 \over 6} \* \Ss(1,3)
          - 9 \* \S(1) \* \z3
  \nonumber\\&& \mbox{}
          - {16 \over 3} \* \Ss(1,-3)
          + {67 \over 3} \* \Ss(1,-2)
          - {23 \over 12} \* \Sss(1,1,1)
          + {7 \over 3} \* \Ssss(1,1,1,1)
          - {7 \over 3} \* \Sss(1,1,2)
          + {32 \over 3} \* \Sss(1,1,-2)
  	  \bigg]
	+ (\Nminus-\Nplus) \* \bigg[
	    2 \* \Ss(4,1)
          - 2 \* \S(5)
  \nonumber\\&& \mbox{}
          - {773 \over 24} \* \S(1)
          - {8 \over 3} \* \Ss(1,1)
          + {163 \over 8} \* \S(2)
          + 6 \* \S(2) \* \z3
          + 4 \* \Ss(2,-3)
          - {32 \over 3} \* \Ss(2,-2)
          - {8 \over 3} \* \Ss(2,1)
          - 8 \* \Sss(2,1,-2)
          + {5 \over 3} \* \Sss(2,1,1)
          + 2 \* \Sss(2,1,2)
  \nonumber\\&& \mbox{}
          - 2 \* \Ssss(2,1,1,1)
          - {11 \over 3} \* \Ss(2,2)
          - 3 \* \Ss(2,3)
          - {23 \over 2} \* \S(3)
          - 4 \* \Ss(3,1)
          + \Sss(3,1,1)
          + {13 \over 6} \* \S(4)
          + {17 \over 2} \* \Ss(1,2)
	\bigg]
	+ (\Nminus-\Nplustwo) \* \bigg[
            {85 \over 12} \* \Ss(1,1)
  \nonumber\\&& \mbox{}
          + {163 \over 12} \* \S(1)
          - 3 \* \Ss(1,2)
          - {9 \over 2} \* \S(2)
          + {8 \over 3} \* \Ss(2,-2)
          - {4 \over 3} \* \Ss(2,1)
          + {4 \over 3} \* \Sss(2,1,1)
          - {4 \over 3} \* \Ss(2,2)
          + {14 \over 3} \* \S(3)
          - {2 \over 3} \* \S(4)
	\bigg]
	+ (1-\Nplus) \* \bigg[
	  4 \* \S(4)
  \nonumber\\&& \mbox{}
          - {191 \over 12} \* \Ss(1,1)
          - 8 \* \Ss(1,2)
          + {20 \over 3} \* \S(2)
          + 8 \* \Ss(2,-2)
          + {11 \over 4} \* \Ss(2,1)
          + \Sss(2,1,1)
          - 3 \* \Ss(2,2)
          - {215 \over 12} \* \S(3)
          - \Ss(3,1)
          + {71 \over 3} \* \S(1)
	\bigg]
  \nonumber\\&& \mbox{}
          + 8 \* (\Nminus-1) \* \Ss(3,-2)
          - {1 \over 16}
          + {11 \over 12} \* \S(1)
          + {4 \over 3} \* \Ss(1,-3)
          - {31 \over 12} \* \Ss(1,1)
          - {8 \over 3} \* \Sss(1,1,-2)
          + {4 \over 3} \* \Sss(1,1,1)
          - {4 \over 3} \* \Ssss(1,1,1,1)
          + {4 \over 3} \* \Sss(1,1,2)
  \nonumber\\&& \mbox{}
          - {5 \over 3} \* \Ss(1,2)
          - {2 \over 3} \* \Ss(1,3)
          \bigg)
 \:\: . \label{eq:ggg2}
\eea
 
Eqs.~(\ref{eq:gps2}) -- (\ref{eq:ggg2}) represent new results of 
this article, with the only exception of the $\ca \n2f$ part of 
Eq.~(\ref{eq:ggg2}) which has been obtained by Bennett and Gracey in 
Ref.~\cite{Bennett:1997ch}.
Our results agree with the even moments $N=2,\dots,12$ computed before 
\cite{Larin:1997wd,Retey:2000nq} using the {\sc Mincer} program 
\cite{Gorishnii:1989gt,Larin:1991fz}. 

The results (\ref{eq:gij0}) -- (\ref{eq:ggg2}) are assembled, after
inserting the QCD values $\cf=4/3$ and $\ca=3$ for the colour factors,
in Figs.~1 and 2 for four active flavours and a typical value 
$\as = 0.2$ for the strong coupling constant. 
The NNLO corrections are markedly smaller than the NLO contributions 
under these circumstances. At $N > 2$ they amount to less than 2\%
and 1\% for the large diagonal quantities $\gamma_{\,\rm qq}$ and 
$\gamma_{\,\rm gg}$, respectively, while for the much smaller 
off-diagonal anomalous dimensions $\gamma_{\,\rm qg}$ and 
$\gamma_{\,\rm gq}$ values of up to 6\% and 4\% are reached. The 
relative NNLO corrections are very large at $N > 2$ for 
$\gamma_{\,\rm ps}$, which is however completely negligible in this 
region of $N$. 

\begin{figure}[p]
\label{pic:gam-diag}
\centerline{\epsfig{file=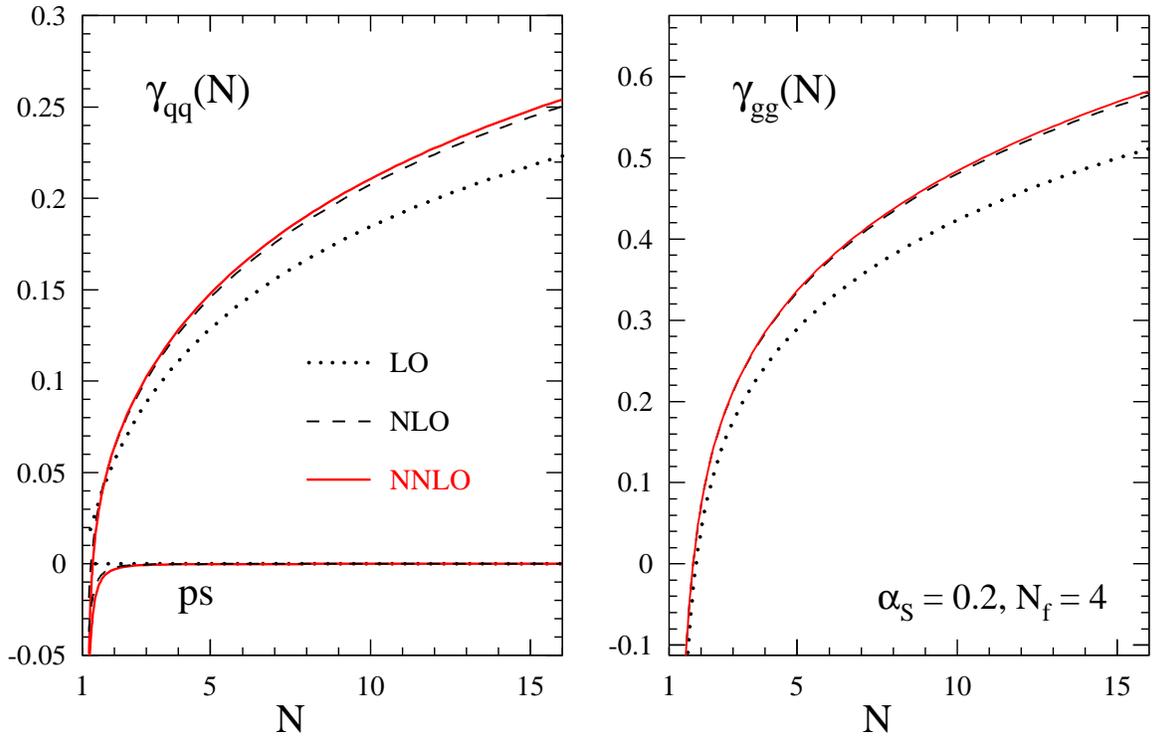,width=15.6cm,angle=0}}
\vspace{-1mm}
\caption{The perturbative expansion of the diagonal anomalous 
 dimensions $\gamma_{\,\rm qq}(N)$ and $\gamma_{\,\rm gg}(N)$ for four 
 flavours at $\as = 0.2$. The pure-singlet (ps) contribution to 
 $\gamma_{\,\rm qq}$ is shown separately.}
\vspace*{2mm}
\end{figure}
\begin{figure}[p]
\label{pic:gam-offd}
\centerline{\epsfig{file=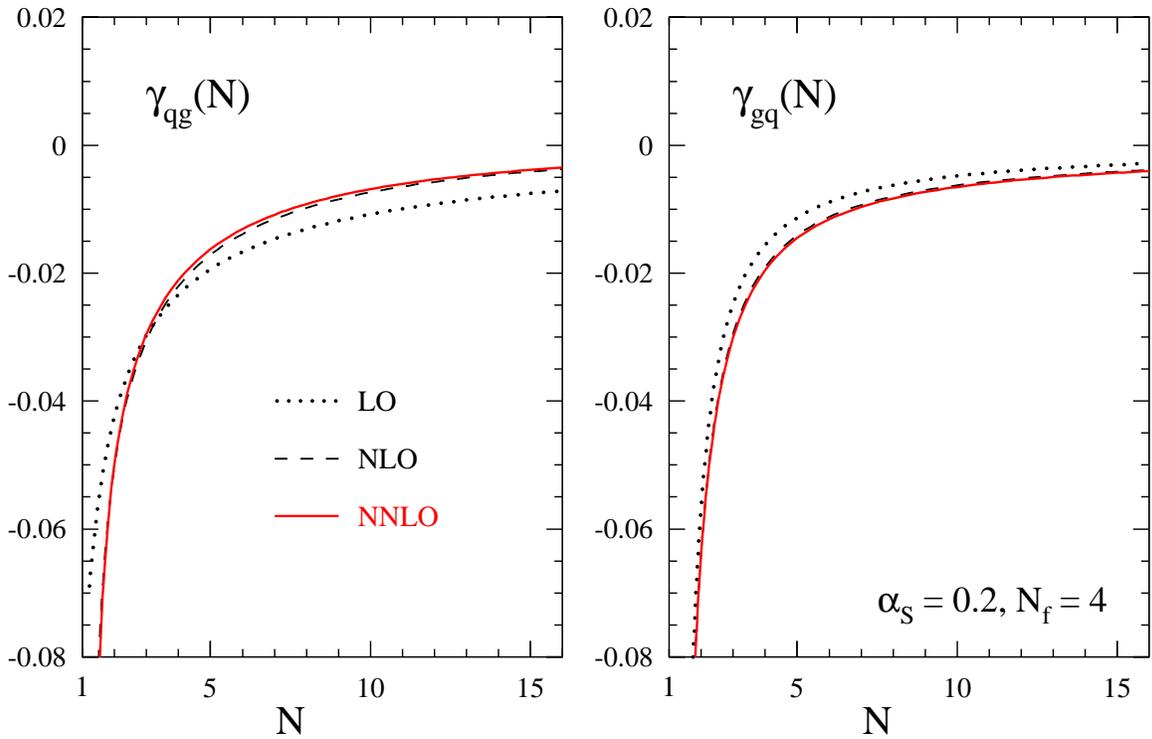,width=15.6cm,angle=0}}
\vspace{-1mm}
\caption{As Fig.~1, but for the off-diagonal anomalous dimensions
 $\gamma_{\,\rm qg}(N)$ and $\gamma_{\,\rm gq}(N)$.}
\vspace*{2mm}
\end{figure}

For $N \ra \infty$ the off-diagonal $n$-loop anomalous dimensions 
vanish like $\displaystyle \frac{1}{N}\ln^{\,2n-2} N$, while the 
diagonal quantities behave as~\cite{Korchemsky:1989si}
\beq
\label{eq:ntoinf}
  \gamma_{\,\rm aa}^{\,(n-1)}(N) \: = \:\: 
  A_n^{\rm a}\: (\ln N +\gamma_e) - B_n^{\,\rm a} - C_n^{\,\rm a} 
  \:\frac{\ln N}{N} + {\cal O} \left( \frac{1}{N} \right) \:\: ,
\eeq
where $\gamma_e$ is the Euler-Mascheroni constant. The leading 
large-$N$ coefficients $A_n^{\rm q}$ of $\gamma_{\,\rm qq}$ have been 
specified up to $n=3$ in Eq.~(3.11) of Ref.~\cite{Moch:2004pa}. As
expected, the constants $A_n^{\rm g}$ are related to those results by 
\beq
\label{eq:Ang}
  A_n^{\rm g} \:\: = \:\:\frac{\ca}{\cf} \: A_n^{\rm q} \:\: .
\eeq
The coefficients $C_n^{\,\rm g}$ in Eq.~(\ref{eq:ntoinf}) can be 
expressed in terms of the $A_n^{\rm g}$ by
\beq
\label{eq:Cng}
  C_1^{\,\rm g} \:\: = \:\: 0 \:\: , \quad
  C_2^{\,\rm g} \:\: = \:\: 4\, C_A\: A_1^{\rm g} 
                  \: = \: (A_1^{\rm g})^2
  \:\: , \quad
  C_3^{\,\rm g} \:\: = \:\: 8\, C_A\: A_2^{\rm g} 
                  \: = \: 2\, A_1^{\rm g}\, A_2^{\rm g}
  \:\: .
\eeq
This result is completely analogous to the corresponding relation for 
$C_n^{\,\rm q}$ in Eq.~(3.12) of Ref.~\cite{Moch:2004pa}.
Finally the $N$-independent contributions $B_n^{\,\rm g}$ can be read 
off directly from the $\delta(1-x)$ terms in Eqs.~(\ref{eq:Pij0}), 
(\ref{eq:Pgg1}) and (\ref{eq:Pgg2}) below.
%
%
\setcounter{equation}{0}
\section{Results in {\bf x}-space}
\label{sec:xresults}
%
%
The N$^{\rm n}$LO singlet splitting functions $P_{\,\rm ab}^{\,(n)}(x)$ 
in
\beq
\label{eq:Pexp}
 P_{\rm ab}\left(\as,x\right) \: = \: \sum_{n=0}\,
  \left(\frac{\as}{4\pi}\right)^{n+1} P^{\,(n)}_{\rm ab}(x)
\eeq
are obtained from the $N$-space results of the previous section by an 
inverse Mellin transformation which expresses these functions in terms 
of harmonic polylogarithms~\cite{Goncharov,Borwein,Remiddi:1999ew}. 
This transformation can be performed by a completely algebraic 
procedure \cite{Moch:1999eb,Remiddi:1999ew} based on the fact that 
harmonic sums occur as coefficients of the Taylor expansion of harmonic 
polylogarithms.

Our notation for the harmonic polylogarithms $H_{m_1,...,m_w}(x)$, 
$m_j = 0,\pm 1$ follows Ref.~\cite{Remiddi:1999ew} to which the reader 
is referred for a detailed discussion. For completeness, we recall the 
basic definitions: The lowest-weight ($w = 1$) functions $H_m(x)$ are 
given by
\beq
\label{eq:hpol1}
  H_0(x)       \: = \: \ln x \:\: , \quad\quad
  H_{\pm 1}(x) \: = \: \mp \, \ln (1 \mp x) \:\: .
\eeq
The higher-weight ($w \geq 2$) functions are recursively defined as
\beq
\label{eq:hpol2}
  H_{m_1,...,m_w}(x) \: = \:
    \left\{ \begin{array}{cl}
    \displaystyle{ \frac{1}{w!}\,\ln^w x \:\: ,}
       & \quad {\rm if} \:\:\: m^{}_1,...,m^{}_w = 0,\ldots ,0 \\[2ex]
    \displaystyle{ \int_0^x \! dz\: f_{m_1}(z) \, H_{m_2,...,m_w}(z)
       \:\: , } & \quad {\rm else}
    \end{array} \right.
\eeq
with
\beq
\label{eq:hpolf}
  f_0(x)       \: = \: \frac{1}{x} \:\: , \quad\quad
  f_{\pm 1}(x) \: = \: \frac{1}{1 \mp x} \:\: .
\eeq
For chains of indices zero we again employ the abbreviated notation
\beq
\label{eq:habbr}
  H_{{\footnotesize \underbrace{0,\ldots ,0}_{\scriptstyle m} },\,
  \pm 1,\, {\footnotesize \underbrace{0,\ldots ,0}_{\scriptstyle n} },
  \, \pm 1,\, \ldots}(x) \: = \: H_{\pm (m+1),\,\pm (n+1),\, \ldots}(x)
  \:\: .
\eeq
Corresponding to the maximal weight $2l\!-\!1$ of the harmonic sums in 
section~3, the $l$-loop splitting functions involve harmonic 
polylogarithms up to weight $2l\!-\!2$. Hence our three-loop results 
cannot be expressed in terms of standard polylogarithms which are 
sufficiently general only for $w \leq 3$. 

For completeness we recall the one- and two-loop non-singlet splitting
functions~\cite{Altarelli:1977zs,Curci:1980uw}
\bea
  P^{\,(0)}_{\rm ps}(x) & \! = \! &  0  \nn \\[1mm] 
  P^{\,(0)}_{\rm qg}(x) & \! = \! &  
            2\, \* \colour4colour{\nf} \, \* \pqg(x)
 \nn \\[1mm]
  P^{\,(0)}_{\rm gq}(x) & \! = \! &  
          2\, \* \colour4colour{\cf} \, \* \pgq(x)
 \nn \\[1mm] 
  P^{\,(0)}_{\rm gg}(x) & \! = \! & 
         \colour4colour{\ca}  \*  \bigg(
            4 \* \pgg(x)
          + {11 \over 3} \* \delta(1 - x)
          \bigg)
          - {2 \over 3} \, \* \colour4colour{\nf} \, \* \delta(1 - x)
\label{eq:Pij0}
\eea
and 
\bea
\label{eq:Pqq1}
  &&P^{\,(1)}_{\,\rm ps}(x) \:\: = \:\:
         4\, \*  \colour4colour{\cf \* \nf}  \*  \bigg(
       {20 \over 9} \* {1 \over x}
       - 2
          + 6 \* x
          - 4 \* \H(0)
       + x^2 \* \bigg[
         {8 \over 3} \* \H(0)
       - {56 \over 9}
          \bigg]
       + (1+x) \* \bigg[
         5 \* \H(0)
          - 2 \* \Hh(0,0)
          \bigg]
          \bigg)
 \\[4mm] 
 \label{eq:Pqg1}
  &&P^{\,(1)}_{\rm qg}(x) \:\: = \:  
         4\, \*  \colour4colour{\ca \* \nf}  \*  \bigg(
      {20 \over 9} \* {1 \over x} - 2 + 25 \* x
          - 2 \* \pqg( - x) \* \Hh(-1,0)
          - 2 \* \pqg(x) \* \Hh(1,1)
         + x^2 \* \bigg[
      {44 \over 3} \* \H(0)
      - {218 \over 9}
          \bigg]
  \nonumber\\&&
         + 4 \* (1-x) \* \bigg[
         \Hh(0,0)
         - 2 \* \H(0)
         + x \* \H(1)
          \bigg]
          - 4 \* \z2 \* x
          - 6 \* \Hh(0,0)
          + 9 \* \H(0)
          \bigg)
       + 4\, \*  \colour4colour{\cf \* \nf}  \*  \bigg(
            2 \* \pqg(x) \* \bigg[
            \Hh(1,0)
          + \Hh(1,1)
          + \H(2)
  \nonumber\\&&
          - \z2
          \bigg]
         +  4 \* x^2 \* \bigg[
          \H(0)
        + \Hh(0,0)
        + {5 \over 2}
          \bigg]
         +  2 \* (1-x) \* \bigg[
          \H(0)
         + \Hh(0,0)
         - 2 \* x \* \H(1)
         + {29 \over 4}
          \bigg]
          - {15 \over 2}
          - \Hh(0,0)
          - {1 \over 2} \* \H(0)
          \bigg)
 \\[4mm] 
 \label{eq:Pgq1}
  &&P^{\,(1)}_{\rm gq}(x) \:\: = \:\:  
         4\, \* \colour4colour{\ca \* \cf}  \*  \bigg(
          {1 \over x}
          + 2 \* \pgq(x) \* \bigg[
            \Hh(1,0)
          + \Hh(1,1)
          + \H(2)
          - {11 \over 6} \* \H(1)
          \bigg]
          - x^2 \* \bigg[
           {8 \over 3} \* \H(0)
         - {44 \over 9}
          \bigg]
          + 4 \* \z2
       - 2
  \nonumber\\&&
          - 7 \* \H(0)
          + 2 \* \Hh(0,0)
          - 2 \* \H(1) \* x
          + (1+x) \* \bigg[
          2 \* \H(0,0)
          - 5 \* \H(0)
          + {37 \over 9}
          \bigg]
          - 2 \* \pgq( - x) \* \Hh(-1,0)
          \bigg)
       - 4\, \* \colour4colour{\cf \* \nf}  \*  \bigg(
            {2 \over 3} \* x
  \nonumber\\&&
          - \pgq(x) \* \bigg[
            {2 \over 3} \* \H(1)
          - {10 \over 9}
          \bigg]
          \bigg)
       + 4\, \* \colour4colour{\cf^2}  \*  \bigg(
          \pgq(x) \* \bigg[
            3 \* \H(1)
          - 2 \* \Hh(1,1)
          \bigg]
          + (1+x) \* \bigg[
          \Hh(0,0)
          - {7 \over 2}
          + {7 \over 2} \* \H(0)
          \bigg]
          - 3 \* \Hh(0,0)
  \nonumber\\&&
          + 1
          - {3 \over 2} \* \H(0)
          + 2 \* \H(1) \* x
          \bigg)
 \\[4mm] 
  &&P^{\,(1)}_{\rm gg}(x) \:\: = \:\:  
         4\, \* \colour4colour{\ca \* \nf}  \*  \bigg(
            1-x
          - {10 \over 9} \* \pgg(x)
          - {13 \over 9} \* \bigg({1 \over x}-x^2\bigg)
          - {2 \over 3} \* (1+x) \* \H(0)
          - {2 \over 3} \* \delta(1 - x)
          \bigg)
       + 4\, \* \colour4colour{\ca^2}  \*  \bigg(
            27
  \nonumber\\&&
          + (1+x) \* \bigg[
         {11 \over 3} \* \H(0)
          + 8 \* \Hh(0,0)
       - {27 \over 2}
          \bigg]
          + 2 \* \pgg( - x) \* \bigg[
            \Hh(0,0)
          - 2 \* \Hh(-1,0)
          - \z2
          \bigg]
          - {67 \over 9} \* \bigg({1 \over x}-x^2\bigg)
          - 12 \* \H(0)
  \nonumber\\&&
       - {44 \over 3} \*x^2\*\H(0)
          + 2 \* \pgg(x) \* \bigg[
            {67 \over 18}
          - \z2
          + \Hh(0,0)
          + 2 \* \Hh(1,0)
          + 2 \* \H(2)
          \bigg]
          + \delta(1 - x) \* \bigg[
            {8 \over 3}
          + 3 \* \z3
          \bigg]
          \bigg)
       + 4\, \* \colour4colour{\cf \* \nf}  \*  \bigg(
            2 \* \H(0)
  \nonumber\\&&
          + {2 \over 3} \* {1 \over x} + {10 \over 3} \* x^2
       - 12
          + (1+x) \* \bigg[
         4
       - 5 \* \H(0)
          - 2 \* \Hh(0,0)
          \bigg]
          - {1 \over 2} \* \delta(1 - x)
          \bigg)
\:\: .\label{eq:Pgg1}
\eea
Here and in Eqs.~(\ref{eq:Pps2}) -- (\ref{eq:Pgg2}) we suppress the
argument $x$ of the polylogarithms and use
\bea
  p_{\rm{qg}}(x) &\! =\! & 1 - 2x + 2x^{\,2}  \nn \\
  p_{\rm{gq}}(x) &\! =\! & 2x^{\,-1} -2 + x  \nn \\
  p_{\rm{gg}}(x) &\! =\! & (1-x)^{-1} + x^{\,-1} - 2 + x - x^{\,2} 
  \:\: .
\eea
Divergences for $x \to 1 $ are understood in the sense of
$+$-distributions.

The third-order pure-singlet contribution to the quark-quark splitting 
function (\ref{eq:Pqq}), corresponding to the anomalous dimension 
(\ref{eq:gps2}), is given by
\bea
  &&P^{\,(2)}_{\rm ps}(x) \:\: = \:\: 
         16\, \*  \colour4colour{\ca \* \cf \* \nf}  \*  \bigg(
             {4 \over 3} \* \big({1 \over x}+x^2\big) \* \bigg[
            {13 \over 3} \* \Hh(-1,0)
	  - {14 \over 9} \* \H(0)
          + {1 \over 2} \* \H(-1) \* \z2
          - \Hhh(-1,-1,0)
          - 2 \* \Hhh(-1,0,0)
  \nonumber\\&&
          - \Hh(-1,2)
          \bigg]
          + {2 \over 3} \* \big({1 \over x}-x^2\big) \* \bigg[
            {16 \over 3} \* \z2
	  + \Hh(2,1)
	  + 9 \* \z3
	  + {9 \over 4} \* \Hh(1,0)
          - {6761 \over 216}
	  + {571 \over 72} \* \H(1)
	  + {10 \over 3} \* \H(2)
          + \H(1) \* \z2
          - {1 \over 6} \* \Hh(1,1)
  \nonumber\\&&
          - 3 \* \Hhh(1,0,0)
          + 2 \* \Hhh(1,1,0)
          + 2 \* \Hhh(1,1,1)
          \bigg]
          + (1-x) \* \bigg[
            {182 \over 9} \* \H(1)
	  + {158 \over 3}
          + {397 \over 36} \* \Hh(0,0)
	  - {13 \over 2} \* \Hh(-2,0)
	  + 3 \* \Hhhh(0,0,0,0)
  \nonumber\\&&
          + {13 \over 6} \* \Hh(1,0)
          + 3 \* x \* \Hh(1,0)
	  + \Hh(-3,0)
          + \H(-2) \* \z2
          + 2 \* \Hhh(-2,-1,0)
          + 3 \* \Hhh(-2,0,0)
          + {1 \over 2} \* \Hh(0,0) \* \z2
          + {1 \over 2} \* \H(1) \* \z2
          - {9 \over 4} \* \Hhh(1,0,0)
  \nonumber\\&&
          - {3 \over 4} \* \Hh(1,1)
          + \Hhh(1,1,0)
          + \Hhh(1,1,1)
          \bigg]
          + (1+x) \* \bigg[
            {7 \over 12} \* \H(0) \* \z2
	  + {31 \over 6} \* \z3
          + {91 \over 18} \* \H(2)
	  + {71 \over 12} \* \H(3)
	  + {113 \over 18} \* \z2
          - {826 \over 27} \* \H(0)
  \nonumber\\&&
          + {5 \over 2} \* \Hh(2,0)
          + {16 \over 3} \* \Hh(-1,0)
          + 6 \* x \* \Hh(-1,0)
	  + {31 \over 6} \* \Hhh(0,0,0)
	  - {17 \over 6} \* \H(2,1)
          + {117 \over 20} \* \z2^2
          + 9 \* \H(0) \* \z3
          + {5 \over 2} \* \H(-1) \* \z2
          + 2 \* \Hhh(2,1,0)
  \nonumber\\&&
          + {1 \over 2} \* \Hhh(-1,0,0)
          - 2 \* \Hh(-1,2)
          + \H(2) \* \z2
          - {7 \over 2} \* \Hhh(2,0,0)
          + \Hhh(-1,-1,0)
          + 2 \* \Hhh(2,1,1)
          + \Hh(3,1)
          - {1 \over 2} \* \H(4)
          \bigg]
	  + 5 \* \Hh(-2,0)
          + \Hh(2,1)
  \nonumber\\&&
          + \Hhhh(0,0,0,0)
          - {1 \over 2} \* \z2^2
          + 4 \* \Hh(-3,0)
          + 4 \* \H(0) \* \z3
	  - {32 \over 9} \* \Hh(0,0)
          - {29 \over 12} \* \H(0)
          - {235 \over 12} \* \z2
	  - {511 \over 12}
	  - {97 \over 12} \* \H(1)
          + {33 \over 4} \* \H(2)
          - \H(3)
  \nonumber\\&&
          - {11 \over 2} \* \H(0) \* \z2
          - {11 \over 2} \* \z3
          - {3 \over 2} \* \Hh(2,0)
          - 10 \* \Hhh(0,0,0)
          + {2 \over 3} \* x^2 \* \bigg[
	    {83 \over 4} \* \Hh(0,0)
	  - {243 \over 4} \* \H(0)
	  + 10 \* \z2
	  + {511 \over 8}
	  + {97 \over 8} \* \H(1)
	  - {4 \over 3} \* \H(2)
  \nonumber\\&&
	  - 4 \* \z3
	  - \H(0) \* \z2
	  + \H(3)
	  + \Hh(2,0)
          - 6 \* \Hh(-2,0)
          \bigg]
          \bigg)
       + 16\, \*  \colour4colour{\cf \* \nf^2}  \*  \bigg(
            {2 \over 27} \* \H(0)
          - 2 
          - \H(2)
          + \z2
          + {2 \over 3} \* x^2 \* \bigg[
	       \H(2)
	     - \z2
	     + 3
  \nonumber\\&&
	     - {19 \over 6} \* \H(0)
          \bigg]
          + {2 \over 9} \* \big({1 \over x}-x^2\big) \* \bigg[
            \Hh(1,1) 
	  + {5 \over 3} \* \H(1)
	  + {2 \over 3} 
          \bigg]
          + (1-x) \* \bigg[
            {1 \over 6} \* \Hh(1,1) 
          - {7 \over 6} \* \H(1)
	  + x \* \H(1)
	  + {35 \over 27} \* \H(0)
	  + {185 \over 54}
          \bigg]
  \nonumber\\&&
          + {1 \over 3} \* (1+x) \* \bigg[
            {4 \over 3} \* \H(2)
          - {4 \over 3} \* \z2
          + \z3
          + \Hh(2,1) 
          - 2 \* \H(3) 
          + 2 \* \H(0) \* \z2
          + {29 \over 6} \* \Hh(0,0) 
          + \Hhh(0,0,0) 
          \bigg]
          \bigg)
       +  16\, \*  \colour4colour{\cf^2 \* \nf}  \*  \bigg(
            {85 \over 12} \* \H(1)
  \nonumber\\&&
          - {25 \over 4} \* \Hh(0,0)
	  - \H(0,0,0)
          + {583 \over 12} \* \H(0)
	  - {101 \over 54}
          + {73 \over 4} \* \z2
          - {73 \over 4} \* \H(2)
          + \H(3)
          - 5 \* \Hh(2,0)
	  - \Hh(2,1)
	  - \H(0) \* \z2
          + x^2 \* \bigg[
	    {55 \over 12} 
  \nonumber\\&&
          - {85 \over 12} \* \H(1)
          - {22 \over 3} \* \Hh(0,0)
          - {109 \over 6}
          - {13 \over 54} \* \H(0)
          + {28 \over 9} \* \z2
          - {28 \over 9} \* \H(2)
	  - {16 \over 3} \* \H(0) \* \z2
          + {16 \over 3} \* \H(3)
          + 4 \* \Hh(2,0)
          + {4 \over 3} \* \Hh(2,1)
	  - {26 \over 3} \* \z3
  \nonumber\\&&
          + {22 \over 3} \* \Hhh(0,0,0)
          \bigg]
          + {4 \over 3} \* \big({1 \over x}-x^2\big) \* \bigg[
	    {23 \over 12} \* \Hh(1,0)
	  - {523 \over 144} \* \H(1)
	  - 3 \* \z3
	  + {55 \over 16} 
          + {1 \over 2} \* \Hhh(1,0,0) 
          + \Hh(1,1) 
          - \Hhh(1,1,0) 
          - \Hhh(1,1,1) 
          \bigg]
  \nonumber\\&&
          + (1-x) \* \bigg[
            {1 \over 2} \* \Hhh(1,0,0) 
          + {7 \over 12} \* \Hh(1,1) 
	  - {2743 \over 72} \* \H(0)
	  - {53 \over 12} \* \H(0,0)
	  - {251 \over 12} \* \H(1)
          - {5 \over 4} \* \z2
          + {5 \over 4} \* \H(2)
	  - {8 \over 3} \* \Hh(1,0)
	  + 3 \* x \* \Hh(1,0)
  \nonumber\\&&
	  + 3 \* \H(0) \* \z2
	  - 3 \* \H(3)
          - \Hhh(1,1,0) 
          - \Hhh(1,1,1) 
          \bigg]
          + (1+x) \* \bigg[
	    {1669 \over 216} 
          + {5 \over 2} \* \H(0,0,0)
	  + 4 \* \Hh(2,1)
	  + 7 \* \Hh(2,0)
          + 10 \* x \* \z3
          - {37 \over 10} \* \z2^2
  \nonumber\\&&
          - 7 \* \H(0) \* \z3
          + 6 \* \Hh(0,0) \* \z2
          - 4 \* \Hhhh(0,0,0,0)
          + \Hhh(2,0,0)
          - 2 \* \Hhh(2,1,0)
          - 2 \* \Hhh(2,1,1)
          - 4 \* \Hh(3,0)
          - \H(3,1)
          - 6 \* \H(4)
          \bigg]
          \bigg)
\:\: . \label{eq:Pps2}
\eea
Due to Eqs.~(\ref{eq:gqg2}) and (\ref{eq:ggq2}) the three-loop 
gluon-quark and quark-gluon splitting functions read
\bea
  &&P^{\,(2)}_{\rm qg}(x) \:\: = \:\: 
       16\, \*  \colour4colour{\ca \* \cf \* \nf}  \*  \bigg(
            \pqg(x) \* \bigg[
            {39 \over 2} \* \H(1) \* \z3
	    - 4 \* \Hhh(1,1,1)
	    + 3 \* \Hhh(2,0,0)
	    - {15 \over 4} \* \Hh(1,2)
	    + {9 \over 4} \* \Hhh(1,1,0)
	    + 3 \* \Hhh(2,1,0)
  \nonumber\\&&
	    + \H(0) \* \z3
	    - 2 \* \Hhh(2,1,1)
	    + 4 \* \H(2) \* \z2
	    - {173 \over 12} \* \H(0) \* \z2
	    - {551 \over 72} \* \Hh(0,0)
	    + {64 \over 3} \* \z3
	    - \z2^2
	    - {49 \over 4} \* \H(2)
            - {3 \over 2} \* \Hhhh(1,0,0,0)
	    - {1 \over 3} \* \Hhh(1,0,0)
  \nonumber\\&&
	    - {385 \over 72} \* \Hh(1,0)
	    - {31 \over 2} \* \Hh(1,1)
	    - {113 \over 12} \* \H(1)
	    + {49 \over 4} \* \Hh(2,0)
	    + {5 \over 2} \* \H(1) \* \z2
	    + {79 \over 6} \* \Hhh(0,0,0)
	    + {173 \over 12} \* \H(3)
	    - {1259 \over 32}
	    + {2833 \over 216} \* \H(0)
  \nonumber\\&&
	    + 6 \* \Hh(2,1)
            + 3 \* \Hhh(1,-2,0)
            + 9 \* \Hh(1,0) \* \z2
            + 6 \* \Hh(1,1) \* \z2
            + \Hhhh(1,1,0,0)
            + 3 \* \Hhhh(1,1,1,0)
            - 4 \* \Hhhh(1,1,1,1)
            - 3 \* \Hhh(1,1,2)
            - 6 \* \Hhh(1,2,1)
  \nonumber\\&&
            - 6 \* \Hh(1,3)
	    + {49 \over 4} \* \z2
          \bigg]
          + \pqg(-x) \* \bigg[
            {17 \over 2} \* \H(-1) \* \z3
	      - {5 \over 2} \* \Hhh(-1,-1,0)
	      - {5 \over 2} \* \Hh(-1,2)
	      - {9 \over 2} \* \Hh(-1,0)
	      + {5 \over 2} \* \Hh(-2,0)
              + {3 \over 2} \* \Hhh(-1,0,0)
  \nonumber\\&&
              - 2 \* \Hh(3,1)
              - 2 \* \H(4)
              - 6 \* \Hh(-2,2)
	      + 6 \* \Hhh(-2,-1,0)
	      - 6 \* \Hhh(-2,0,0)
	      + 2 \* \Hh(0,0) \* \z2
	      + 9 \* \H(-2) \* \z2
              + 3 \* \Hhh(-1,-2,0)
              - 2 \* \Hhh(-1,2,1)
  \nonumber\\&&
              - 6 \* \Hhhh(-1,-1,-1,0)
              + 6 \* \Hhhh(-1,-1,0,0)
              + 6 \* \Hhh(-1,-1,2)
              + 9 \* \Hh(-1,0) \* \z2
              - 9 \* \Hh(-1,-1) \* \z2
              - 2 \* \Hhh(-1,2,0)
              - {11 \over 2} \* \Hhhh(-1,0,0,0)
  \nonumber\\&&
              - 6 \* \Hh(-1,3)
          \bigg]
          + \big({1 \over x}-x^2\big) \* \bigg[
            {55 \over 12} 
          - 4 \* \z3
          + {23 \over 9} \* \Hh(1,0)
          - {4 \over 3} \* \Hhh(1,1,0)
          \bigg]
          + \big({1 \over x}+x^2\big) \* \bigg[
            {2 \over 3} \* \Hhh(1,0,0)
          - {371 \over 108} \* \H(1)
          + {23 \over 9} \* \Hh(1,1)
  \nonumber\\&&
          - {2 \over 3} \* \Hhh(1,1,1)
          \bigg]
          + (1-x) \* \bigg[
              6 \* \Hhh(2,1,0)
	    + 3 \* \Hhh(2,1,1)
	    - {5 \over 6} \* \Hhh(1,1,1)
	    - 7 \* \Hhh(2,0,0)
	    - 2 \* \Hh(1,2)
	    + 39 \* \H(0) \* \z3
	    - 4 \* \H(2) \* \z2 
	    - {16 \over 3} \* \z3 
  \nonumber\\&&
	    + \Hhh(1,1,0) 
	    + {154 \over 3} \* \H(0) \* \z2 
	    + {899 \over 24} \* \Hh(0,0) 
	    + {121 \over 10} \* \z2^2 
	    + {607 \over 36} \* \H(2) 
	    - {5 \over 2} \* \H(1) \* \z2 
	    + {65 \over 6} \* \Hhh(1,0,0) 
	    - {29 \over 12} \* \Hh(1,0) 
	    - {13 \over 18} \* \Hh(1,1) 
  \nonumber\\&&
	    - {1189 \over 108} \* \H(1) 
	    - {67 \over 3} \* \Hh(2,1) 
	    - 29 \* \Hh(2,0) 
	    - {949 \over 36} \* \z2 
	    - {67 \over 2} \* \Hhh(0,0,0) 
	    - {142 \over 3} \* \H(3) 
	    + {215 \over 32} 
	    - {3989 \over 48} \* \H(0) 
            + 2 \* \Hh(-3,0)
          \bigg]
  \nonumber\\&&
          + (1+x) \* \bigg[
              \Hhh(-1,0,0)
	    - 10 \* \H(-2) \* \z2
	    + 6 \* \Hhh(-2,0,0) 
	    + 2 \* \Hh(0,0) \* \z2
	    - 9 \* \Hhh(-1,-1,0)
	    - 7 \* \Hh(-1,2)
	    - 9 \* \Hh(-2,0)
            - 2 \* \Hh(3,1)
  \nonumber\\&&
            - 4 \* \Hhh(-2,-1,0)
	    - 4 \* \H(4)
	    - 4 \* \Hh(3,0)
            - 4 \* \Hhhh(0,0,0,0)
	    + {37 \over 2} \* \Hh(-1,0)
            + {5 \over 2} \* (1+x) \* \H(-1) \* \z2
          \bigg]
	   - 4 \* \Hhh(-2,0,0) 
	   + 2 \* \Hh(0,0) \* \z2 
  \nonumber\\&&
	   + \H(2) \* \z2 
	   - 3 \* \Hhh(1,1,0) 
          + 2 \* \Hhhh(0,0,0,0)
          + \Hh(-3,0)
	  - 9 \* \Hhh(2,1,0) 
          - {9 \over 2} \* \Hhh(2,1,1)
          + {11 \over 3} \* \Hhh(1,1,1) 
          + {19 \over 2}  \* \Hhh(2,0,0)
          + {9 \over 2} \* \Hh(1,2) 
  \nonumber\\&&
	  - {91 \over 2} \* \H(0) \* \z3 
          + 8 \* \H(-2) \* \z2 
          + {5 \over 2} \* \Hhh(-1,-1,0) 
          + {5 \over 2} \* \Hh(-1,2) 
          + {9 \over 2} \* \Hh(-1,0) 
          + {39 \over 2} \* \Hh(-2,0)
          - {473 \over 12} \* \H(0) \* \z2 
          - {1853 \over 48} \* \Hh(0,0) 
  \nonumber\\&&
          - {217 \over 12} \* \z3
          - {59 \over 4} \* \z2^2 
          - {169 \over 18} \* \H(2) 
          - {13 \over 4} \* \H(1) \* \z2 
          - {2 \over 3} \* \Hhh(1,0,0) 
          + {167 \over 24} \* \Hh(1,0) 
          + {191 \over 18} \* \Hh(1,1) 
          + {1283 \over 108} \* \H(1) 
          + {185 \over 12} \* \Hh(2,1) 
  \nonumber\\&&
          + {75 \over 4} \* \Hh(2,0) 
          + {170 \over 9} \* \z2 
          + {85 \over 4} \* \Hhh(0,0,0) 
          + {425 \over 12} \* \H(3) 
          + {7693 \over 192} 
          + {3659 \over 48} \* \H(0) 
      - 2 \* x \* \bigg[
            x \* \Hh(2,2)
           + 4  \* \Hh(3,0)
           - 4 \* \Hh(-2,2)
      \bigg]
          \bigg)
  \nonumber\\&&
       +  16\, \*  \colour4colour{\ca \* \nf^2}  \*  \bigg(
            {1 \over 6} \* \pqg(x) \* \bigg[
            \Hh(1,2)
          - \H(1) \* \z2
          - \Hhh(1,0,0)
          - \Hhh(1,1,0)
          - \Hhh(1,1,1)
	  - {229 \over 18} \* \H(0)
	  + {4 \over 3} \* \Hh(0,0)
	  + {11 \over 2} 
          \bigg]
          + x \* \bigg[
            {1 \over 6} \* \H(2)
  \nonumber\\&&
          - {53 \over 18} \* \H(0)
          + {17 \over 6} \* \H(0,0)
          - \z3
          + {11 \over 18} \* \z2
          - {139 \over 108}
          \bigg]
          + {1 \over 3} \* \pqg(-x) \* \Hhh(-1,0,0)
          - {53 \over 162} \* \big({1 \over x}-x^2\big) 
          - {2 \over 9} \* (1-x) \* \bigg[
            6 \* \Hhh(0,0,0)
  \nonumber\\&&
	  - {7 \over 6} \* x \* \H(1)
	  - \Hh(0,0)
          + {7 \over 2} \* x \* \Hh(1,1)
          \bigg]
          + {7 \over 9} \* x \* (1+x) \* \Hh(-1,0)
          + {7 \over 4} \* \H(0)
          - {19 \over 54} \* \H(1)
          + \Hhh(0,0,0)
          + {5 \over 9} \* \Hh(1,1)
          + {5 \over 9} \* \Hh(-1,0)
  \nonumber\\&&
          - {85 \over 216}
          \bigg)
       +  16\, \*  \colour4colour{\ca^2 \* \nf}  \*  \bigg(
           \pqg(x) \* \bigg[
            3 \* \Hh(1,3)
	    + {31 \over 6} \* \Hhh(1,0,0)
	    - {17 \over 2} \* \Hh(2,1)
	    + {7 \over 5} \* \z2^2
	    - {55 \over 12} \* \Hhh(1,1,0)
	    + {31 \over 12} \* \H(3)
            - {31 \over 2} \* \H(1) \* \z3
  \nonumber\\&&
	    - {5 \over 12} \* \Hh(2,0)
	    - {63 \over 8} \* \Hh(1,0)
	    - {23 \over 12} \* \Hh(1,2)
	    - {155 \over 6} \* \z3
	    + {25 \over 24} \* \H(2)
	    - {2537 \over 27} \* \H(0)
	    + {867 \over 8}
	    - {23 \over 2} \* \Hhh(-1,0,0)
	    + 3 \* \H(4)
	    - \Hhh(1,1,1)
  \nonumber\\&&
	    + {383 \over 72} \* \Hh(1,1)
	    - {25 \over 2} \* \Hh(-2,0)
	    - {3 \over 8} \* \z2
	    - {7 \over 4} \* \H(1) \* \z2
	    - 3 \* \Hh(0,0) \* \z2
	    - {31 \over 12} \* \H(0) \* \z2
	    + {103 \over 216} \* \H(1)
            + {5 \over 2} \* \Hhhh(1,0,0,0)
	    + {2561 \over 72} \* \Hh(0,0)
  \nonumber\\&&
	    + \Hhh(1,1,1)
	    - 2 \* \Hhh(2,0,0)
            - 3 \* \Hhh(1,-2,0)
            - 5 \* \Hh(1,0) \* \z2
	    + 3 \* \Hhh(0,0,0)
            - \Hh(1,1) \* \z2
            - \Hhhh(1,1,0,0)
            - 4 \* \Hhhh(1,1,1,0)
            + 2 \* \Hhhh(1,1,1,1)
  \nonumber\\&&
            - 2 \* \Hhh(1,1,2)
            - 2 \* \Hhh(1,2,0)
          \bigg]
          + \pqg(-x) \* \bigg[
            \Hh(-1,-1) \* \z2
	    - 2 \* \Hh(-1,2)
	    - 6 \* \Hhh(-1,-1,0)
	    + \Hhh(1,1,1) 
	    + 2 \* \H(-2) \* \z2
	    - \Hhh(-2,0,0)
  \nonumber\\&&
	    + {727 \over 36} \* \Hh(-1,0)
	    - \H(-1) \* \z2
            - 2 \* \Hh(-2,2)
            - {5 \over 2} \* \H(-1) \* \z3
            - \Hhh(-1,-2,0)
            + 2 \* \Hhhh(-1,-1,0,0)
            + 2 \* \Hhh(-1,-1,2)
            - {3 \over 2} \* \Hhhh(-1,0,0,0)
  \nonumber\\&&
            + 6 \* \Hhhh(-1,-1,-1,0)
            - 2 \* \Hh(-1,3)
            + 2 \* \Hhh(-1,2,1)
          \bigg]
          + \big({1 \over x}-x^2\big) \* \bigg[
            {2 \over 3} \* \Hh(2,1)
            + {32 \over 9} \* \z2
            - 2 \* \Hhh(1,0,0)
            + {4 \over 3} \* \Hhh(1,1,0)
            - {10 \over 9} \* \Hh(1,1)
  \nonumber\\&&
            - {8 \over 3} \* \Hhh(-1,0,0)
            + {3 \over 2} \* \Hh(1,0)
            + 6 \* \z3
            + {161 \over 36} \* \H(1)
            - {2351 \over 108} 
          \bigg]
          + {2 \over 3} \* \big({1 \over x}+x^2\big) \* \bigg[
            {26 \over 3} \* \Hh(-1,0)
          - {28 \over 9} \* \H(0)
          - 2 \* \Hhh(-1,-1,0)
  \nonumber\\&&
          - 2 \* \Hh(-1,2)
          + \H(1) \* \z2 
          + \H(-1) \* \z2
          + {10 \over 3} \* \H(2)
          + \Hhh(1,1,1)
          \bigg]
          + (1-x) \* \bigg[
	      15 \* \Hhhh(0,0,0,0)
	    - 5\* \H(2) \* \z2
	    - {65 \over 6} \* \z3
	    + {11 \over 6} \* \Hhh(1,1,1)
  \nonumber\\&&
	    - {3 \over 2} \* \H(4)
	    + {5 \over 2} \* \Hh(0,0) \* \z2
	    + \Hhh(1,1,0)
	    - {31 \over 6} \* \Hh(2,0)
	    + {17 \over 12} \* \Hh(1,0)
	    - {551 \over 20} \* \z2^2
	    - {29 \over 4} \* \Hhh(1,0,0)
	    - {113 \over 4} \* \H(2)
	    + {18691 \over 72} \* \H(0)
  \nonumber\\&&
	    + {2243 \over 108}
	    + {265 \over 6} \* \Hhh(-1,0,0)
	    + {33 \over 2} \* \Hhh(2,0,0)
	    + 19 \* \Hh(2,1)
	    + {31 \over 12} \* \Hh(1,1)
	    + {23 \over 2} \* \Hh(-2,0)
	    - {497 \over 36} \* \z2
	    + {29 \over 6} \* \H(1) \* \z2
	    - {143 \over 12} \* \H(3)
  \nonumber\\&&
	    - {11 \over 6} \* \Hhh(1,1,1)
	    - {19 \over 12} \* \H(0) \* \z2
	    + {1223 \over 72} \* \H(1)
	    - {43 \over 6} \* \Hhh(0,0,0)
	    - {3011 \over 36} \* \Hh(0,0)
          \bigg]
          + (1+x) \* \bigg[
	      8 \* \Hhh(2,1,0)
	    - 4 \* \Hh(-1,2)
  \nonumber\\&&
	    + 7 \* \Hhh(-1,-1,0)
	    - {35 \over 6} \* \Hhh(1,1,1)
	    - 5 \* \H(-2) \* \z2
	    - 11 \* \Hhh(-2,0,0)
	    + {1 \over 3} \* \Hh(-1,0)
	    + {15 \over 2} \* \H(-1) \* \z2
	    + 8 \* \Hh(3,1)
	    - 10 \* \Hhh(-2,-1,0)
  \nonumber\\&&
	    + 5 \* \H(2) \* \z2
	    + 4 \* \Hhh(2,1,1)
	    - \Hh(-3,0)
	    + 36 \* \H(0) \* \z3
	    - 5 \* \H(2) \* \z2
            \bigg]
              + 2 \* \Hh(-1,2) 
	      + 6 \* \Hhh(-1,-1,0)
	      - 6 \* \Hhh(2,1,0) 
	      - 3 \* \Hhh(2,1,1) 
  \nonumber\\&&
	      - 11 \* \Hhhh(0,0,0,0) 
	      - 5 \* \Hh(3,1) 
	      + {25 \over 4} \* \Hhh(1,1,1)
	      + {13 \over 2} \* \H(-2) \* \z2
	      + {27 \over 2} \* \Hhh(-2,0,0)
	      + {11 \over 2} \* \Hh(-3,0) 
	      + {13 \over 2} \* \H(2) \* \z2 
	      - {17 \over 4} \* \Hhh(1,0,0) 
  \nonumber\\&&
	      + 13 \* \Hhh(-2,-1,0) 
	      - {17 \over 12} \* \Hhh(1,1,1) 
	      - {3 \over 4} \* \H(4) 
	      - {1 \over 4} \* \Hh(0,0) \* \z2 
	      + \Hh(1,2) 
	      + {11 \over 2} \* \Hhh(1,1,0) 
	      + {79 \over 12} \* \Hh(2,0) 
	      + {67 \over 8} \* \Hh(1,0) 
	      + {263 \over 8} \* \z2^2 
  \nonumber\\&&
	      + {119 \over 3} \* \z3 
	      + {967 \over 24} \* \H(2) 
              - {305 \over 12} \* \Hh(-1,0) 
              - 24 \* \H(0) \* \z3 
              + \H(-1) \* \z2 
              - {13375 \over 72} \* \H(0) 
              - {1889 \over 18} 
              - 38 \* \Hhh(-1,0,0) 
              - {21 \over 2} \* \Hh(2,1) 
  \nonumber\\&&
              - {79 \over 4} \* \Hhh(2,0,0) 
              - {217 \over 24} \* \Hh(1,1) 
              - {7 \over 2} \* \Hh(-2,0) 
              + {79 \over 72} \* \z2 
              + {4 \over 3} \* \H(1) \* \z2 
              + {17 \over 12} \* \Hhh(1,1,1) 
              + {17 \over 12} \* \H(0) \* \z2
              + {31 \over 18} \* \H(1) 
              + 3 \* \Hhh(0,0,0) 
  \nonumber\\&&
              + {145 \over 12} \* \H(3) 
              + {1553 \over 24}  \* \Hh(0,0)
          \bigg)
       +  16\, \* \colour4colour{\cf \* \nf^2}  \*  \bigg(
            {7 \over 6} \* \Hhh(0,0,0) 
          + {11 \over 36} \* \H(1) 
	  - {739 \over 96}
          + {163 \over 24} \* \H(0)
          + {7 \over 24} \* \Hh(0,0)
          + 2 \* \Hhhh(0,0,0,0)
  \nonumber\\&&
          - {5 \over 9} \* \Hh(1,1)
          - {5 \over 9} \* \H(2)
          - {5 \over 18} \* \Hh(1,0)
          + {5 \over 9} \* \z2
          + {1 \over 6} \* \pqg(x) \* \bigg[
            \Hh(2,1)
	  + {91 \over 2} 
	  - {35 \over 3} \* \H(0)
	  - {22 \over 3} \* \Hh(0,0)
          + \Hhh(1,1,1)
	  + 6 \* \Hhh(0,0,0)
  \nonumber\\&&
          - \z3
          - 2 \* \Hhh(1,0,0)
	  + {7 \over 9} \* \H(1)
          \bigg]
          + {77 \over 81} \* \big({1 \over x}-x^2\big) 
          + (1-x) \* \bigg[
	    {1 \over 12} \* \H(1) 
          - {6463 \over 432} 
	  - 4 \* \Hhhh(0,0,0,0)
	  - {16 \over 3} \* \Hhh(0,0,0) 
	  + {7 \over 9} \* x \* \Hh(1,1)
  \nonumber\\&&
          + {7 \over 9} \* x \* \H(2)
          + {8 \over 9} \* x \* \Hh(1,0)
	  - {7 \over 9} \* x \* \z2
          \bigg]
          - (1+x) \* \bigg[
            {3475 \over 216} \* \H(0) 
          + {103 \over 12} \* \Hh(0,0)
          \bigg]
          \bigg)
       +  16\, \*  \colour4colour{\cf^2 \* \nf}  \*  \bigg(
          \pqg(x) \* \bigg[
              7 \* \Hh(1,3)
	    + 7 \* \H(4)
  \nonumber\\&&
            - 2 \* \Hh(-3,0)
            - 7 \* \H(1) \* \z3
	    + 5 \* \Hh(2,2)
	    + 6 \* \Hh(3,0)
	    + 6 \* \Hh(3,1)
	    + \Hhh(2,1,0)
	    + 4 \* \Hhh(2,0,0)
	    + 3 \* \Hh(2,1)
	    + 2 \* \Hhh(2,1,1)
	    + {5 \over 2} \* \Hh(2,0)
  \nonumber\\&&
	    + {61 \over 8} \* \H(2)
	    - {61 \over 8} \* \z2
	    + {87 \over 8} \* \H(1)
	    + {11 \over 2} \* \Hh(1,2)
	    + {61 \over 8} \* \Hh(1,1)
	    + {17 \over 2} \* \Hh(1,0)
	    - 7 \* \Hh(0,0) \* \z2
	    + {5 \over 2} \* \Hhh(1,0,0)
	    + {5 \over 2} \* \Hhh(1,1,0)
	    - {19 \over 2} \* \z3
  \nonumber\\&&
	    + {81 \over 32}
	    + {11 \over 2} \* \H(3)
	    - {11 \over 2} \* \H(0) \* \z2
	    - {7 \over 2} \* \H(1) \* \z2
	    + {15 \over 2} \* \Hhh(0,0,0)
	    + {87 \over 8} \* \H(0)
	    + {11 \over 5} \* \z2^2
	    + 3 \* \Hhh(1,1,1)
	    - 5 \* \H(2) \* \z2
	    - 7 \* \H(0) \* \z3
  \nonumber\\&&
	    + 11\* \Hh(0,0)
            - 2 \* \Hhh(1,-2,0)
            - 7 \* \Hh(1,0) \* \z2
            + 3 \* \Hhhh(1,0,0,0)
            - 5 \* \Hh(1,1) \* \z2
            + 4 \* \Hhhh(1,1,0,0)
            + \Hhhh(1,1,1,0)
            + 2 \* \Hhhh(1,1,1,1)
            + 5 \* \Hhh(1,1,2)
  \nonumber\\&&
            + 6 \* \Hhh(1,2,0)
            + 6 \* \Hhh(1,2,1)
          \bigg]
          + 4 \* \pqg(-x) \* \bigg[
	      \Hhhh(0,0,0,0)
	    - \Hh(-2,0)
	    + \Hhh(-1,-1,0)
            - \Hhh(-2,0,0)
            + {1 \over 2} \* \Hhh(-1,-2,0)
	    - {5 \over 8} \* \Hh(-1,0)
  \nonumber\\&&
	    - {5 \over 4} \* \Hhh(-1,0,0)
            - {1 \over 2} \* \Hh(-3,0)
	    + {1 \over 2} \* \H(-1) \* \z2
            + \Hhhh(-1,-1,0,0)
            - {1 \over 4} \* \Hhhh(-1,0,0,0)
          \bigg]
          + 2 \* (1-x) \* \bigg[
	       \Hhh(2,1,0)
	     - \Hhh(2,0,0)
	     - \Hh(2,2)
  \nonumber\\&&
	     - \Hh(3,1)
	     - 2 \* \Hh(3,0)
	     - 2 \* \H(-1) \* \z2
	     + \Hh(1,2)
	     - \Hhh(1,0,0)
	     - \Hhh(1,1,0)
	     + \H(2) \* \z2
	     - \z2^2
	     + {43 \over 8} \* \H(2)
	     + {49 \over 8} \* \z2
	     + {13 \over 8} \* \Hh(1,1)
  \nonumber\\&&
	     - {33 \over 16} \* \H(1)
	     + {5 \over 2} \* \Hh(1,0)
	     + {7 \over 2} \* \Hh(0,0) \* \z2
	     + {21 \over 4} \* \z3
	     + {479 \over 64}
	     - {1 \over 2} \* \Hhh(1,1,1)
	     - {1 \over 2} \* \H(3)
	     + {1 \over 4} \* \Hh(2,1)
	     + {1 \over 2} \* \Hhh(2,1,1)
	     + {3 \over 2} \* \H(0) \* \z2
  \nonumber\\&&
	     + {1 \over 2} \* \H(0) \* \z3
	     - {7 \over 2} \* \H(4)
	     + \H(1) \* \z2
	     - {19 \over 2} \* \Hhh(0,0,0)
	     - {239 \over 16}\* \Hh(0,0)
	     - {405 \over 32} \* \H(0)
          \bigg] 
          + 8 \* (1+x) \* \bigg[
	        \Hhh(-1,-1,0)
	      - \Hhh(-1,0,0)
  \nonumber\\&&
	      - \Hhhh(0,0,0,0)
	      + {3 \over 4} \* \Hh(-2,0)
	      - {9 \over 4} \* \Hh(-1,0)
          \bigg]
          - 4 \* \Hhh(-1,-1,0)
          + 5 \* \Hhhh(0,0,0,0)
          + 5 \* \Hhh(-1,0,0) 
          - 13 \* \Hh(-2,0) 
          + {1 \over 2} \* \Hh(-1,0) 
  \nonumber\\&&
          + 4 \* x \* \Hhh(-2,0,0)
          - {113 \over 8} \* \H(2) 
          - {71 \over 8} \* \z2 
          - {35 \over 4} \* \H(1) 
          - {11 \over 2} \* \Hh(1,2) 
          - {33 \over 8} \* \Hh(1,1) 
          - {7 \over 2} \* \Hh(1,0) 
          - {7 \over 2} \* \Hh(0,0) \* \z2 
          - {5 \over 2} \* \Hhh(1,0,0) 
  \nonumber\\&&
          - {5 \over 2} \* \Hhh(1,1,0) 
          - {5 \over 2} \* \z3 
          - {157 \over 64} 
          - {9 \over 4} \* \Hhh(1,1,1) 
          - {9 \over 4} \* \H(3)
          - {5 \over 4} \* \Hh(2,1) 
          - {1 \over 2}  \* \Hhh(2,1,1) 
          + {1 \over 4} \* \H(0) \* \z2 
          + \H(2) \* \z2 
          + {5 \over 2} \* \H(0) \* \z3
  \nonumber\\&&
          + {9 \over 5} \* \z2^2  
          + {7 \over 2} \* \H(4) 
          + {7 \over 2} \* \H(1) \* \z2 
          + {49 \over 4}  \* \Hhh(0,0,0)
          + {391 \over 16} \* \Hh(0,0) 
          + {401 \over 16} \* \H(0) 
          - \Hhh(2,0,0)
          - \Hhh(2,1,0) 
          + \Hh(2,2) 
          + \Hh(3,1) 
  \nonumber\\&&
          + 2 \* \Hh(3,0) 
          + 6 \* \H(-1) \* \z2
          + {1 \over 2} \* \Hh(2,0)
          + 2 \* \H(-2) \* \z2
          + 4 \* \Hhh(-2,-1,0)
          \bigg)
\label{eq:Pqg2}
\eea
and
\bea
  &&P^{\,(2)}_{\rm gq}(x) \:\: = \:\: 
       16\, \*  \colour4colour{\ca \* \cf \* \nf}  \*  \bigg(
           {2 \over 9} \* x^2 \* \bigg[
	      {25 \over 6} \* \H(1)
	    - {131 \over 4}
	    + 3 \* \z2
 	    - \Hh(-1,0)
	    - 3 \* \H(2)
	    + \Hh(1,1)
	    + {125 \over 6} \* \H(0)
	    - \Hh(0,0)
          \bigg]
  \nonumber\\&&
          + {5 \over 6} \* \pgq(x) \* \bigg[
            \Hh(1,2)
          + \Hh(2,1)
          + {967 \over 120}
          + {251 \over 90} \* \H(1)
          - {39 \over 10} \* \Hh(1,1)
          - 3 \* \z3
          - {2 \over 5} \* \H(0) \* \z2
          - {1 \over 5} \* \H(1) \* \z2
          - {4 \over 3} \* \Hh(1,0)
          + \Hhh(1,1,0)
  \nonumber\\&&
          - {2 \over 5} \* \Hhh(1,0,0)
          + \Hhh(1,1,1)
          + {2 \over 5} \* \Hh(2,0)
          \bigg]
          + {2 \over 3} \* \pgq(-x) \* \bigg[
             2 \* \H(-1) \* \z2
          + {7 \over 4} \* \z2
          + {41 \over 12} \* \Hh(-1,0)
          - {151 \over 72} \* \H(0)
          + {1 \over 2} \* \Hh(-2,0)
  \nonumber\\&&
          + {5 \over 3} \* \H(2)
          + 2 \* \Hhh(-1,-1,0)
          - \Hhh(-1,0,0)
          - \Hh(-1,2)
          \bigg]
          + {2 \over 3} \* (1-x) \* \bigg[
            \Hh(-2,0)
          + 2 \* \z3
          - \H(3)
          \bigg]
          + (1+x) \* \bigg[
            {179 \over 108} \* \H(1)
  \nonumber\\&&
          + {5 \over 9} \* \z2
          + {25 \over 9} \* \Hh(-1,0)
          - {5 \over 36} \* \Hh(1,1)
          - {167 \over 36} \* \Hh(0,0)
          - {1 \over 3} \* \Hh(2,1)
          - {4 \over 3} \* \H(0) \* \z2
          \bigg]
          - {193 \over 72} 
	  + {1 \over 4} \* \H(1)
          + {1 \over 9} \* \Hh(-1,0)
          + 4 \* \H(2)
  \nonumber\\&&
          - {1 \over 4} \* \Hh(1,1)
          + {227 \over 18} \* \H(0)
          - {35 \over 12} \* \Hh(0,0)
          - \Hh(2,1)
          - {2 \over 3} \* \H(0) \* \z2
          + {10 \over 3} \* \Hh(-2,0)
          + 3 \* \z3
          + 2 \* \H(3)
          + {2 \over 3} \* \Hhh(0,0,0)
          + x \* \bigg[
            {11 \over 4} \* \z2
  \nonumber\\&&
	  - {523 \over 144} 
          - {19 \over 36} \* \H(2)
          + {271 \over 108} \* \H(0)
          - {5 \over 6} \* \Hh(1,0)
          \bigg]
          \bigg)
       +  16\, \*  \colour4colour{\ca \* \cf^2}  \*  \bigg(
            x^2 \* \bigg[
	      {7 \over 2} 
	    + {173 \over 54}  \* \H(1)
	    - 2  \* \z3
	    - {2 \over 3}  \* \Hhh(1,1,1)
	    - {26 \over 9}  \* \Hh(1,1)
  \nonumber\\&&
	    - 6  \* \H(2)
	    + 2  \* \Hh(2,1)
	    + 6  \* \z2
	    + {335 \over 54} \* \H(0)
          - {28 \over 9} \* \Hh(0,0)
          - {8 \over 3} \* \Hhh(0,0,0)
          \bigg]
          + \pgq(x) \* \bigg[
            {3 \over 2} \* \H(1) \* \z3
          + {163 \over 32}
          - 5 \* \z2
          + {27 \over 4} \* \z3
  \nonumber\\&&
          + {6503 \over 432} \* \H(1)
          + {2 \over 9} \* \Hh(1,1)
          + {35 \over 3} \* \Hhh(1,1,1)
          + 4 \* \H(2)
          + {9 \over 2} \* \Hh(2,1)
	    + 4 \* \Hhh(1,0,0)
	    + 2 \* \Hhh(2,0,0)
	    - \H(2) \* \z2
	    + {41 \over 12} \* \Hh(1,2)
	    + \Hh(2,2)
  \nonumber\\&&
	    + {191 \over 24} \* \Hh(1,0)
	    + 3 \* \Hh(2,0)
	    - 2 \* \Hhh(2,1,1)
	    - {3 \over 2} \* \H(-1) \* \z2
	    - {59 \over 12} \* \H(1) \* \z2
          + 5 \* \Hhh(1,-2,0)
          + \Hh(1,0) \* \z2
          + {5 \over 2} \* \Hhhh(1,0,0,0)
          - 2 \* \Hh(1,1) \* \z2
  \nonumber\\&&
          + {1 \over 12} \* \Hhh(1,1,0)
          + 5 \* \Hhhh(1,1,0,0)
          - 3 \* \Hhhh(1,1,1,0)
          - 4 \* \Hhhh(1,1,1,1)
          - \Hhh(1,1,2)
          - 2 \* \Hhh(1,2,1)
          + \Hhh(2,1,0)
          \bigg]
          + \pgq(-x) \* \bigg[
            \Hh(-1,0) 
  \nonumber\\&&
	   + \Hh(-1,0) \* \z2
	   + {3 \over 2} \* \Hhh(-1,0,0)
	   + {27 \over 10} \* \z2^2
	   - 3 \* \Hhh(-1,-1,0)
          - {11 \over 2} \* \H(-1) \* \z3
          - 3 \* \Hhh(-1,-2,0)
          - {3 \over 2} \* \Hhhh(-1,0,0,0)
          - 3 \* \Hh(-1,2)
  \nonumber\\&&
          + 5 \* \Hh(-1,-1) \* \z2
          - 4 \* \Hhhh(-1,-1,0,0)
          - 2 \* \Hhh(-1,-1,2)
          + 6 \* \Hhhh(-1,-1,-1,0)
          + 2 \* \Hhh(-1,2,1)
          \bigg]
          + (1-x) \* \bigg[
            \H(2) \* \z2
          - \Hh(2,2)
  \nonumber\\&&
          + {23 \over 12} \* \Hh(1,0)
          - {7061 \over 432} \* \H(0)
          - {4631 \over 144} \* \Hh(0,0)
          - {38 \over 3} \* \Hhh(0,0,0)
          - \Hh(-3,0)
          - 2 \* \Hh(3,0)
          - {4433 \over 432} \* \H(1)
          - 2 \* \Hhh(2,0,0)
          - {21 \over 2} \* \Hhh(1,0,0)
  \nonumber\\&&
          - {2 \over 5} \* \z2^2
          - {7 \over 2} \* \Hh(1,2)
          + {23 \over 2} \* \H(1) \* \z2
          - 4 \* \H(0) \* \z3
          \bigg]
          + (1+x) \* \bigg[
            {49 \over 6} \* \H(3)
          - \Hh(-2,0)
          - {55 \over 6} \* \H(0) \* \z2
          - {1 \over 2} \* \Hh(3,1)
          - {1159 \over 36} \* \z2
  \nonumber\\&&
          + {655 \over 576} 
          - {151 \over 6} \* \z3
          - {185 \over 18} \* \Hh(1,1)
          + {1 \over 6} \* \Hhh(1,1,1)
          - {95 \over 9} \* \H(2)
          + {29 \over 6} \* \Hh(2,1)
          - {171 \over 4} \* \Hh(-1,0)
          - 12 \* \Hhh(-1,0,0)
	  + 7 \* \H(-1) \* \z2
  \nonumber\\&&
          + 16 \* \Hhh(-1,-1,0)
          + {5 \over 3} \* \Hh(2,0)
          + {3 \over 2} \* \Hhh(2,1,1)
          + 4 \* \Hhhh(0,0,0,0) 
          \bigg]
          - 35 \* \Hh(-2,0)
          - {179 \over 27} \* \H(0)
          + {2041 \over 144} \* \Hh(0,0)
          - {19 \over 6} \* \Hhh(0,0,0)
  \nonumber\\&&
          - 2 \* \Hh(3,0)
          - {13 \over 2} \* \H(0) \* \z2
          - 13 \* \Hh(-3,0)
          - {13 \over 2} \* \Hh(3,1)
          + {15 \over 2} \* \H(3)
	  - {2005 \over 64}
          + {157 \over 4} \* \z2
          + 8 \* \z3
          + {1291 \over 432} \* \H(1)
          + {55 \over 12} \* \Hh(1,1)
  \nonumber\\&&
          + {3 \over 2}  \* \H(2)
          + {1 \over 2} \* \Hh(2,1)
          + {27 \over 4} \* \Hh(-1,0)  
          - {11 \over 2} \* \Hhh(1,0,0)
          - 8 \* \Hhh(2,0,0)
          - 4 \* \z2^2
          + {3 \over 2} \* \Hh(1,2)
          - \Hh(2,2)
          + {5 \over 2} \* \H(1) \* \z2
          + 8 \* \Hhh(-1,-1,0)
  \nonumber\\&&
          + 4 \* \Hh(2,0)
          + {3 \over 2} \* \Hhh(2,1,1)
          - \H(-1) \* \z2
          + 7 \* \H(2) \* \z2
          + 6 \* \H(-2) \* \z2
          + 12 \* \Hhh(-2,-1,0)
          - 6 \* \Hhh(-2,0,0)
          + x \* \bigg[
	    3 \* \Hhh(1,1,1)
          - \Hh(0,0) \* \z2
  \nonumber\\&&
          + {9 \over 2} \* \Hhh(-1,0,0)
          - {35 \over 8} \* \Hh(1,0)
          + 2 \* \H(4)
          + 3 \* \Hhh(1,1,0)
          + \Hh(-1,2)
          \bigg]
          \bigg)
       +  16\, \*  \colour4colour{\ca^2 \* \cf}  \*  \bigg(
            x^2 \* \bigg[
 	    {2 \over 3} \* \H(1) \* \z2
          - {2105 \over 81}
	  - {77 \over 18} \* \Hh(0,0)
  \nonumber\\&& 
	  - 6 \* \H(3)
	  + {16 \over 3} \* \z3
	  - 10 \* \Hh(-1,0)
	  - {14 \over 3} \* \Hh(2,0)
	  - {2 \over 3} \* \H(-1) \* \z2
	  - {14 \over 3} \* \Hhh(0,0,0)
	  + {104 \over 9} \* \H(2)
	  - {4 \over 3} \* \Hhh(1,0,0)
	  + {37 \over 9} \* \Hh(1,1)
  \nonumber\\&&
	  + {4 \over 3} \* \Hhh(-1,-1,0)
	  - {104 \over 9} \* \z2
	  - {8 \over 3} \* \Hh(2,1)
	  + {145 \over 18} \* \Hh(1,0)
	  + {4 \over 3} \* \Hh(-1,2)
	  + {2 \over 3} \* \Hhh(1,1,1)
	  - {109 \over 27} \* \H(1)
	  + {8 \over 3} \* \Hhh(-1,0,0)
	  + 6 \* \H(0) \* \z2
  \nonumber\\&&
	  + 4 \* \Hh(-2,0)
	  + {584 \over 27} \* \H(0)
          \bigg]
          + \pgq(x) \* \bigg[
            {7 \over 2} \* \H(1) \* \z3
	    + {138305 \over 2592} 
	    - {1 \over 3} \* \Hh(2,0)
	    + {13 \over 4} \* \H(-1) \* \z2
	    + 2 \* \Hhh(2,1,1)
	    + {11 \over 2} \* \Hhh(1,0,0)
  \nonumber\\&&
	    + 4 \* \Hh(3,1)
	    - {43 \over 6} \* \Hhh(1,1,1)
	    - {109 \over 12} \* \z2
	    - {17 \over 3} \* \Hh(2,1)
	    - {71 \over 24} \* \Hh(1,0)
	    - {11 \over 6} \* \Hh(-2,0)
	    - {21 \over 2} \* \z3
          + {3 \over 2} \* \Hhhh(1,0,0,0)
          - \Hhh(1,-2,0)
  \nonumber\\&&
	    + {395 \over 54} \* \H(0)
          - 2 \* \Hh(1,0) \* \z2
          - \Hh(1,1) \* \z2
          - {55 \over 12} \* \Hhh(1,1,0)
          + 2 \* \Hhhh(1,1,0,0)
          + 4 \* \Hhhh(1,1,1,0)
          + 2 \* \Hhhh(1,1,1,1)
          + 4 \* \Hhh(1,1,2)
          - {55 \over 12} \* \Hh(1,2)
  \nonumber\\&&
          + 6 \* \Hhh(1,2,0)
          + 4 \* \Hhh(1,2,1)
          + 4 \* \Hh(1,3)
          + 3 \* \Hhh(2,1,0)
          + 3 \* \Hh(2,2)
          \bigg]
          + \pgq(-x) \* \bigg[
            {23 \over 2} \* \H(-1) \* \z3
	    + 5 \* \H(-2) \* \z2
	    + 2 \* \Hhh(-2,-1,0)
  \nonumber\\&&
	    + {109 \over 12} \* \Hh(-1,0)
	    + \H(0) \* \z3
	    + {17 \over 5} \* \z2^2
	    + {1 \over 6} \* \H(1) \* \z2
	    + 2 \* \H(2) \* \z2
	    - {65 \over 24} \* \Hh(1,1)
	    - {19 \over 2} \* \Hhh(-1,-1,0)
	    - 4 \* \Hh(3,0)
	    - 3 \* \Hhh(2,0,0)
  \nonumber\\&&
	    - 7 \* \Hhh(-2,0,0)
	    - {3 \over 2} \* \Hh(-1,2)
	    + {3379 \over 216} \* \H(1)
	    - 4 \* \H(-2,2)
	    - {49 \over 6} \* \Hhh(-1,0,0)
          - {11 \over 2} \* \Hhhh(-1,0,0,0)
          - 13 \* \Hh(-1,-1) \* \z2
          - 8 \* \Hh(-1,3)
  \nonumber\\&&
          - 6 \* \Hhhh(-1,-1,-1,0)
          + 12 \* \Hhhh(-1,-1,0,0)
          + 10 \* \Hhh(-1,-1,2)
          + 10 \* \Hh(-1,0) \* \z2
          + 5 \* \Hhh(-1,-2,0)
          - 2 \* \Hhh(-1,2,0)
          - 2 \* \Hhh(-1,2,1)
  \nonumber\\&&
	    + {11 \over 6} \* \H(0) \* \z2
          \bigg]
          + (1-x) \* \bigg[
	    {41699 \over 2592}
          - 3 \* \Hhh(-2,-1,0)
          - {3 \over 2} \* \H(-2) \* \z2
          - {128 \over 9} \* \z2
          - 4 \* \Hh(3,0)
          + {26 \over 3} \* \z3
          - {5 \over 2} \* \Hhh(-2,0,0)
  \nonumber\\&&
          - 7 \* \H(1) \* \z2
          + {97 \over 12} \* \Hhh(1,0,0)
          + {10 \over 3} \* \Hhh(-1,0,0)
          + {245 \over 12} \* \H(3)
          - 8 \* \Hhhh(0,0,0,0)
          \bigg]
          + (1+x) \* \bigg[
            4 \* \Hh(3,1)
          - \Hhh(2,1,1)
          + {29 \over 6} \* \Hh(-1,2)
  \nonumber\\&&
          + {17 \over 6} \* \Hh(-2,0)
          - 12 \* \H(2,0)
          - {31 \over 12} \* \Hh(2,1)
          + {1 \over 2} \* \Hhh(2,0,0)
          - \H(2) \* \z2
          + {61 \over 36} \* \Hh(1,0)
          - 4 \* \H(0) \* \z3
          - {13 \over 3} \* \H(-1) \* \z2
          - {46 \over 3} \* \Hhh(-1,-1,0)
  \nonumber\\&&
          + {25 \over 4} \* \H(4)
          + {93 \over 4} \* \H(0) \* \z2
          - {55 \over 9} \* \Hh(1,1)
          - {71 \over 18} \* \H(2)
          + {49 \over 18} \* \Hh(0,0)
          - {13 \over 2} \* \Hh(0,0) \* \z2
          - {47 \over 40} \* \z2^2
          \bigg]
          + {6131 \over 2592} 
          - {31 \over 2} \* \H(-2) \* \z2
  \nonumber\\&&
          - 15 \* \Hhh(-2,-1,0)
          + {9 \over 2} \* \Hhh(-1,0,0)
          - 3 \* \Hhh(2,1,1)
          - {9 \over 4} \* \Hh(2,1)
          + {53 \over 3} \* \Hh(-2,0)
          - {1 \over 2} \* \Hhh(-2,0,0)
          - 5 \* \Hh(2,0)
          - {7 \over 6} \* \Hhh(1,1,1)
          - 8 \* \H(0) \* \z3
  \nonumber\\&&
          - {67 \over 40} \* \z2^2
          + {29 \over 6} \* \Hh(-1,2)
          - \Hh(-1,0)
          + 8 \* \Hh(-2,2)
          + 25 \* \H(0) \* \z2
          + {412 \over 9} \* \H(1)
          + {928 \over 9} \* \H(0)
          + {1 \over 4} \* \H(4)
          - 65 \* \H(3)
	  - 38 \* \Hh(0,0)
  \nonumber\\&&
          - 9 \* \Hh(-3,0)
          - {17 \over 3} \* \Hhh(0,0,0)
          + x \* \bigg[
            {27 \over 2} \* \Hh(-1,0)
          - {1 \over 2} \* \Hhhh(0,0,0,0)
          + {3 \over 4} \* \Hh(0,0) \* \z2
          + {1 \over 2} \* \Hh(-3,0)
          - 14 \* \H(0,0,0)
          + {1 \over 12} \* \Hhh(1,1,1)
  \nonumber\\&&
          - {43 \over 36} \* \z2
          - {1 \over 2} \* \H(2) \* \z2
          + {7 \over 72} \* \H(0)
          + {749 \over 54} \* \H(1)
          + {135 \over 4} \* \z3
          + {97 \over 24} \* \Hh(1,0)
          + {43 \over 12} \* \H(1) \* \z2
          - {85 \over 12} \* \H(-1) \* \z2
          - {13 \over 3} \* \Hhh(1,0,0)
  \nonumber\\&&
	  + {53 \over 12} \* \H(2)
          + {39 \over 4} \* \Hh(1,1)
          - 2 \* \Hh(3,1)
          + {13 \over 6} \* \Hhh(-1,-1,0)
          + {7 \over 4} \* \Hhh(2,0,0)
          - 4 \* \Hhh(1,1,0)
          - 4 \* \Hh(1,2)
          \bigg]
          \bigg)
       +  16\, \*  \colour4colour{\cf \* \nf^2}  \*  \bigg(
            {1 \over 9} - {1 \over 9} \* {1 \over x} 
  \nonumber\\&&
	    + {2 \over 9} \* x
          - {1 \over 6} \* x \* \H(1)  
          + {1 \over 6} \* \pgq(x) \* \bigg[
            \Hh(1,1)
          - {5 \over 3} \* \H(1)
          \bigg]
          \bigg)
       +  16\, \*  \colour4colour{\cf^2 \* \nf}  \*  \bigg(
            {4 \over 9} \* x^2 \* \bigg[
	      \Hh(0,0)
	    - {11 \over 6} \* \H(0)
	    - {7 \over 2}
	    + \Hh(-1,0)
          \bigg]
  \nonumber\\&&
          + {1 \over 3} \* \pgq(x) \* \bigg[
            \Hh(1,2)
          - \Hh(1,0)
          - \H(1) \* \z2
          + 9 \* \z3
          + {83 \over 12} \* \Hh(1,1)
          + 2 \* \Hh(-2,0)
          - {7 \over 36} \* \H(1)
          + 2 \* \H(0) \* \z2
          - {1625 \over 48} 
          + {3 \over 2} \* \Hhh(1,0,0)
  \nonumber\\&&
          + 2 \* \Hhh(1,1,0)
          - {5 \over 2} \* \Hhh(1,1,1)
          \bigg]
          + {31 \over 18} \* \pgq(-x) \* \bigg[
            {95 \over 93} \* \H(0)
          - \z2
          - \H(-1,0)
          \bigg]
          + {1 \over 3} \* (2-x) \* \bigg[
            6 \* \Hhhh(0,0,0,0)
          - \H(3)
	  - {13051 \over 288} 
  \nonumber\\&&
          - {13 \over 2} \* \z3
          - 4 \* \Hh(-2,0)
          - \Hh(2,0)
	  - {1 \over 2} \* \Hh(1,0)
          - {1 \over 2} \* \Hh(2,1)
          + 2 \* \Hhh(0,0,0)
	  - {653 \over 24} \* \Hh(0,0)
          \bigg]
          + (1+x) \* \bigg[
            \H(0) \* \z2
	  - {1187 \over 216} \* \H(0)
  \nonumber\\&&
          + {8 \over 9} \* \H(2) 
	  - {85 \over 18} \* \Hh(-1,0)
	  - {101 \over 18} \* \z2
          \bigg]
          - {80 \over 27} \* \H(0)
          + {23 \over 18} \* \z2
          - {1 \over 3} \* \Hh(1,1) 
	  + {5 \over 4} \* x \* \Hh(1,1)
          - {1 \over 9} \* \H(1) 
	  - {37 \over 12} \* x \* \H(1)
          + {23 \over 18} \* \Hh(-1,0)
  \nonumber\\&&
          + {1501 \over 54} 
          + \H(0) \* \z2
	  - \Hhh(0,0,0)
          + {101 \over 3} \* \Hh(0,0)
	  - {1 \over 3} \* \Hh(1,0)
          \bigg)
       + 16\, \*  \colour4colour{\cf^3}  \*  \bigg(
           \pgq(x) \* \bigg[
            3 \* \Hh(1,1) \* \z2
	    + 3 \* \H(1) \* \z2
	    + {7 \over 2} \* \z2
  \nonumber\\&&
	    - {23 \over 8} \* \Hh(1,1)
            - 8 \* \H(1) \* \z3
            - 6 \* \Hhh(1,-2,0)
            - 2 \* \Hh(1,0) \* \z2
            + 3 \* \Hhh(1,1,0)
            - 3 \* \Hhhh(1,1,0,0)
            - \Hhhh(1,1,1,0)
            + 2 \* \Hhhh(1,1,1,1)
            - 3 \* \Hhh(1,1,2)
  \nonumber\\&&
            - 2 \* \Hhh(1,2,0)
            - 2 \* \Hhh(1,2,1)
	    - {9 \over 2} \* \Hhh(1,1,1)
	    - {3 \over 2} \* \Hhh(1,0,0)
	    - {47 \over 16} 
	    - {47 \over 16} \* \H(1)
	    - {15 \over 2} \* \z3
          \bigg]
          + \pgq(-x) \* \bigg[
            2 \* \Hhh(-1,-2,0)
  \nonumber\\&&
	    + 6 \* \H(-1,-1,0)
	    + 3 \* \H(-1) \* \z2
	    + {7 \over 4} \* \Hh(1,0)
	    - {16 \over 5} \* \z2^2
	    - 6 \* \Hhh(-1,0,0)
	    - {7 \over 2} \* \Hh(-1,0)
            + 4 \* \Hhhh(-1,-1,0,0)
            - 2 \* \Hh(-1,0) \* \z2
  \nonumber\\&&
            - \Hhhh(-1,0,0,0)
          \bigg]
          + (1-x) \* \bigg[
	      9 \* \Hhh(1,0,0)
	    + \Hhh(1,1,1)
            - 10 \* \H(1) \* \z2
            + 3 \* \H(0) \* \z3
            + \Hh(2,2)
            - \H(2) \* \z2
            + \Hhh(0,0,0)
            + 5 \* \Hhh(2,0,0)
  \nonumber\\&&
	    - 4 \* \H(3)
            + \Hhh(2,1,1)
            + 3 \* \Hh(0,0) \* \z2
            + 3 \* \Hh(3,1)
            - 3 \* \H(4)
	    + {211 \over 16} \* \H(1)
	    + {49 \over 20} \* \z2^2 
         \bigg]
          + (1+x) \* \bigg[
              11 \* \z3
            + {1 \over 4} \* \Hh(1,1)
            + {1 \over 4} \* \Hh(1,0)
  \nonumber\\&&
            + {91 \over 16} \* \H(0)
            + 36 \* \Hh(-1,0)
            + 8 \* \Hhh(-1,0,0)
            - 14 \* \Hhh(-1,-1,0)
            - 7 \* \H(-1) \* \z2
            + 2 \* \Hh(1,2)
            + 4 \* \H(0) \* \z2
            - \Hh(2,1)
            + 2 \* \Hhh(-2,0,0)  
  \nonumber\\&&
            + 5 \* \Hh(-2,0)
	    + {11 \over 2} \* \H(2)
            - 2 \* \Hhhh(0,0,0,0) 
          \bigg]
          - 2 \* \Hhh(-1,-1,0)
          - \H(-1) \* \z2 
          - {13 \over 4} \* \z2 
          + {9 \over 4} \* \Hh(1,0)
          + {9 \over 20} \* \z2^2
          + {287 \over 32} 
          + {11 \over 16}  \* \H(1) 
  \nonumber\\&&
          + 4 \* \Hhh(-1,0,0)
          + 16 \* \Hh(-3,0)
          - 4 \* \H(-2) \* \z2
          - 8 \* \Hhh(-2,-1,0)
          - 5 \* \H(2) \* \z2
          + {19 \over 4} \* \H(2)
          + \H(2,2)
	  - {35 \over 8} \* \Hh(0,0)
          + 9 \* \H(0) \* \z3
  \nonumber\\&&
          + 25 \* \Hh(-2,0)
          + 6 \* \Hhh(-2,0,0)
          + {3 \over 2} \* x \* \bigg[
	    {58 \over 3} \* \z2
          - {7 \over 3} \* \H(1) \* \z2
          + 4 \* \Hh(1,1)
          - {3 \over 2} \* \Hhh(1,1,1)
          + {5 \over 2} \* \Hhh(1,0,0)
	  - {175 \over 96} 
          + \Hh(3,1)
          + {19 \over 3} \* \z3
  \nonumber\\&&
          + 2 \* \Hh(2,0)
	  - 14 \* \H(0)
          + \Hh(0,0) \* \z2
          - \Hh(-1,0)
          - \H(4)
          - {3 \over 2} \* \Hh(2,1)
          + {1 \over 3} \* \Hhh(2,1,1)
          + 3 \* \Hhh(2,0,0)
          - {5 \over 6} \* \H(3)
          - \Hh(1,2)
          - {7 \over 6} \* \H(0) \* \z2
  \nonumber\\&&
          + {2 \over 3} \* \Hhh(1,1,0)
          - {29 \over 6} \* \Hhh(0,0,0)
          - {185 \over 8} \* \Hh(0,0)
          \bigg]
          \bigg)
\:\ .\label{eq:Pgq2}
\eea
Finally the Mellin inversion of Eq.~(\ref{eq:ggg2}) yields the NNLO
gluon-gluon splitting function
\bea
  &&P^{\,(2)}_{\rm gg}(x) \:\: = \:\: 
       16\, \*  \colour4colour{\ca \* \cf \* \nf}  \*  \bigg(
            x^2 \* \bigg[
	    {4 \over 9} \* \H(2)
	  + 3 \* \Hh(1,0)
	  - {97 \over 12} \* \H(1)
	  + {8 \over 3} \* \Hh(-2,0)
	  - {2 \over 3} \* \H(0) \* \z2
	  + {103 \over 27} \* \H(0)
	  - {16 \over 3} \* \z2
	  + 2 \* \H(3)
  \nonumber\\&&
          - 6 \* \Hh(-1,0)
	  + 2 \* \Hh(2,0)
	  + {127 \over 18} \* \Hh(0,0)
	  - {511 \over 12}
          \bigg]
          + \pgg(x) \* \bigg[
            2 \* \z3
          - {55 \over 24} 
          \bigg]
          + {4 \over 3} \* \big({1 \over x}-x^2\big) \* \bigg[
            {17 \over 24} \* \Hh(1,0)
          - {43 \over 18} \* \H(0)
  \nonumber\\&&
          - {521 \over 144} \* \H(1)
          - {6923 \over 432} 
	  - {1 \over 2} \* \Hh(2,1)
          + 2 \* \H(1) \* \z2
          + \H(0) \* \z2
          - 2 \* \Hhh(1,0,0)
          + {1 \over 12} \* \Hh(1,1)
          - \Hhh(1,1,0)
          - \Hhh(1,1,1)
          \bigg]
          - {175 \over 12} \* \H(2)
  \nonumber\\&&
          + 6 \* \Hh(-1,0)
          + 8 \* \H(0) \* \z3
          - 6 \* \Hh(-2,0)
          - {53 \over 6} \* \H(0) \* \z2
          - {49 \over 2} \* \H(0)
          + {185 \over 4} \* \z2
          + {511 \over 12} 
          - {1 \over 2} \* \Hh(2,0)
          - 3 \* \H(1,0)
          - 4 \* \Hhhh(0,0,0,0)
  \nonumber\\&&
          - {67 \over 12} \* \Hh(0,0)
          + {43 \over 2} \* \z3
          - \Hh(2,1)
          + {97 \over 12} \* \H(1)
          - 4 \* \z2^2
          - {9 \over 2} \* \H(3)
          - 8 \* \Hh(-3,0)
          + {33 \over 2} \* \Hhh(0,0,0)
          + {4 \over 3} \* \big({1 \over x}+x^2\big) \* \bigg[
            {1 \over 2} \* \H(2)
          - \Hh(2,0)
  \nonumber\\&&
          + {11 \over 3} \* \Hh(-1,0)
          + \Hh(-2,0)
          + {19 \over 6} \* \z2
          + 2 \* \z3
          - \H(-1) \* \z2 
          - 4 \* \Hhh(-1,-1,0) 
          - {1 \over 2} \* \Hhh(-1,0,0) 
          - \Hh(-1,2) 
          \bigg]
        + (1-x) \* \bigg[
            9 \* \H(1) \* \z2
  \nonumber\\&&
          + 12 \* \Hhhh(0,0,0,0)
	  - {293 \over 108}
	  + {61 \over 6} \* \H(0) \* \z2 
	  - {7 \over 3} \* \Hh(1,0)
          - {857 \over 36} \* \H(1)
          - 9 \* \H(0) \* \z3
          + 16 \* \Hhh(-2,-1,0)
          - 4 \* \Hhh(-2,0,0)
          + 8 \* \H(-2) \* \z2
  \nonumber\\&&
          - {13 \over 2} \* \Hhh(1,0,0)
          + {3 \over 4} \* \Hh(1,1)
          - \Hhh(1,1,0)
          - \Hhh(1,1,1)
          \bigg]
          + (1+x) \* \bigg[
	    {1 \over 6} \* \Hh(2,0)
          - {95 \over 3} \* \Hh(-1,0)
          - {149 \over 36} \* \H(2)
	  + {3451 \over 108} \* \H(0)
  \nonumber\\&&
	  - 7 \* \Hh(-2,0)
	  + {302 \over 9} \* \Hh(0,0)
	  + {19 \over 6} \* \H(3)
	  - {991 \over 36} \* \z2
          - {163 \over 6} \* \z3
          - {35 \over 3} \* \Hhh(0,0,0)
	  + {17 \over 6} \* \Hh(2,1)
          - {43 \over 10} \* \z2^2
          + 13 \* \H(-1) \* \z2
  \nonumber\\&&
          + 18 \* \Hhh(-1,-1,0)
          - \Hh(3,1)
          - 6 \* \H(4)
          - 4 \* \Hh(-1,2)
          + 6 \* \Hh(0,0) \* \z2
          + 8 \* \H(2) \* \z2
          - 7 \* \Hhh(2,0,0)
          - 2 \* \Hhh(2,1,0)
          - 2 \* \Hhh(2,1,1)
          - 4 \* \Hh(3,0)
  \nonumber\\&&
          - 9 \* \Hhh(-1,0,0)
          \bigg]
          - {241 \over 288} \* \delta(1 - x)
          \bigg)
       + 16\, \*  \colour4colour{\ca \* \nf^2}  \*  \bigg(
            {19 \over 54} \* \H(0) - {1 \over 24} \* x \* \H(0)
          - {1 \over 27} \* \pgg(x)
          + {13 \over 54} \* \big({1 \over x}-x^2\big) \* \bigg[
            {5 \over 3} 
          - \H(1) 
          \bigg]
  \nonumber\\&&
          + (1-x) \* \bigg[
            {11 \over 72} \* \H(1) 
          - {71 \over 216} 
          \bigg]
          + {2 \over 9} \* (1+x) \* \bigg[
            \z2
	  + {13 \over 12} \* x \* \H(0)
          - {1 \over 2} \* \Hh(0,0)
          - \H(2)
          \bigg]
          + {29 \over 288} \* \delta(1 - x)
          \bigg)
  \nonumber\\&&
       + 16\, \*  \colour4colour{\ca^2 \* \nf}  \*  \bigg(
            x^2 \* \bigg[
            \z3
	   + {11 \over 9} \* \z2
	   + {11 \over 9} \* \Hh(0,0)
	   - {2 \over 3} \* \H(3)
	   + {2 \over 3} \* \H(0) \* \z2
	   + {1639 \over 108} \* \H(0)
	   - 2 \* \Hh(-2,0)
          \bigg]
          + {1 \over 3} \* \pgg(x) \* \bigg[
            {10 \over 3} \* \z2
  \nonumber\\&&
          - {209 \over 36} 
          - 8 \* \z3
          - 2 \* \Hh(-2,0)
          - {1 \over 2} \* \H(0)
          - {10 \over 3} \* \Hh(0,0)
          - {20 \over 3} \* \Hh(1,0)
          - \Hhh(1,0,0)
          - {20 \over 3} \* \H(2)
          - \H(3)
          \bigg]
          + {10 \over 9} \* \pgg(-x) \* \bigg[
            \z2
  \nonumber\\&&
          + 2 \* \Hh(-1,0)
          + {3 \over 10} \* \H(0) \* \z2
          - \Hh(0,0)
          \bigg]
          + {1 \over 3} \* \big({1 \over x}-x^2\big) \* \bigg[
            \H(3)
          - \H(0) \* \z2
          - {13 \over 3} \* \H(2)
          + {5443 \over 108} 
          - 3 \* \H(1) \* \z2
          + {205 \over 36} \* \H(1)
  \nonumber\\&&
          - {13 \over 3} \* \Hh(1,0)
          + \Hhh(1,0,0)
          \bigg]
          + \big({1 \over x}+x^2\big) \* \bigg[
	    {151 \over 54} \* \H(0)
          - {8 \over 3} \* \z2
          + {1 \over 3} \* \H(-1) \* \z2
	  - \z3
          + 2 \* \Hhh(-1,-1,0)
          - {2 \over 3} \* \Hhh(-1,0,0)
  \nonumber\\&&
          - {37 \over 9} \* \Hh(-1,0)
          + {2 \over 3} \* \Hh(-1,2)
          \bigg]
          + (1-x) \* \bigg[
            {5 \over 6} \* \Hh(-2,0)
          + \Hh(-3,0)
          + 2 \* \Hhh(0,0,0)
          - {269 \over 36} \* \z2
          - {4097 \over 216} 
          - 3 \* \H(-2) \* \z2
  \nonumber\\&&
          - 6 \* \Hhh(-2,-1,0)
          + 3 \* \Hhh(-2,0,0)
          - {7 \over 2} \* \H(1) \* \z2
          + {677 \over 72} \* \H(1)
          + \Hh(1,0)
          + {7 \over 4} \* \Hhh(1,0,0)
          \bigg]
          + (1+x) \* \bigg[
            {193 \over 36} \* \H(2)
          - {11 \over 2} \* \H(-1) \* \z2
  \nonumber\\&&
          + {39 \over 20} \* \z2^2
          - {7 \over 12} \* \H(3)
          - {53 \over 9} \* \H(0,0)
          + {7 \over 12} \* \H(0) \* \z2
          - {5 \over 2} \* \Hh(0,0) \* \z2
          + 5 \* \z3
          - 7 \* \Hhh(-1,-1,0)
          + {77 \over 6} \* \Hh(-1,0)
          + {9 \over 2} \* \Hhh(-1,0,0)
  \nonumber\\&&
          + 2 \* \Hh(-1,2)
          - 3 \* \H(2) \* \z2
          - {2 \over 3} \* \Hh(2,0)
          + {3 \over 2} \* \Hhh(2,0,0)
          + {3 \over 2} \* \H(4)
          \bigg]
          + {1 \over 9} \* \z2
          + 7 \* \Hh(-2,0)
          + 2 \* \H(2)
	  + {458 \over 27} \* \H(0)
          + \Hh(0,0) \* \z2
  \nonumber\\&&
          + {3 \over 2} \* \z2^2
          + 4 \* \Hh(-3,0)
          - x \* \bigg[
            {131 \over 12}  \* \H(0,0)
	  - {8 \over 3} \* \H(0) \* \z2
          + {7 \over 2}  \* \H(3)
          - \Hhhh(0,0,0,0)
          + {7 \over 6}  \* \Hhh(0,0,0)
	  + {1943 \over 216}  \* \H(0)
          + 6  \* \H(0) \* \z3
          \bigg]
  \nonumber\\&&
          - \delta(1 - x) \* \bigg[
            {233 \over 288} 
          + {1 \over 6} \* \z2
          + {1 \over 12} \* \z2^2
          + {5 \over 3} \* \z3
          \bigg]
          \bigg)
       + 16\, \*  \colour4colour{\ca^3}  \*  \bigg(
            x^2 \* \bigg[
	       33 \* \Hh(-2,0)
	     + 33 \* \H(0) \* \z2
             - {1249 \over 18} \* \Hh(0,0)
  \nonumber\\&&
	     - 44 \* \Hhh(0,0,0)
	     - {110 \over 3} \* \H(3)
	     - {44 \over 3} \* \Hh(2,0)
	     + {85 \over 6} \* \z2
	     + {6409 \over 108} \* \H(0)
          \bigg]
          + \pgg(x) \* \bigg[
            {245 \over 24} 
          - {67 \over 9} \* \z2
          - {3 \over 10} \* \z2^2
          + {11 \over 3} \* \z3
  \nonumber\\&&
          - 4 \* \Hh(-3,0)
          + 6 \* \H(-2) \* \z2
          + 4 \* \Hhh(-2,-1,0)
          + {11 \over 3} \* \Hh(-2,0)
          - 4 \* \Hhh(-2,0,0)
          - 4 \* \Hh(-2,2)
          + {1 \over 6} \* \H(0)
          - 7 \* \H(0) \* \z3
          + {67 \over 9} \* \Hh(0,0)
  \nonumber\\&&
          - 8 \* \Hh(0,0) \* \z2
          + 4 \* \Hhhh(0,0,0,0)
          - 6 \* \H(1) \* \z3
          - 4 \* \Hhh(1,-2,0)
          + 10 \* \Hhh(2,0,0)
          - 6 \* \Hh(1,0) \* \z2
          + 8 \* \Hhhh(1,0,0,0)
          + 8 \* \Hhhh(1,1,0,0)
          + 8 \* \H(4)
  \nonumber\\&&
          + {134 \over 9} \* \Hh(1,0)
          + {11 \over 6} \* \Hhh(1,0,0)
          + 8 \* \Hhh(1,2,0)
          + 8 \* \Hh(1,3)
          + {134 \over 9} \* \H(2)
          - 4 \* \H(2) \* \z2
          + 8 \* \Hh(3,1)
          + 8 \* \Hh(2,2)
          + {11 \over 6} \* \H(3)
          + 10 \* \Hh(3,0)
  \nonumber\\&&
          + 8 \* \Hhh(2,1,0)
          \bigg]
          + \pgg(-x) \* \bigg[
            {11 \over 2} \* \z2^2
          - {11 \over 6} \* \H(0) \* \z2
          - 4 \* \Hh(-3,0)
          + 16 \* \H(-2) \* \z2
          - 12 \* \Hh(-2,2)
          - {134 \over 9} \* \Hh(-1,0)
          + 2 \* \H(2) \* \z2
  \nonumber\\&&
          + 8 \* \Hhh(-2,-1,0)
          + 12 \* \H(-1) \* \z3
          - 18 \* \Hhh(-2,0,0)
          + 8 \* \Hhh(-1,-2,0)
          - 16 \* \Hh(-1,-1) \* \z2
          + 24 \* \Hhhh(-1,-1,0,0)
          + 16 \* \Hhh(-1,-1,2)
  \nonumber\\&&
          + 18 \* \Hh(-1,0) \* \z2
          - 16 \* \Hhhh(-1,0,0,0)
          - 4 \* \Hhh(-1,2,0)
          - 16 \* \Hh(-1,3)
          - 5 \* \H(0) \* \z3
          - 8 \* \Hh(0,0) \* \z2
          + 4 \* \Hhhh(0,0,0,0)
          + 2 \* \Hh(3,0)
  \nonumber\\&&
          - {67 \over 9} \* \z2
          + {67 \over 9} \* \Hh(0,0)
          + 8 \* \H(4)
          \bigg]
          + \bigg({1 \over x}-x^2\bigg) \* \bigg[
            {16619 \over 162} 
          + {22 \over 3} \* \Hh(2,0)
          - {55 \over 2} \* \z3
          - {11 \over 2} \* \H(0) \* \z2
	  - {67 \over 9} \* \H(2)
          - {67 \over 9} \* \Hh(1,0)
  \nonumber\\&&
          - {413 \over 108} \* \H(1) 
          - {11 \over 2} \* \H(1) \* \z2 
          + {33 \over 2} \* \Hhh(1,0,0) 
          \bigg]
          + 11 \* \big({1 \over x}+x^2\big) \* \bigg[
            {71 \over 54} \* \H(0)
          - {1 \over 6} \* \H(3)
          - {389 \over 198} \* \z2
	  - {2 \over 3} \* \Hh(-2,0)
          - {1 \over 2} \* \H(-1) \* \z2
  \nonumber\\&&
          + \Hhh(-1,-1,0) 
          - {523 \over 198} \* \Hh(-1,0) 
          + {8 \over 3} \* \Hhh(-1,0,0) 
          + \Hh(-1,2) 
          \bigg]
          + (1-x) \* \bigg[
            {31 \over 36} \* \H(1)
          + {27 \over 2} \* \H(1,0)
          - {25 \over 2} \* \H(1,0,0)
          -  4 \* \H(-3,0)
  \nonumber\\&&
	  - {263 \over 12} \* \H(0,0)
          - {29 \over 3} \* \H(0,0,0)
          - {19 \over 3} \* \H(-2,0)
          - {11317 \over 108} 
          - 4 \* \H(-2) \* \z2
          - 8 \* \H(-2,-1,0)
          - 12 \* \H(-2,0,0)
          - {3 \over 2} \* \H(1) \* \z2
          \bigg]
  \nonumber\\&&
          + (1+x) \* \bigg[
            {27 \over 2} \* \H(0) \* \z2
          - {43 \over 6} \* \H(3)
          + {29 \over 3} \* \Hh(2,0)
	  + {4651 \over 216} \* \H(0)
          - {329 \over 18} \* \z2
          + {11 \over 2} \* (1+x) \* \z3
	  - {43 \over 5} \* \z2^2
          - {215 \over 6} \* \Hh(-1,0)
  \nonumber\\&&
          - 22 \* \Hh(0,0) \* \z2
          - 8 \* \H(0) \* \z3
          - 3 \* \Hhh(-1,-1,0)
          + 38 \* \Hhh(-1,0,0)
          + 25 \* \Hh(-1,2)
          + 10 \* \Hhh(2,0,0)
          - 4 \* \H(2) \* \z2
          + 16 \* \Hh(3,0)
          + 26 \* \H(4)
  \nonumber\\&&
	  - {158 \over 9} \* \H(2)
          - {53 \over 2} \* \H(-1) \* \z2
          \bigg]
          - 29 \* \Hh(0,0)
          - {40 \over 3} \* \Hhh(0,0,0)
          + 27 \* \Hh(-2,0)
          + {41 \over 3} \* \H(0) \* \z2
          - 20 \* \H(3)
          - 24 \* \Hh(2,0)
          + {53 \over 6} \* \z2
  \nonumber\\&&
          + {601 \over 12} \* \H(0)
          + 24 \* \z3
          + 2 \* \z2^2
          + 27 \* \H(2)
          - 4 \* \Hh(0,0) \* \z2
          - 16 \* \H(0) \* \z3
          - 16 \* \Hh(-3,0)
          + 28 \* x \* \Hhhh(0,0,0,0)
          + \delta(1 - x) \* \bigg[
            {79 \over 32} 
  \nonumber\\&&
          - \z2 \* \z3
          + {1 \over 6} \* \z2
          + {11 \over 24} \* \z2^2
          + {67 \over 6} \* \z3
          - 5 \* \z5
          \bigg]
          \bigg)
       + 16\, \*  \colour4colour{\cf \* \nf^2}  \*  \bigg(
            {2 \over 9} \* x^2 \* \bigg[
	      {11 \over 6} \* \H(0)
	    + \H(2)
	    - \z2
	    + 2 \* \Hh(0,0)
	    - 9 
          \bigg]
          + {1 \over 3} \* \H(2)
  \nonumber\\&&
          - {1 \over 3} \* \z2
          - {10 \over 3} \* \H(0)
          - {1 \over 3} \* \Hh(0,0)
          + 2 
          + {2 \over 9} \* \big({1 \over x}-x^2\big) \* \bigg[
	    {8 \over 3} \* \H(1)
          - 2 \* \Hh(1,0) 
          - \Hh(1,1) 
	  - {77 \over 18}
          \bigg]
          - (1-x) \* \bigg[
            {1 \over 3} \* \Hh(1,0) 
          + {1 \over 6} \* \Hh(1,1) 
  \nonumber\\&&
	  + {4 \over 9} 
          + {13 \over 6} \* \H(1)
          + x \* \H(1)
          \bigg]
          + {1 \over 3} \* (1+x) \* \bigg[
            {68 \over 9} \* \H(0)
          - {4 \over 3} \* \H(2)
          + {4 \over 3} \* \z2
	  + {29 \over 6} \* \Hh(0,0)
          - \z3
          + 2 \* \H(0) \* \z2
          - \Hhh(0,0,0)
          - 2 \* \H(3)
  \nonumber\\&&
          - \Hh(2,1)
          - 2 \* \Hh(2,0)
          \bigg]
          + {11 \over 144} \* \delta(1 - x)
          \bigg)
       + 16\, \*  \colour4colour{\cf^2 \* \nf}  \*  \bigg(
            {4 \over 3} \* x^2 \* \bigg[
	      {163 \over 16}
	    + {27 \over 8} \* \H(0)
	    + {7 \over 2} \* \Hh(0,0)
	    - \Hh(2,0)
	    - \z2
	    + {9 \over 4} \* \Hh(1,0)
  \nonumber\\&&
	    - \Hh(2,1)
	    + {1 \over 2} \* \Hhh(0,0,0)
	    + {85 \over 16} \* \H(1)
	    + \H(2)
	    - 2 \* \Hh(-2,0)
            - {3 \over 2} \* \z3
          \bigg]
          + {4 \over 3} \* \big({1 \over x}-x^2\big) \* \bigg[
	    {31 \over 16} \* \H(1)
	  - {11 \over 16}
	  - {5 \over 4} \* \Hh(1,0)
          + {1 \over 2} \* \Hhh(1,0,0) 
  \nonumber\\&&
          - \H(1) \* \z2 
          - \Hh(1,1) 
          + \Hhh(1,1,0) 
          + \Hhh(1,1,1) 
          + \z3
          \bigg]
          + {4 \over 3} \* \big({1 \over x}+x^2\big) \* \bigg[
            \H(-1) \* \z2
          + 2 \* \Hhh(-1,-1,0)
          - \Hhh(-1,0,0)
          \bigg]
             + {215 \over 12} \* \Hh(0,0)
  \nonumber\\&&
             + {20 \over 3} \* \H(0)
             - {131 \over 6} 
             + 3 \* \Hh(2,0)
             + {205 \over 12} \* x \* \z2
             - 3 \* \Hh(1,0)
	     + \Hh(2,1)  
             - {85 \over 12} \* \H(1)
             + {11 \over 4} \* \H(2)
             + 8 \* \Hh(-2,0)
             + 2 \* \z2^2
             - \H(0) \* \z2
  \nonumber\\&&
             + \H(3)
             + 6 \* \H(0) \* \z3
             + 8 \* \Hh(-3,0)
             - 4 \* x \* \Hhh(0,0,0)
          + (1-x) \* \bigg[
              {107 \over 12} \* \H(1)
            - {5 \over 6}  \* \Hh(1,0) 
            - 4 \* \z2
	    + \H(0) \* \z3
            - 8 \* \Hhh(-2,-1,0)
  \nonumber\\&&
            - 4 \* \H(-2) \* \z2
            + 4 \* \Hhh(-2,0,0)
            - 4 \* \H(1) \* \z2
            + {7 \over 2} \* \Hhh(1,0,0)
            - {7 \over 12} \* \Hh(1,1)
            + \Hhh(1,1,0)
            + \Hhh(1,1,1)
          \bigg]
          + (1+x) \* \bigg[
	      {5 \over 4} \* \H(2)
	    + {33 \over 8} 
  \nonumber\\&&
            - {99 \over 4} \* \Hh(0,0)
            - 8 \* \Hh(2,0)
	    - {541 \over 24} \* \H(0)
	    - 4 \* \Hh(2,1)
	    - {3 \over 2} \* \Hhh(0,0,0)
	    - 2 \* x \* \z3
            + {9 \over 2} \* \z2^2
            + 5 \* \H(0) \* \z2
            - 5 \* \H(3)
            - 4 \* \H(-1) \* \z2
  \nonumber\\&&
            - 8 \* \Hhh(-1,-1,0)
            + {67 \over 3} \* \Hh(-1,0)
            + 4 \* \Hhh(-1,0,0)
            + 2 \* \Hh(0,0) \* \z2
            - 2 \* \Hhhh(0,0,0,0)
            - 4 \* \H(2) \* \z2
            + 3 \* \Hhh(2,0,0)
            + 2 \* \Hhh(2,1,0)
  \nonumber\\&&
            + 2 \* \Hhh(2,1,1)
            + \Hh(3,1)
            - 2 \* \H(4)
          \bigg]
          + {1 \over 16} \* \delta(1 - x)
          \bigg)
\:\: .\label{eq:Pgg2}
\eea

The large-$x$ behaviour of the gluon-gluon splitting function 
$P^{\,(2)}_{gg}(x)$ is given by
\beq
\label{eq:Pgglx}
  P^{(2)}_{{\rm gg},x\ra 1}(x) \: = \: 
    \frac{A_3^{\rm g}}{(1-x)_+} \: + \: B_3^{\,\rm g}\: \delta(1-x) 
    \: + \: C_3^{\,\rm g}\: \ln (1-x) \: + \: {\cal O}(1) \:\: .
\eeq
The constants $A_3^{\rm g}$ and $C_3^{\,\rm g}$ have been specified in 
Eqs.~(\ref{eq:Ang}) and (\ref{eq:Cng}), respectively, while the 
coefficients of $\delta(1-x)$ are explicit in Eq.~(\ref{eq:Pgg2}). 
The corresponding limit of the gluon-quark and quark-gluon splitting 
functions is
\beq
\label{eq:Pablx}
  P^{(2)}_{ab,x\ra 1}(x) \: = \: \sum_{i=0}^{3}\, D_i^{\,\rm ab}\, 
 \ln^{\,4-i} (1-x) \: + \: {\cal O}(1) 
\eeq
with
\bea
\label{eq:Pqglx}
  D_0^{\,\rm qg} & = & 
        \frac{4}{3}\: C_A^{\, 2} \nf 
      - \frac{8}{3}\: \ca \cf \nf
      + \frac{4}{3}\: C_F^{\, 2} \nf  \nonumber \\[1.5mm]
  D_1^{\,\rm qg} & = &  
      - \frac{22}{9}\: C_A^{\, 2} \nf
      + \frac{40}{9}\: \ca \cf \nf
      - 2\: C_F^{\, 2} \nf
      + \frac{4}{9}\: \ca \n2f 
      - \frac{4}{9}\: \cf \n2f  \nn \\[1.5mm]
  D_2^{\,\rm qg} & = & 
      \left[\, - \frac{268}{9} + 8 \,\z2 \right]\: C_A^{\, 2} \nf
      + \frac{16}{9}\: \ca \cf \nf
      + \left[\, 28 - 8\,\z2 \right]\: C_F^{\, 2} \nf
      + \frac{40}{9}\: \ca \n2f
      - \frac{40}{9}\: \cf \n2f \nn \\[1.5mm]
  D_3^{\,\rm qg} & = &  
      \left[\, - \frac{950}{27} + \frac{44}{3}\,\z2 + 80\,\z3 \right]\: 
         C_A^{\, 2} \nf
      + \left[\, \frac{1904}{27} - 12\,\z2 - 208\,\z3 \right]\: 
         \ca \cf \nf
\nn \\[1mm] & & \mbox{}
      - \left[\, 34 - 128\,\z3 \right]\: C_F^{\, 2} \nf
      + \left[\, \frac{152}{27} - \frac{8}{3}\,\z2 \right]\: \ca \n2f
      - \frac{188}{27}\: \cf \n2f
\eea
and
\bea
\label{eq:Pgqlx}
  D_0^{\,\rm gq} & = & 
        \frac{4}{3}\: C_A^{\, 2} \cf 
      - \frac{8}{3}\: \ca C_F^{\, 2}
      + \frac{4}{3}\: C_F^{\, 3}   \nonumber \\[1.5mm]
  D_1^{\,\rm gq} & = &  
        \frac{182}{9}\: C_A^{\, 2} \cf
      - \frac{344}{9}\: \ca C_F^{\, 2}
      + 18\: C_F^{\, 3} 
      - \frac{20}{9}\: \ca \cf \nf 
      + \frac{20}{9}\: C_F^{\, 2} \nf  \nn \\[1.5mm]
  D_2^{\,\rm gq} & = & 
      \left[\, \frac{1093}{9} - 8 \,\z2 \right]\: C_A^{\, 2} \cf
      - \frac{1342}{9}\: \ca C_F^{\, 2}
      + \left[\, 29 + 8\,\z2 \right]\: C_F^{\, 3} 
      - \frac{256}{9}\: \ca \cf \nf
\nn \\ & & \mbox{}
      + \frac{232}{9}\: C_F^{\, 2} \nf 
      + \frac{4}{3} \cf \n2f \\[1.5mm]
  D_3^{\,\rm gq} & = &  
      \left[\,\frac{9766}{27} - \frac{164}{3}\,\z2 + 16\,\z3\right]\: 
         C_A^{\, 2} \cf
      - \left[\, \frac{9178}{27} - \frac{28}{3}\,\z2 - 64\,\z3\right]\: 
         \ca C_F^{\, 2}
      + \frac{64}{9} \cf \n2f
\nn \\[1mm] & & \mbox{}
      + \left[\, 36 + 32\,\z2 - 80\,\z3 \right]\: C_F^{\, 3} 
      - \left[\, \frac{2944}{27} - \frac{8}{3}\,\z2 \right]\: 
         \ca \cf \nf 
      + \left[\, \frac{1408}{27} + \frac{32}{3}\,\z2 \right]\: 
         C_F^{\, 2} \nf \:\: .\nn
\eea
 
It is worthwhile to notice that all the coefficients in 
Eqs.~(\ref{eq:Pqglx}) and (\ref{eq:Pgqlx}) except $D_3^{\,\rm gq}$ 
vanish for the choice 
\beq
\label{eq:n1col}
  C_A \: \equiv \: n_c \: =\: C_F \: =\: \nf \nonumber
\eeq
of the colour factors leading to a $N\! =\! 1$ supersymmetric theory.
This is part of a general structure. The combination
\beq
\label{eq:n1rel}
  \Delta_{\,\rm S}^{}(x) \:\equiv\:  
  P_{\rm qq}^{(n)}(x) + P_{\rm gq}^{(n)}(x) -
  P_{\rm qg}^{(n)}(x) - P_{\rm gg}^{(n)}(x)
\eeq
of the $(n\!+\!1)$-loop \MSb\ splitting functions is found to be much
simpler than the functions $P_{\rm ab}^{(n)}(x)$ themselves. In fact,
after transforming to the dimensional reduction (DR) scheme respecting 
the supersymmetry, $\Delta_{\,\rm S}^{}(x)$ vanishes for both the
unpolarized~\cite{Furmanski:1980cm} and polarized (spin-dependent)
\cite{Mertig:1996ny,Vogelsang:1996vh,Vogelsang:1996im}
two-loop splitting functions. We are not (yet) in a position to 
present this scheme transformation at the third order. However, we do
obtain the above-mentioned simplification within the \MSb\ scheme;
especially all harmonic polylogarithms of weight four cancel in the 
combination (\ref{eq:n1rel}) for choice (\ref{eq:n1col}) of the colour 
factors. We plan to return to this issue in a later publication.

We now return to the end-point behaviour. At small $x$ the three-loop 
splitting functions read
\beq
\label{eq:Pabsx}
  P^{(2)}_{\rm ab,x\ra 0}(x) \: = \: E_1^{\,\rm ab}\: \frac{\ln x}{x}
  \: +\: E_2^{\,\rm ab}\: \frac{1}{x} \: +\: {\cal O}( \ln^{\,4} x)
  \:\: .
\eeq
The coefficients of the $1/x$ terms of $P^{(2)}_{\rm qq}$ (which are, 
of course, entirely due the pure-singlet contribution given in 
Eq.~(\ref{eq:Pps2})$\,$) are given by
\bea
\label{eq:Pqqsx}
  E_1^{\,\rm qq} &\!\! =\!\! & \mbox{}
     - \frac{896}{27}\: \ca \cf \nf \nn \\[1mm]
  E_2^{\,\rm qq} &\!\! =\!\! & \mbox{} 
     \left[\,- \frac{27044}{81} + \frac{512}{9}\,\z2 + 96\,\z3 \right]\:
       \ca \cf \nf
     + \left[\, \frac{220}{3} - 64\,\z3 \right]\: C_F^{\, 2} \nf
     + \frac{64}{27} \cf \n2f \:\: , \quad
\eea
or, after inserting $C_A=3$ and $C_F=4/3$ and the numerical values of
$\z2$ and $\z3$,
\bea
\label{eq:Pqqsx2}
  E_1^{\,\rm qq} &\! \cong\! & - 132.741 \: \nf \nn \\
  E_2^{\,\rm qq} &\! \cong\! & - 505.999 \: \nf + 3.16049\: \n2f
  \:\: .
\eea
The corresponding results for the gluon-quark splitting function
(\ref{eq:Pqg2}) are
\bea
\label{eq:Pqgsx}
  E_1^{\,\rm qg} &\!\! =\!\! & \mbox{}
     - \frac{896}{27}\:  C_A^{\, 2} \nf 
     \:\: = \:\: \frac{\ca}{\cf}\: E_1^{\,\rm qq} \\[1.5mm]
  E_2^{\,\rm qg} &\!\! =\!\! & \mbox{} 
     \left[\,- \frac{9404}{27} + \frac{512}{9}\,\z2 + 96\,\z3 \right]\:
       C_A^{\, 2} \nf
     + \left[\, \frac{220}{3} - 64\,\z3 \right]\: \ca \cf \nf
     - \frac{424}{81} \ca \n2f 
     + \frac{1232}{81} \cf \n2f \nn
\eea
and
\bea
\label{eq:Pqgsx2}
  E_1^{\,\rm qg} &\! \cong\! & - 298.667\: \nf \nn \\
  E_2^{\,\rm qg} &\! \cong\! & - 1268.28\: \nf + 4.57613\: \n2f
  \:\: .
\eea
The coefficients $E_1$ in Eqs.~(\ref{eq:Pqqsx}) and (\ref{eq:Pqgsx})
agree with those obtained by Catani and Hautmann in 
Ref.~\cite{Catani:1994sq} from the small-$x$ resummation. 

The small-$x$ coefficients of the quark-gluon splitting function
(\ref{eq:Pgq2}) are given by
\bea
\label{eq:Pgqsx}
  E_1^{\,\rm gq} &\!\! =\!\! & \mbox{}
     \left[\, \frac{6320}{27} - \frac{176}{3}\,\z2 - 32\,\z3 \right]\:
       C_A^{\, 2} \cf
     + \left[\, \frac{1208}{27} - \frac{32}{3}\,\z2 \right]\: \ca\cf\nf
     - \left[\, \frac{1520}{27} - \frac{64}{3}\,\z2 \right]\: 
       C_F^{\, 2}\nf \nn \\[1.5mm]
  E_2^{\,\rm gq} &\!\! =\!\! & \mbox{} 
     \left[\, \frac{138305}{81} - \frac{872}{3}\,\z2 - 336\,\z3 
            - \frac{544}{5} \zeta_2^{\, 2} \right]\: C_A^{\, 2} \cf
     + \left[\, \frac{1934}{9} - \frac{112}{3}\,\z2 - 80\,\z3 \right]\: 
       \ca\cf\nf
\nn \\[1mm] & & \mbox{}
     + \left[\, 163 - 160\,\z2 + 216\,\z3 
     - \frac{432}{5}\,\zeta_2^{\, 2} \right]\: \ca C_F^{\, 2}
     - \left[\, 94 - 112\,\z2 + 240\,\z3 
     - \frac{512}{5}\,\zeta_2^{\, 2} \right]\: C_F^{\, 3}
\nn \\[1mm] & & \mbox{}
     - \left[\, \frac{3250}{9} - \frac{496}{9}\,\z2 - 96\,\z3 \right]\: 
       C_F^{\, 2}\nf 
     - \frac{16}{9} \cf \n2f  \:\: ,
\eea
or
\bea
\label{eq:Pgqsx2}
  E_1^{\,\rm gq} &\! \cong\! & 1189.27\: + 71.0825\: \nf \nn \\
  E_2^{\,\rm gq} &\! \cong\! & 6163.11\: - 46.4075\: \nf
    - 2.37037\: \n2f
  \:\: .
\eea
Finally the corresponding coefficients of the three-loop gluon-gluon
splitting function (\ref{eq:Pgg2}) read
\bea
\label{eq:Pggsx}
  E_1^{\,\rm gg} &\!\! =\!\! & \mbox{}
     \left[\, \frac{6320}{27} - \frac{176}{3}\,\z2 - 32\,\z3 \right]\:
       C_A^{\, 3} 
     + \left[\, \frac{1136}{27} - \frac{32}{3}\,\z2 \right]\: 
       C_A^{\, 2}\nf
     - \left[\, \frac{1376}{27} - \frac{64}{3}\,\z2 \right]\: 
       \ca \cf \nf \nn \\[1mm]
    &\!\! =\!\! & \frac{\ca}{\cf}\: E_1^{\,\rm gq}
    - \frac{8}{3}\, \ca \nf\, ( \,\ca -2\, \cf ) 
    \:\: \ra \:\: \frac{\ca}{\cf}\: E_1^{\,\rm gq}
    - \frac{8}{3}\, \nf  \quad \mbox{for SU(N)} \nn \\[1.5mm]
  E_2^{\,\rm gg} &\!\! =\!\! & \mbox{} 
     \left[\, \frac{146182}{81} - \frac{3112}{9}\,\z2 
            - \frac{1144}{3}\,\z3 - \frac{464}{5} \zeta_2^{\,2} \right]
       \: C_A^{\, 3} 
     + \left[\, \frac{19264}{81} - \frac{128}{3}\,\z2 
              - \frac{176}{3}\,\z3 \right]\: C_A^{\, 2} \nf
\nn \\[1mm] & & \mbox{}
     - \left[\, \frac{30662}{81} - \frac{608}{9}\,\z2 
              - \frac{224}{3}\,\z3 \right]\: \ca \cf \nf 
     - \left[\, \frac{44}{3} - \frac{64}{3}\,\z3 \right] C_F^{\, 2}\nf 
     + \frac{472}{81}\, \ca \n2f 
\nn \\[1mm] & & \mbox{}
     - \frac{1232}{81}\, \cf \n2f 
\eea
and
\bea
\label{eq:Pggsx2}
  E_1^{\,\rm gg} &\! \cong\! & 2675.85 + 157.269\: \nf \nn \\
  E_2^{\,\rm gg} &\! \cong\! & 14214.2 + 182.958\: \nf - 2.79835\: \n2f
  \:\: .
\eea
The coefficient $E_1^{\,\rm gg}$ is identical to the result obtained
from the next-to-leading logarithmic BFKL equation by Fadin and Lipatov
in Ref.~\cite{Fadin:1998py} after transformation to the \MSb\ scheme
(as given, e.g., in Eq.~(4.7) of Ref.~\cite{vanNeerven:2000uj}). 
Numerically the simple relation $\cf E_1^{\,\rm gg}=\ca E_1^{\,\rm gq}$ 
is broken by less than 2\% in the $\nf$ part and less than 0.5\% for 
the complete coefficients at $\nf = 3,\ldots,6$. 

The three-loop splitting functions (\ref{eq:Pps2}) -- (\ref{eq:Pgg2})
are shown in Figs.~3 -- 6 for $\nf = 4$ together with the approximate
expressions inferred in Ref.~\cite{vanNeerven:2000wp} from the 
fixed-$N$ results of Refs.~\cite{Larin:1997wd,Retey:2000nq} and the 
small-$x$ limits of Refs.~\cite{Catani:1994sq,Fadin:1998py}. 
Also displayed are the respective leading small-$x$ contributions 
$E_1^{\,\rm ab} x^{\,-1} \ln x$. Notice that all splitting functions
have been multiplied by $x$ for display purposes.

With the exception of $P^{(2)}_{\rm gq}$, where no small-$x$ `anchor'
was available, our exact results comply with the error bands of 
Ref.~\cite{vanNeerven:2000wp} for the full range of $x$ shown in the
figures. Hence it is reasonable to expect that an extension of the
results of Refs.~\cite{Larin:1997wd, Retey:2000nq} to the next order, 
using a future four-loop generalization of the {\sc Mincer} program 
\cite{Gorishnii:1989gt,Larin:1991fz}, would, together with small-$x$
constraints, facilitate relevant estimates of $P^{(3)}_{\rm ab}(x)$.
We expect that such an extension, while still a formidable task, will 
be performed much earlier than the fourth-order version of the present 
calculation.

\begin{figure}[p]
\label{pic:P2ps4}
\vspace{-4mm}
\centerline{\epsfig{file=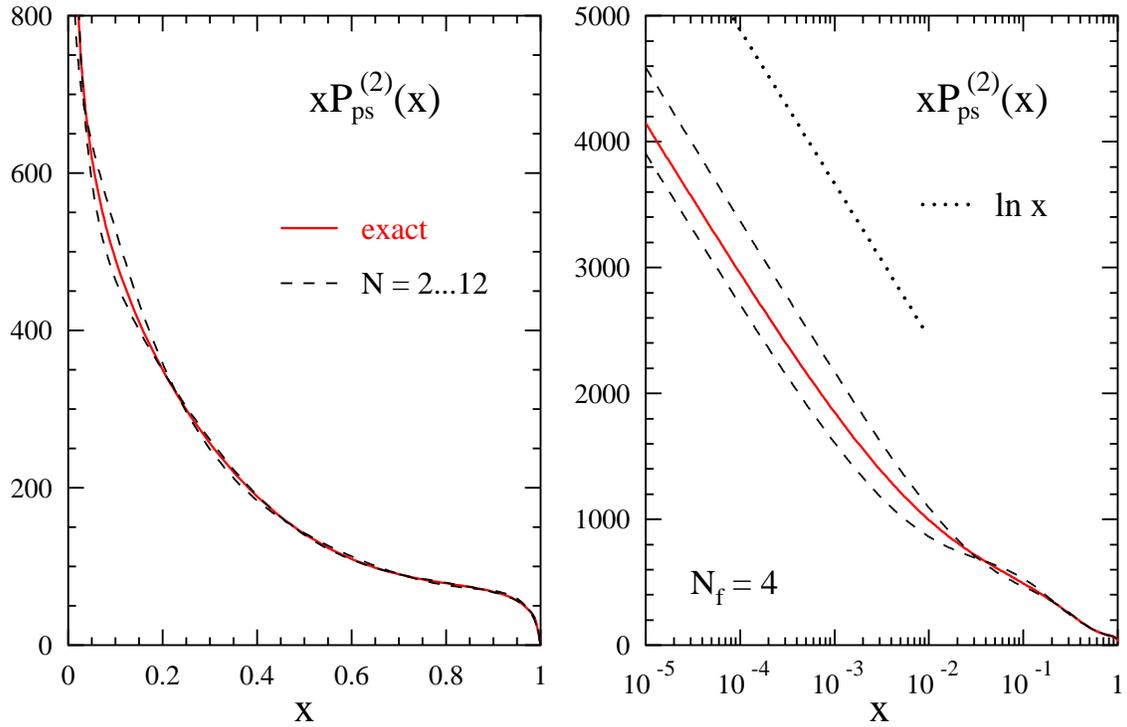,width=15.0cm,angle=0}}
\vspace{-2mm}
\caption{The three-loop pure-singlet splitting function (\ref{eq:Pps2})
 for four flavours, multiplied by $x$ for display purposes. Also shown
 is the uncertainty band derived in Ref.~\cite{vanNeerven:2000wp}
 using the lowest six even-integer moments
 \cite{Larin:1997wd, Retey:2000nq} and the leading small-$x$ term
 \cite{Catani:1994sq}. The latter contribution is shown separately
 on the right-hand-side (dotted line) for $x<0.01$.}
\vspace{1mm}
\end{figure}
\begin{figure}[p]
\label{pic:P2qg4}
\vspace{-2mm}
\centerline{\epsfig{file=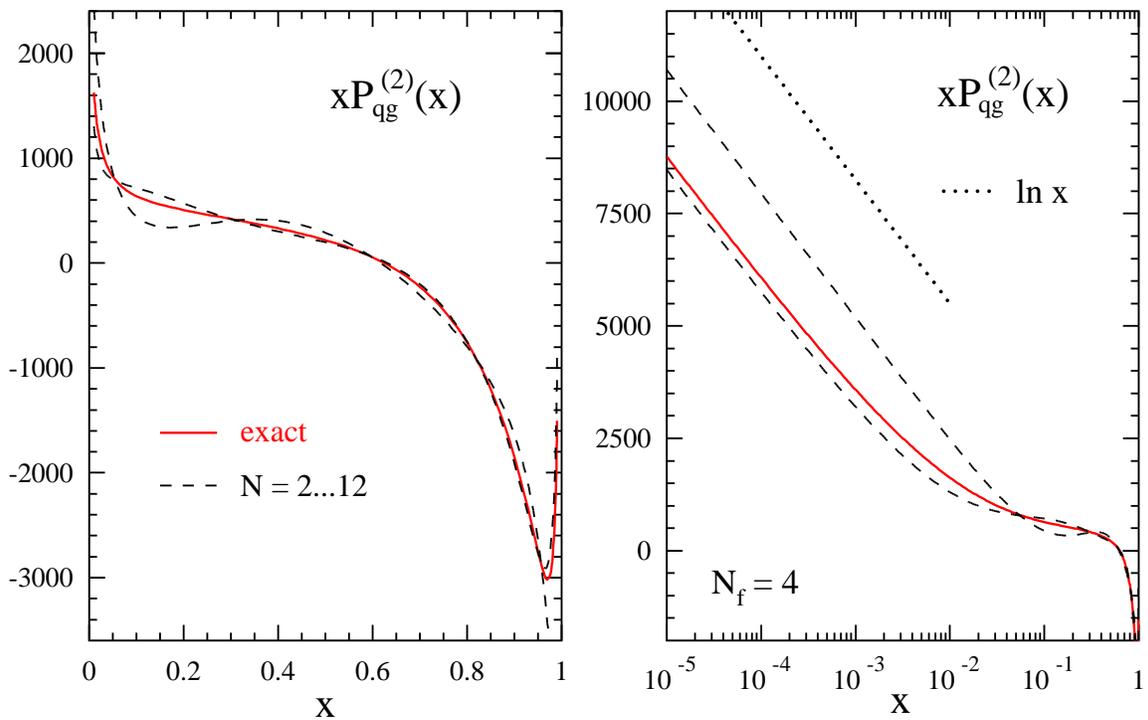,width=15.0cm,angle=0}}
\vspace{-2mm}
\caption{As Fig.~3, but for the third-order gluon-quark splitting 
 function specified in Eq.~(\ref{eq:Pqg2}).}
\vspace{1mm}
\end{figure}

\begin{figure}[p]
\label{pic:P2gq4}
\vspace{-2mm}
\centerline{\epsfig{file=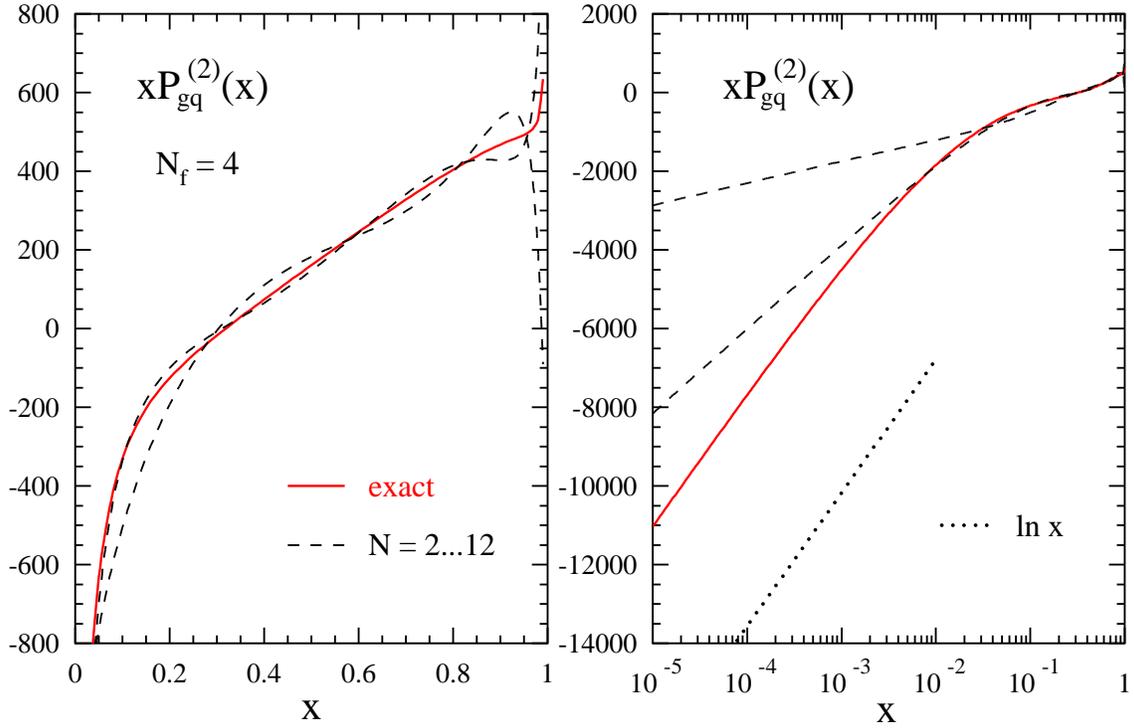,width=15.0cm,angle=0}}
\vspace{-2mm}
\caption{As Fig.~3, but for the three-loop quark-gluon splitting
 function (\ref{eq:Pgq2}). Note that in this case the leading small-$x$ 
 contribution was unknown before the present calculation.}
\vspace{1mm}
\end{figure}
\begin{figure}[p]
\label{pic:P2gg4}
\vspace{-2mm}
\centerline{\epsfig{file=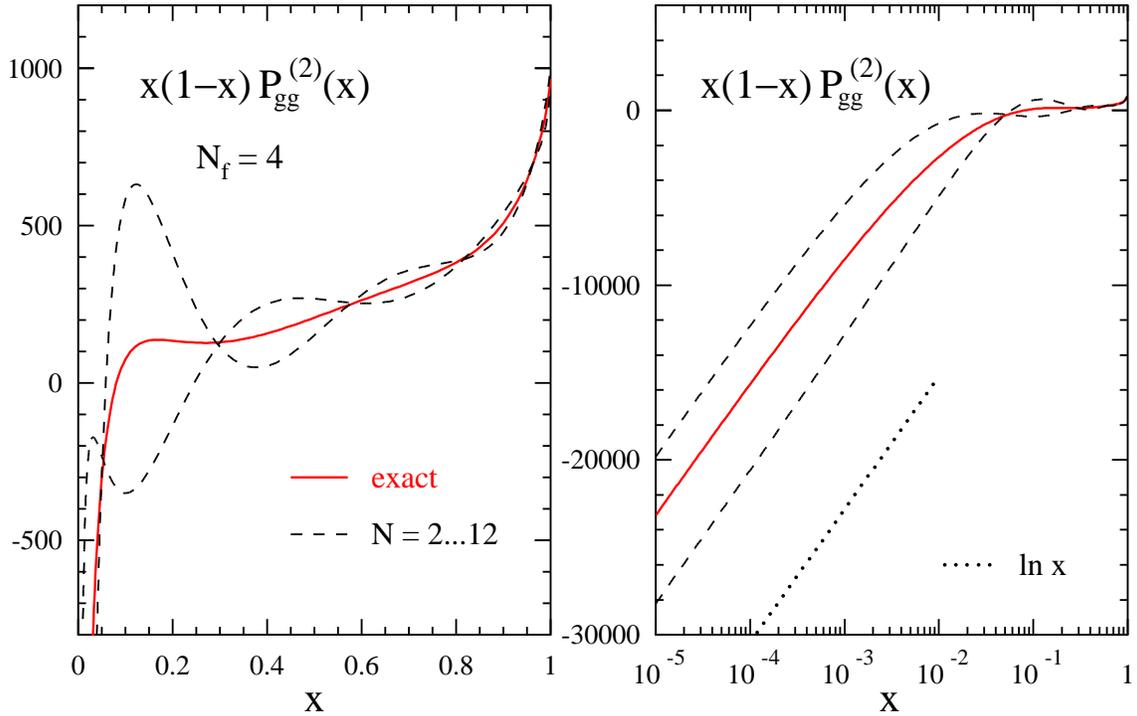,width=15.0cm,angle=0}}
\vspace{-2mm}
\caption{As Fig.~3, but for the third-order gluon-gluon splitting
 function specified in Eq.~(\ref{eq:Pgg2}). This diagonal quantity has 
 been additionally multiplied by $(1-x)$. The leading small-$x$ term
 (again shown by the dotted line on the right-hand-side) has been first 
 obtained in Ref.~\cite{Fadin:1998py}.}
\vspace{1mm}
\end{figure}

\begin{figure}[thb]
\label{pic:PggCnv}
\vspace{-2mm}
\centerline{\epsfig{file=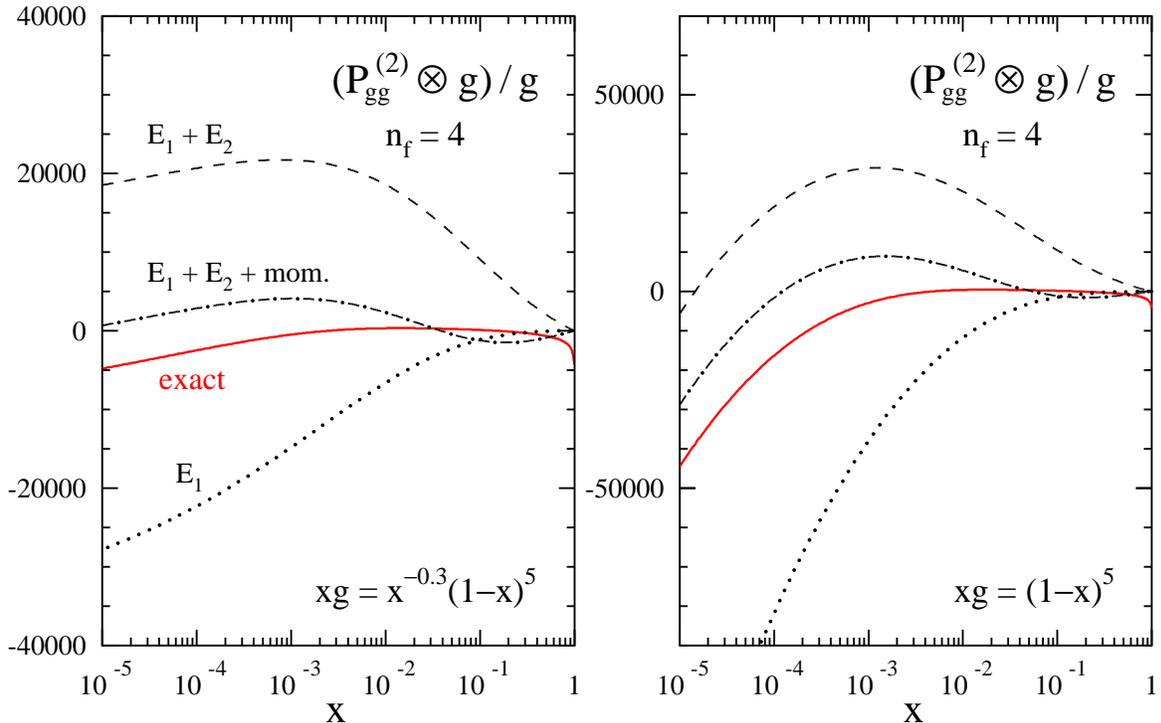,width=15.0cm,angle=0}}
\vspace{-2mm}
\caption{The convolution of the three-loop gluon-gluon splitting 
 function (\ref{eq:Pgg2}) with schematic `steep' (left) and `flat' 
 (right) gluon 
 distributions. Also shown the results obtained by instead using only 
 the leading ($E_1$) and the leading and next-to-leading ($E_1+E_2$)
 small-$x$ terms in Eq.~(\ref{eq:Pabsx}), and by supplementing the 
 latter by a constant term restoring the correct second moment.}
\vspace{1mm}
\end{figure}

As also illustrated in Figs.~3 -- 6, the leading small-$x$ terms $\sim
x^{\,-1} \ln x$ alone do not provide good approximations of the full
results (\ref{eq:Pps2})--(\ref{eq:Pgg2}) at experimentally relevant
small values of~$x$. At $x=10^{-4}$, for example, they exceed the exact
values of $P_{\rm ab}^{(2)}(x)$ by factors between 1.6 and 2.0 for
$\nf=4$. Good small-$x$ approximations of these quantities are obtained
by including all $x^{\,-1}$ contributions as specified in
Eq.~(\ref{eq:Pabsx}) -- (\ref{eq:Pggsx2}). However this does not apply,
as obvious from Fig.~7, to the convolution [$P_{\rm gg}^{(2)} \otimes
g$](x) by which $P_{\rm gg}^{(2)}$ enters the evolution equations
(\ref{eq:evol}). Even if the two terms explicit in Eq.~(\ref{eq:Pabsx})
are (non-uniquely) supplemented by an $x$-independent contribution
restoring the correct second moment, even the sign of the convolution
remains wrong down to $x \simeq 10^{-5}$ for the simplified, but not
unrealistic gluon distribution $xg \sim x^{\,-0.3}(1-x)^5$.

As our exact expressions (\ref{eq:Pps2}) -- (\ref{eq:Pgg2}) for the
the functions $P^{\, (2)}_{\rm ab}(x)$ are neither particularly short
nor especially simple, we also provide compact approximate 
representations built up, besides powers of $x$, only from the 
$+$-distribution (for $P^{\, (2)}_{\rm gg}(x)$) and the end-point 
logarithms
\beq
\label{eq:logs}
  \DD_{\,0} \: = \: 1/(1-x)_+ \: ,
  \quad L_1 \: = \: \ln (1-x) \: ,
  \quad L_0 \: = \: \ln x \:\: .
\eeq
Inserting the numerical values of the QCD colour factors, 
$P^{\,(2)}_{\rm ps}$ in Eq.~(\ref{eq:Pps2}) can be represented~by 
\bea
\label{eq:Pps-ap}
  P^{(2)}_{\rm ps}(x)\!\!\! & \cong & \!\!\!\Big\{ \: \nf
     \Big( - 5.926\: L_1^3 - 9.751\: L_1^2 - 72.11\: L_1
     + 177.4 + 392.9\: x - 101.4\: x^2 - 57.04\: L_0 L_1
  \nn \\[-0.5mm] & & \mbox{} \quad
     - 661.6\: L_0 + 131.4\: L_0^2 - 400/9\: L_0^3 + 160/27\: L_0^4
     - 506.0\: x^{\,-1} - 3584/27\: x^{\,-1} L_0  \, \Big) 
  \nn \\ &+& \mbox{} \n2f \: \Big(
     1.778\: L_1^2 + 5.944\: L_1 + 100.1 - 125.2\: x + 49.26\: x^2
     - 12.59\: x^3 - 1.889\: L_0 L_1
  \nn \\[-0.5mm] & & \mbox{} \quad
     + 61.75\: L_0 + 17.89\: L_0^2 + 32/27\: L_0^3 + 256/81\: x^{\,-1} 
     \, \Big) \Big\} (1-x)
  \:\: .
\eea
Correspondingly the off-diagonal quantities (\ref{eq:Pqg2}) and
(\ref{eq:Pgq2}) can be parametrized by
\bea
\label{eq:Pqg-ap}
  P^{(2)}_{\rm qg}(x)\!\!\! & \cong & \mbox{} \nf
     \Big( \: 100/27\: L_1^4 - 70/9\: L_1^3 - 120.5\: L_1^2 
     + 104.42\: L_1 + 2522 - 3316\: x + 2126\: x^2 
  \nn \\ & & \mbox{} \quad
     + L_0 L_1\, (1823 - 25.22\: L_0)
     - 252.5\: x L_0^3 + 424.9\: L_0 + 881.5\: L_0^2 - 44/3\: L_0^3
  \nn \\ & & \mbox{} \quad
     + 536/27\: L_0^4 - 1268.3\: x^{\,-1} - 896/3\: x^{\,-1} L_0 \,\Big)
  \nn \\ &+& \mbox{} \n2f \: \Big( \:
     20/27\: L_1^3 + 200/27\: L_1^2 - 5.496\: L_1 - 252.0 + 158.0\: x
     + 145.4\: x^2  
  \nn \\ & & \mbox{} \quad
     - 139.28\: x^3 - L_0 L_1\, ( 53.09  + 80.616\: L_0 ) 
     - 98.07\: xL_0^2 + 11.70\: xL_0^3 
  \nn \\ & & \mbox{} \quad
     - 254.0\: L_0 - 90.80\: L_0^2 - 376/27\: L_0^3
     - 16/9\: L_0^4 + 1112/243\: x^{\,-1} \, \Big)
\eea
and
\bea
\label{eq:Pgq-ap}
  P^{(2)}_{\rm gq}(x)\!\!\! & \cong & \mbox{} 
     + 400/81\: L_1^4 + 2200/27\: L_1^3 + 606.3\: L_1^2 + 2193\: L_1
     - 4307 + 489.3\: x + 1452\: x^2 
  \nn \\[0.5mm] & & \mbox{} 
     + 146.0\: x^3 - 447.3\: L_0^2 L_1 
     - 972.9\: xL_0^2 + 4033\: L_0 - 1794\: L_0^2 + 1568/9\: L_0^3
  \nn \\[1mm] & & \mbox{} 
     - 4288/81\: L_0^4 + 6163.1\: x^{\,-1} + 1189.3\: x^{\,-1} L_0
  \nn \\ &+& \mbox{} \nf \: \Big( \:
     - 400/81\: L_1^3 - 68.069\: L_1^2 - 296.7\: L_1 - 183.8 
     + 33.35\: x - 277.9\: x^2
  \nn \\ & & \mbox{} \quad
     + 108.6\: xL_0^2 - 49.68\: L_0 L_1 + 174.8\: L_0 + 20.39\: L_0^2 
     + 704/81\: L_0^3 
  \nn \\ & & \mbox{} \quad
     + 128/27\: L_0^4 - 46.41\: x^{\,-1} + 71.082\: x^{\,-1} L_0 
     \,\Big)
  \nn \\[-0.5mm] &+& \mbox{} \n2f \: \Big( \:
     96/27\: L_1^2\: (\, x^{\,-1} - 1 + 1/2\:\, x\, ) 
     + 320/27\: L_1\: (\, x^{\,-1} - 1 + 4/5\:\, x\, ) 
  \nn \\[-1mm] & & \mbox{} \quad
     - 64/27\: (\, x^{\,-1} - 1 - 2\,x \, ) \,\Big) \:\: ,
\eea
where the $\n2f$ part is exact.  Finally the gluon-gluon splitting 
function (\ref{eq:Pgg2}) can be approximated~by
\bea
\label{eq:Pgg-ap}
  P^{(2)}_{\rm gg}(x)\!\!\! & \cong & \mbox{} \!\!
     + 2643.521\: {\cal D}_0 + 4425.894\: \delta(1-x) + 3589\: L_1
     - 20852\: + 3968\: x - 3363\: x^2 
  \nn \\[0.5mm] & & \mbox{} 
     + 4848\: x^3 + L_0 L_1\: ( 7305 + 8757\: L_0 ) 
     + 274.4\: L_0 - 7471\: L_0^2 + 72\: L_0^3 - 144\: L_0^4
  \nn \\[1mm] & & \mbox{} 
     + 14214\: x^{\,-1} + 2675.8\: x^{\,-1} L_0
  \nn \\ &+& \mbox{} \nf \: \Big( \:
     - 412.172\: {\cal D}_0 - 528.723\: \delta(1-x) - 320\: L_1
     - 350.2 + 755.7\: x - 713.8\: x^2 
  \nn \\ & & \mbox{} \quad
     + 559.3\: x^3 + L_0 L_1\: ( 26.15 - 808.7\: L_0 ) + 1541\: L_0 
     + 491.3\: L_0^2 + 832/9\: L_0^3 
  \nn \\ & & \mbox{} \quad
     + 512/27\: L_0^4 + 182.96\: x^{\,-1} + 157.27\: x^{\,-1} L_0 
     \,\Big) 
  \nn \\ &+& \mbox{} \n2f \: \Big( \:
     - 16/9\: {\cal D}_0 + 6.4630\: \delta(1-x) - 13.878 + 153.4\: x 
     - 187.7\: x^2 + 52.75\: x^3 
  \nn \\ & & \mbox{} \quad
     - L_0 L_1\: ( 115.6 - 85.25\: x + 63.23\: L_0 )
     - 3.422\: L_0 + 9.680\: L_0^2 - 32/27\: L_0^3 
  \nn \\ & & \mbox{} \quad
     - 680/243\: x^{\,-1}
     \,\Big) \:\: .
\eea
The coefficients of $1/x$, $(\ln x)/x$, $\ln^{\,3} x$ and $\ln^{\,4} x$ 
are exact in Eqs.~(\ref{eq:Pps-ap}) -- (\ref{eq:Pgg-ap}), up to a 
truncation of the irrational numbers. The same holds for the 
coefficients of $\ln^{\,3} (1-x)$ and $\ln^{\,4} (1-x)$ in 
Eqs.~(\ref{eq:Pqg-ap}) and (\ref{eq:Pgq-ap}), and those of ${\cal D}_0$ 
and $\ln (1-x)$ in Eq.~(\ref{eq:Pgg-ap}). The remaining terms (except, 
or course, for the $\delta(1-x)$ parts in Eq.~(\ref{eq:Pgg-ap})$\,$) 
have been obtained by fits to the exact results (\ref{eq:Pps2}) -- 
(\ref{eq:Pgg2}) at $10^{-6} \leq x \leq 1\! -\! 10^{-6}$ which we 
evaluated using the {\sc Fortran} code of Ref.~\cite{Gehrmann:2001pz}. 
Smaller values of $x$ are not needed, as all $1/x$ terms are exact.
Except for values of $x$ very close to zeros of $P^{(2)}_{\rm ab}(x)$, 
the parametrizations (\ref{eq:Pps-ap}) -- (\ref{eq:Pgg-ap}) deviate 
from the exact expressions by less than one part in a thousand, which 
should be amply sufficient for foreseeable numerical applications.

Finally the coefficients of $\delta (1-x)$ in Eq.~(\ref{eq:Pgg-ap}) 
have been slightly adjusted from their exact values using the lowest 
integer moments. This is a somewhat tricky point, so let us briefly 
elaborate on it. For $P_{\rm qq}$ and $P_{\rm gg}$ the low moments, 
and partly also the convolutions with the parton distributions, involve 
large cancellations between the integrals over the (fitted) regular 
parts and the $\delta(1-x)$ contributions. The second moment of the 
$\nf$-independent part of $P_{\rm gg}$, for example, vanishes due to 
the momentum sum rule (recall that $P_{\rm qg}$ has no $n_{\! f}^{\,0}$ 
contribution, cf.~Eq.~(\ref{eq:Poffd})$\,$), while the respective 
third-order coefficient of $\delta(1-x)$ is as large as 
$4\cdot 10^{\,3}$. A maximal accuracy of the parametrization 
(\ref{eq:Pgg-ap}), and of the convolutions with the gluon distribution,
is thus achieved by `fitting' this coefficient to the second moment.
For the case under consideration this actually leads to a very small
adjustment of about 0.01\%.

One important approach to implementing higher-order results into the
numerical evolution of the parton distributions and the analysis of
general hard processes is the moment-space technique
\cite{Diemoz:1988xu,Gluck:1990ze,Graudenz:1996sk,Kosower:1998vj,%
Stratmann:2001pb}, which requires the analytic continuation of the
anomalous dimensions~(\ref{eq:Pdef}) to certain complex values of $N$.
Also these complex-$N$ moments can be readily obtained to a perfectly 
sufficient accuracy using Eqs.~(\ref{eq:Pps-ap}) -- (\ref{eq:Pgg-ap})
together with the corresponding non-singlet results in Eqs.~(4.22) --
(4.24) of Ref.~\cite{Moch:2004pa}. The Mellin transform of these 
parametrizations involve only simple harmonic sums $S_{m>0}(N)$
of which the analytic continuations in terms of logarithmic derivatives 
of Euler's $\Gamma$-function are well known. The reader is referred
to Refs.~\cite{Blumlein:1998if,Blumlein:2000hw} for a more 
mathematical approach to the analytic continuations.
%
%
\setcounter{equation}{0}
\section{Numerical implications}
\label{sec:sresults}
%
%
We are now ready to illustrate the numerical effect of our new
three-loop splitting functions $P^{(2)}_{\rm ab}(x)$ on the evolution 
(\ref{eq:evol}) of the singlet-quark and gluon distributions 
$q_{\rm s}(x,\mu_f^{\,2})$ and $g(x,\mu_f^{\,2})$.  For all figures 
we choose a reference scale $\mu_f^{\,2} \: =\: \mu_0^{\,2}\:\simeq\:
30$ GeV$^2$ --  a scale relevant, for example, for deep-inelastic 
scattering both at fixed-target experiments and the {\it ep} collider 
HERA -- and employ the sufficiently realistic model distributions
\bea
\label{eq:shapes}
  xq_{\rm s}(x,\mu_{0}^{\,2}) &\! = \! &
  0.6\: x^{\, -0.3} (1-x)^{3.5}\, (1 + 5.0\: x^{\, 0.8\,})  \nn \\
  xg (x,\mu_{0}^{\,2})\:\: &\! = \! &
  1.6\: x^{\, -0.3} (1-x)^{4.5}\, (1 - 0.6\: x^{\, 0.3\,})
\eea
irrespective of the order of the expansion. This order-independence 
does not hold for actual data-fitted parton distributions like those 
in Refs.~\cite{Martin:2002dr,Alekhin:2002fv}, 
but here it facilitates direct comparisons of the various contributions 
to the scale derivatives $\dot{f} \equiv d \ln f / d\ln \mu_f^{\,2\,}$ 
for $\, f= q_{\rm s}, \: g$. 
For the same reason we employ an order-independent value for the strong 
coupling constant,
\beq
\label{eq:asvalue}
  \as (\mu_{0}^{\,2}) \: = \: 0.2 \:\: ,
\eeq
corresponding to a fairly standard value at the $Z$ mass, 
$\as (M_Z^{\, 2}) \simeq 0.116$, beyond the leading 
order. Finally our default for the number of effectively massless
flavours is $\nf =4$. 

The respective scale derivatives of the singlet-quark and gluon 
distributions are graphically displayed in Figs.~8 and 9 over a wide 
range of $x$. Numerical values can be found for four characteristic
$x$-values in Tables 2 and 3, where we also show the dependence on $\nf$
and the break-up into the quark- and gluon-initiated contributions. 
As these two terms can occur with different signs, and since the LO 
and NLO results partly display a somewhat anomalous behaviour (see
below), the picture is much less clear-cut here than in the non-singlet 
sector discussed in Ref.~\cite{Moch:2004pa}.

\begin{figure}[htp]
\label{pic:dq+exp}
\vspace{-1mm}
\centerline{\epsfig{file=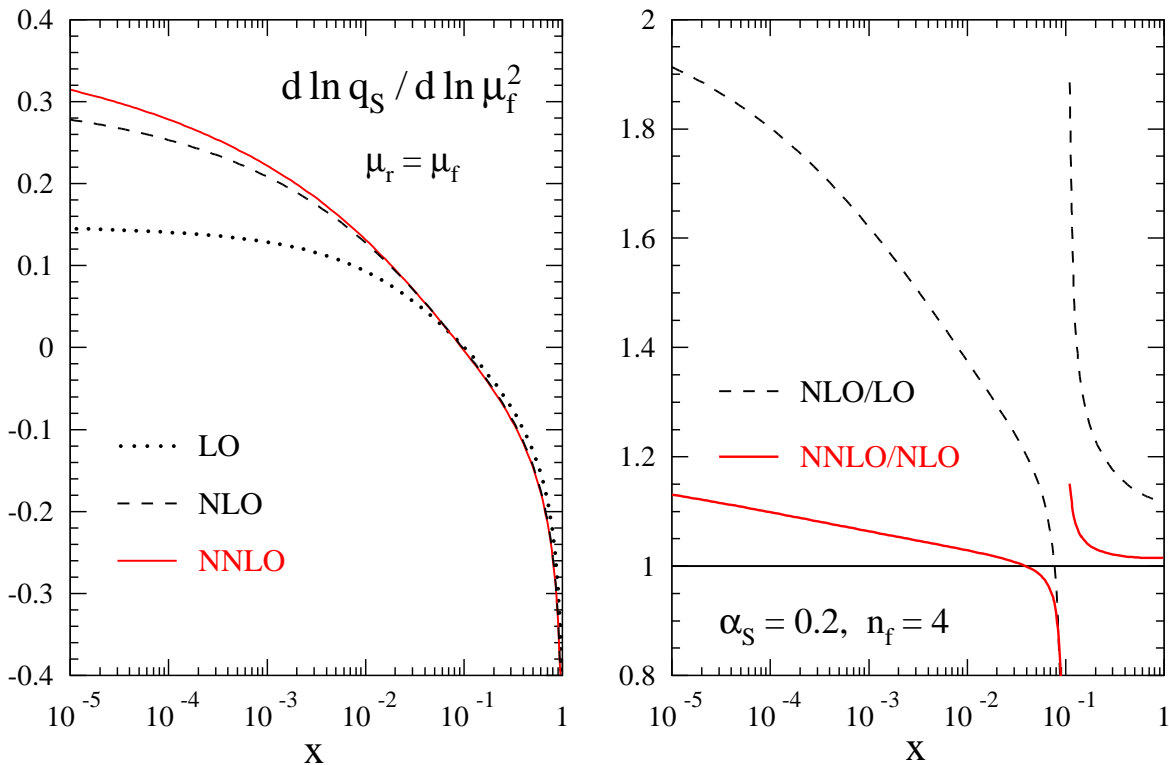,width=16cm,angle=0}}
\vspace{-2mm}
\caption{The perturbative expansion of the scale derivative $\,\dot{q}_
 {\,\rm s} \,\equiv\, d \ln q_{\rm s} / d\ln \mu_f^{\,2}\,$ of the
 singlet quark density at $\,\mu_f^{\,2} \: =\: \mu_0^{\,2\,}$ for the
 initial conditions specified in Eqs.~(\ref{eq:shapes}) and
 (\ref{eq:asvalue}).}
\end{figure}

The scale derivative of the quark distribution (Fig.~8 and Table 2) is 
dominated at large $x$ (small~$x$) by the $P_{\rm qq}\otimes 
q_{\rm s}$ ($ P_{\rm qg}\otimes g) $ contributions. The former 
(latter) is actually negligible for very small (large) values of~$x$. 
The NNLO corrections are small at large $x$ with respect to both the 
total derivative and the NLO contributions.  At small-$x$ all NLO 
contributions are very large (or the LO terms are abnormally small, 
recall that $xP_{\rm qq}^{(0)}$ and $xP_{\rm qg}^{(0)}$vanish for 
$x\ra 0$).  Consequently the total NNLO corrections, while reaching 
10\% at $x = 10^{\,-4}$, remain smaller than the NLO results by a 
factor of eight or more over the full $x$-range.

\begin{figure}[htb]
\label{pic:dq-exp}
\centerline{\epsfig{file=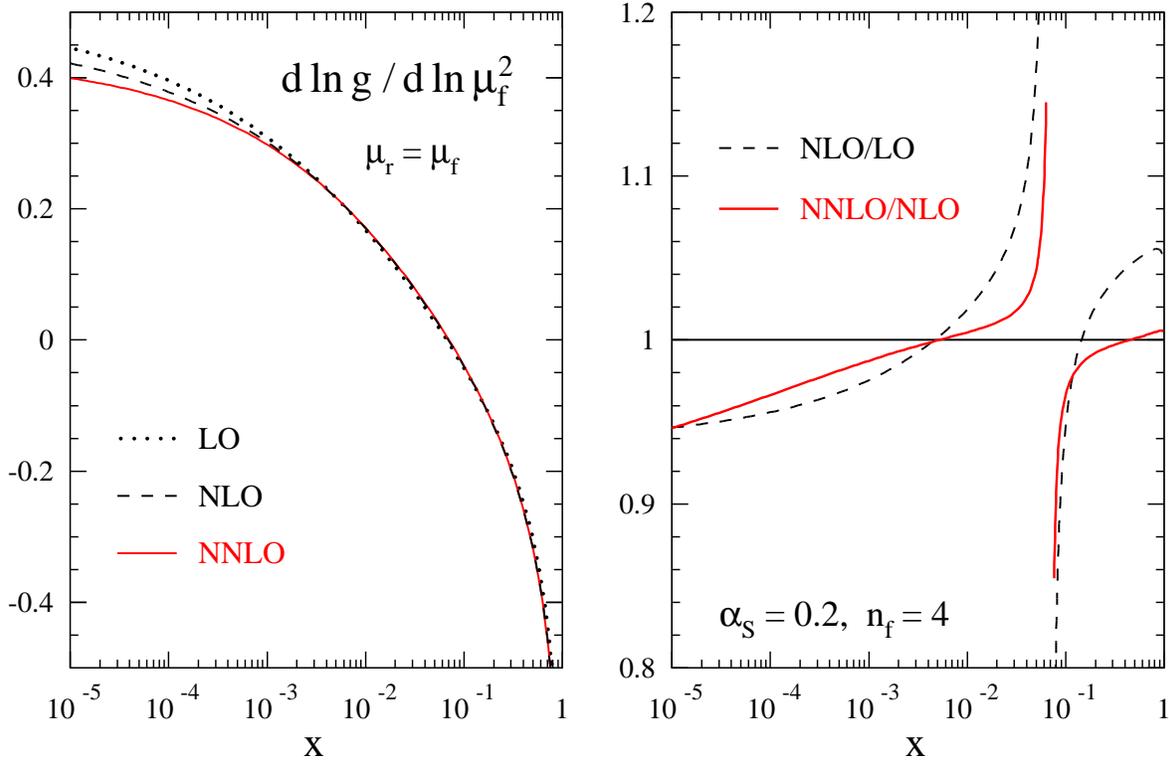,width=16cm,angle=0}}
\vspace{-2mm}
\caption{As Fig.~8, but for the gluon density.
 The spikes close to $x = 0.1$ in the right parts of both figures
 are due to zeros of the LO and NLO predictions and do not represent
 large corrections.}
\end{figure}

The situation is rather different for the evolution of the gluon
density (Fig.~9 and Table 3). The contribution from $ P_{\rm gg}
\otimes g$ dominates for all $x$ (except for extremely large values
not considered here), but the $ P_{\rm gq} \otimes q_{\rm s}$ part is 
nowhere negligible.
Already the NLO corrections are small especially at small $x$ and 
furthermore the $g$- and $q_{\rm s}$-initiated terms cancel each 
other to some extent. Thus the ratio $r_2/r_1$ of the relative NNLO and 
NLO corrections is rather large at small values of $x$, despite the 
NNLO contribution amounting to only 3\% for $x$ as low as $10^{\,-4}$.

\begin{table}[p]
\label{table2}
\begin{center}
\begin{tabular}{||c||r|r|r||r|r||r||}
\hline \hline
 & & & & & & \\[-0.3cm]
\multicolumn{1}{||c||}{$x$} &
\multicolumn{1}{c|} {LO} &
\multicolumn{1}{c|} {NLO} &
\multicolumn{1}{c||} {NNLO} &
\multicolumn{1}{c|} {$r_1$} &
\multicolumn{1}{c||}{$r_{\,2}$} &
\multicolumn{1}{c||}{$r_{\,2}/r_1$} \\[0.5mm] \hline \hline
\multicolumn{7}{||c||}{} \\[-3mm]
\multicolumn{7}{||c||}{ complete (Fig.~8) } \\
\multicolumn{7}{||c||}{} \\[-0.3cm] \hline \hline
 & & & & & & \\[-0.2cm]
$ 10^{-4}$ &
$ 1.405\cdot 10^{-1}$ &$ 2.532\cdot 10^{-1}$ &$ 2.781\cdot 10^{-1}$ &
  0.802 & 0.099 & 0.12 \\
  0.002  &
$ 1.218\cdot 10^{-1}$ &$ 1.890\cdot 10^{-1}$ &$ 1.991\cdot 10^{-1}$ &
  0.552 & 0.053 & 0.10 \\
  0.25  &
$-5.783\cdot 10^{-2}$ &$-6.919\cdot 10^{-2}$ &$-7.093\cdot 10^{-2}$ &
  0.196 & 0.025 & 0.13 \\
  0.75  &
$-2.056\cdot 10^{-1}$ &$-2.311\cdot 10^{-1}$ &$-2.346\cdot 10^{-1}$ &
  0.124 & 0.015 & 0.12 \\[1mm]
\hline \hline
\multicolumn{7}{||c||}{} \\[-3mm]
\multicolumn{7}{||c||}{ $P_{\rm qg} \otimes g$ contribution } \\
\multicolumn{7}{||c||}{} \\[-0.3cm] \hline \hline
 & & & & & & \\[-0.2cm]
$ 10^{-4}$ &
$ 1.624\cdot 10^{-1}$ &$ 2.610\cdot 10^{-1}$ &$ 2.787\cdot 10^{-1}$ &
  0.607 & 0.068 & 0.11 \\
 0.002   &
$ 1.421\cdot 10^{-1}$ &$ 1.996\cdot 10^{-1}$ &$ 2.053\cdot 10^{-1}$ &
  0.404 & 0.029 & 0.07 \\
  0.25  &
$ 1.146\cdot 10^{-2}$ &$ 1.087\cdot 10^{-2}$ &$ 1.037\cdot 10^{-2}$ &
 -0.051 &-0.046 & 0.89 \\
  0.75  &
$ 3.773\cdot 10^{-4}$ &$ 1.682\cdot 10^{-4}$ &$ 1.578\cdot 10^{-4}$ &
 -0.554 &-0.062 & 0.11 \\[1mm]
\hline \hline
\multicolumn{7}{||c||}{} \\[-3mm]
\multicolumn{7}{||c||}{ $P_{\rm qq} \otimes q $ contribution } \\
\multicolumn{7}{||c||}{} \\[-0.3cm] \hline \hline
 & & & & & & \\[-0.2cm]
$ 10^{-4}$ &
$-2.185\cdot 10^{-2}$ &$-7.838\cdot 10^{-3}$ &$-5.308\cdot 10^{-4}$ &
 -0.641 &-0.932 & 1.45 \\
  0.002  &
$-2.036\cdot 10^{-2}$ &$-1.056\cdot 10^{-2}$ &$-6.182\cdot 10^{-3}$ &
 -0.481 &-0.415 & 0.86 \\
  0.25  &
$-6.929\cdot 10^{-2}$ &$-8.006\cdot 10^{-2}$ &$-8.130\cdot 10^{-2}$ &
  0.156 & 0.015 & 0.10 \\
  0.75  &
$-2.060\cdot 10^{-1}$ &$-2.313\cdot 10^{-1}$ &$-2.347\cdot 10^{-1}$ &
  0.123 & 0.015 & 0.12 \\[1mm]
\hline \hline
\multicolumn{7}{||c||}{} \\[-3mm]
\multicolumn{7}{||c||}{ complete, but $\nf = 3$ } \\
\multicolumn{7}{||c||}{} \\[-0.3cm] \hline \hline
 & & & & & & \\[-0.2cm]
$ 10^{-4}$ &
$ 9.993\cdot 10^{-2}$ &$ 1.831\cdot 10^{-1}$ &$ 2.018\cdot 10^{-1}$ &
  0.832 & 0.102 & 0.12 \\
  0.002  &
$ 8.625\cdot 10^{-2}$ &$ 1.354\cdot 10^{-1}$ &$ 1.429\cdot 10^{-1}$ &
  0.570 & 0.055 & 0.10 \\
  0.25  &
$-6.070\cdot 10^{-2}$ &$-7.293\cdot 10^{-2}$ &$-7.527\cdot 10^{-2}$ &
  0.202 & 0.032 & 0.16 \\
  0.75  &
$-2.057\cdot 10^{-1}$ &$-2.344\cdot 10^{-1}$ &$-2.397\cdot 10^{-1}$ &
  0.139 & 0.023 & 0.16 \\[1mm]
\hline \hline
\multicolumn{7}{||c||}{} \\[-3mm]
\multicolumn{7}{||c||}{ low scale: $\nf = 3\:$, $\:\as = 0.4$ 
 and modified input (see Table 2)} \\
\multicolumn{7}{||c||}{} \\[-0.3cm] \hline \hline
 & & & & & & \\[-0.2cm]
$ 10^{-4}$ &
$ 2.132\cdot 10^{-1}$ &$ 9.317\cdot 10^{-1}$ &$ 1.416\cdot 10^{-0}$ &
  3.37\phantom 0 & 0.520 & 0.15 \\
  0.002  &
$ 2.047\cdot 10^{-1}$ &$ 6.047\cdot 10^{-1}$ &$ 7.762\cdot 10^{-1}$ &
  1.95\phantom 0 & 0.284 & 0.15 \\
$ 0.25   $ &
$-8.394\cdot 10^{-2}$ &$-1.227\cdot 10^{-1}$ &$-1.384\cdot 10^{-1}$ &
  0.462 & 0.128 & 0.28 \\
$ 0.735   $ &
$-3.870\cdot 10^{-1}$ &$-4.962\cdot 10^{-1}$ &$-5.362\cdot 10^{-1}$ &
  0.282 & 0.081 & 0.29 \\[1mm]
\hline \hline
\end{tabular}
\end{center}
\caption{The LO, NLO and NNLO logarithmic derivatives of the singlet
 quark distribution at four representative values of $x$, together with 
 the ratios $r_n = {\rm N^nLO} / {\rm N^{n-1}LO}- 1\,$ for the default 
 input parameters specified in the first paragraph of this section and 
 some variations thereof.}
\end{table}

\begin{table}[p]
\label{table3}
\begin{center}
\begin{tabular}{||c||r|r|r||r|r||r||}
\hline \hline
 & & & & & & \\[-0.3cm]
\multicolumn{1}{||c||}{$x$} &
\multicolumn{1}{c|} {LO} &
\multicolumn{1}{c|} {NLO} &
\multicolumn{1}{c||} {NNLO} &
\multicolumn{1}{c|} {$r_1$} &
\multicolumn{1}{c||}{$r_{\,2}$} &
\multicolumn{1}{c||}{$r_{\,2}/r_1$} \\[0.5mm] \hline \hline
\multicolumn{7}{||c||}{} \\[-3mm]
\multicolumn{7}{||c||}{ complete (Fig.~9) } \\
\multicolumn{7}{||c||}{} \\[-0.3cm] \hline \hline
 & & & & & & \\[-0.2cm]
$ 10^{-4}$ &
$ 3.956\cdot 10^{-1}$ &$ 3.782\cdot 10^{-1}$ &$ 3.655\cdot 10^{-1}$ &
 -0.044 &-0.034 & 0.76 \\
  0.002  &
$ 2.730\cdot 10^{-1}$ &$ 2.689\cdot 10^{-1}$ &$ 2.670\cdot 10^{-1}$ &
 -0.015 &-0.007 & 0.47 \\
  0.25  &
$-1.614\cdot 10^{-1}$ &$-1.661\cdot 10^{-1}$ &$-1.653\cdot 10^{-1}$ &
  0.029 &-0.005 &-0.18 \\
  0.75  &
$-4.721\cdot 10^{-1}$ &$-4.980\cdot 10^{-1}$ &$-5.000\cdot 10^{-1}$ &
  0.055 & 0.004 & 0.08 \\[1mm]
\hline \hline
\multicolumn{7}{||c||}{} \\[-3mm]
\multicolumn{7}{||c||}{ $P_{\rm gg} \otimes g$ contribution } \\
\multicolumn{7}{||c||}{} \\[-0.3cm] \hline \hline
 & & & & & & \\[-0.2cm]
$ 10^{-4}$ &
$ 3.085\cdot 10^{-1}$ &$ 2.895\cdot 10^{-1}$ &$ 2.814\cdot 10^{-1}$ &
 -0.062 &-0.028 & 0.45 \\
 0.002   &
$ 1.898\cdot 10^{-1}$ &$ 1.825\cdot 10^{-1}$ &$ 1.825\cdot 10^{-1}$ &
 -0.038 & 0.000 & 0.00 \\
  0.25  &
$-2.129\cdot 10^{-1}$ &$-2.287\cdot 10^{-1}$ &$-2.295\cdot 10^{-1}$ &
  0.074 & 0.003 & 0.04 \\
  0.75  &
$-5.226\cdot 10^{-1}$ &$-5.667\cdot 10^{-1}$ &$-5.717\cdot 10^{-1}$ &
  0.084 & 0.009 & 0.11 \\[1mm]
\hline \hline
\multicolumn{7}{||c||}{} \\[-3mm]
\multicolumn{7}{||c||}{ $P_{\rm gq} \otimes q $ contribution } \\
\multicolumn{7}{||c||}{} \\[-0.3cm] \hline \hline
 & & & & & & \\[-0.2cm]
$ 10^{-4}$ &
$ 8.715\cdot 10^{-2}$ &$ 8.873\cdot 10^{-2}$ &$ 8.409\cdot 10^{-2}$ &
  0.018 &-0.052 & 2.89 \\
  0.002  &
$ 8.323\cdot 10^{-2}$ &$ 8.642\cdot 10^{-2}$ &$ 8.454\cdot 10^{-2}$ &
  0.038 &-0.022 & 0.57 \\
  0.25  &
$ 5.154\cdot 10^{-2}$ &$ 6.264\cdot 10^{-2}$ &$ 6.424\cdot 10^{-2}$ &
  0.215 & 0.026 & 0.12 \\
  0.75  &
$ 5.047\cdot 10^{-2}$ &$ 6.871\cdot 10^{-2}$ &$ 7.168\cdot 10^{-2}$ &
  0.361 & 0.043 & 0.12 \\[1mm]
\hline \hline
\multicolumn{7}{||c||}{} \\[-3mm]
\multicolumn{7}{||c||}{ complete, but $\nf = 3$ } \\
\multicolumn{7}{||c||}{} \\[-0.3cm] \hline \hline
 & & & & & & \\[-0.2cm]
$ 10^{-4}$ &
$ 4.062\cdot 10^{-1}$ &$ 4.053\cdot 10^{-1}$ &$ 3.973\cdot 10^{-1}$ &
 -0.002 &-0.020 & 8.29 \\
  0.002  &
$ 2.836\cdot 10^{-1}$ &$ 2.915\cdot 10^{-1}$ &$ 2.927\cdot 10^{-1}$ &
  0.028 & 0.004 & 0.15 \\
  0.25  &
$-1.508\cdot 10^{-1}$ &$-1.578\cdot 10^{-1}$ &$-1.579\cdot 10^{-1}$ &
  0.047 & 0.001 & 0.01 \\
  0.75  &
$-4.615\cdot 10^{-1}$ &$-4.971\cdot 10^{-1}$ &$-5.028\cdot 10^{-1}$ &
  0.077 & 0.012 & 0.15 \\[1mm]
\hline \hline
\multicolumn{7}{||c||}{} \\[-3mm]
\multicolumn{7}{||c||}{ low scale: $\nf = 3\:$, $\:\as = 0.4$ and } 
\\[1.5mm]
\multicolumn{7}{||c||}{ $xq_{\rm s}(x,\mu_{0}^{\,2}) \: = \:
  0.6\: x^{\, -0.1} (1-x)^{3}\, (1 + 10\: x^{\, 0.8\,})$ } \\[1mm]
\multicolumn{7}{||c||}{ $ \!\! xg(x,\mu_{0}^{\,2})\:\:  = \:
  1.2\: x^{\, -0.1} (1-x)^{4}\, (1 + 1.5\: x) \:\:\: $ } \\
\multicolumn{7}{||c||}{} \\[-0.3cm] \hline \hline
 & & & & & & \\[-0.2cm]
$ 10^{-4}$ &
$ 2.135\cdot 10^{-0}$ &$ 2.015\cdot 10^{-0}$ &$ 1.536\cdot 10^{-0}$ &
 -0.056 &-0.238 & 4.23 \\
  0.002  &
$ 1.412\cdot 10^{-0}$ &$ 1.430\cdot 10^{-0}$ &$ 1.376\cdot 10^{-0}$ &
  0.013 &-0.038 &-3.00 \\
$ 0.25   $ &
$-1.663\cdot 10^{-1}$ &$-1.631\cdot 10^{-1}$ &$-1.523\cdot 10^{-1}$ &
 -0.019 &-0.067 & 3.46 \\
$ 0.75   $ &
$-9.055\cdot 10^{-1}$ &$-1.064\cdot 10^{-0}$ &$-1.116\cdot 10^{-0}$ &
  0.175 & 0.049 & 0.28 \\[1mm]
\hline \hline
\end{tabular}
\end{center}
\caption{As Table 2, but for the scale derivative $\,d \ln g / 
 d\ln \mu_f^{\,2}$ of the gluon distribution.}
\end{table}

We now turn to the stability of the perturbative expansions in Figs.~8
and 9 under variations of the renormalization scale $\mu_{\,r}$. For 
$\mu_{\,r} \neq \mu_f$ the expansion of the splitting functions in 
Eq.~(\ref{eq:Pexp}) is, using the abbreviation $a_{\rm s} \equiv 
\as/(4\pi)\,$, replaced by 
\bea
  P_{\rm ab}(\mu_f,\mu_{\,r})
  &\! =\! & \quad
    a_{\rm s}(\mu_{\,r}^{\,2}) \, P_{\rm ab}^{(0)}  \:\: + \:\: 
    a_{\rm s}^{\,2}(\mu_{r}^{\,2}) \, \left( P_{\rm ab}^{(1)} 
    - \beta_0\, P_{\rm ab}^{(0)} \ln \frac{\mu_f^{\,2}}{\mu_{r}^{\,2}} 
    \right) \: \\ 
 & & \mbox{}\!\!\!
    + \:  a_{\rm s}^{\,3}(\mu_{r}^{\,2}) \, \left( P_{\rm ab}^{(2)}
    - \bigg\{ \beta_1 P_{\rm ab}^{(0)} + 2\beta_0\, P_{\rm ab}^{(1)}
   \bigg\} \ln \frac{\mu_f^{\,2}}{\mu_{r}^{\,2}} 
    + \beta_0^2\, P_{\rm ab}^{(0)} \ln^2 
   \frac{\mu_f^{\,2}}{\mu_{r}^{\,2}} \, \right) + \ldots \nn \:\: ,
\eea
where $\beta_k$ represent the \MSb\ expansion coefficients of the 
$\beta$-function of QCD \
\cite{Caswell:1974gg,Jones:1974mm,Tarasov:1980au,Larin:1993tp}.

In Figs.~10 and 11 the respective consequences of varying $\mu_{\,r}$
over the rather wide range $\frac{1}{8}\,\mu_f^{\,2}\,\leq\,\mu_{r}
^{\,2}\, \leq\, 8 \mu_f^{\,2}$ are displayed for the logarithmic
$\mu_f$-derivatives of the singlet-quark and gluon distributions
(\ref{eq:shapes}) at six representative values of $x$. In both cases
the scale dependence is considerably reduced over the full $x$-range by
including the third-order corrections. With the exception of the
smallest $x$-value considered, $x = 10^{\,-5}$ (and of $x= 0.05$ in
Fig.~11, where the derivative is very small anyway), the points of
fastest apparent convergence and of minimal $\mu_{\,r}$-sensitivity,
$\partial\dot{f}/\partial\mu_{\,r} =0$, are rather close to the
`natural' choice $\mu_{\,r} = \mu_f$ for the renormalization scale.

\begin{figure}[htb]
\label{pic:scales1}
\vspace{+1mm}
\centerline{\epsfig{file=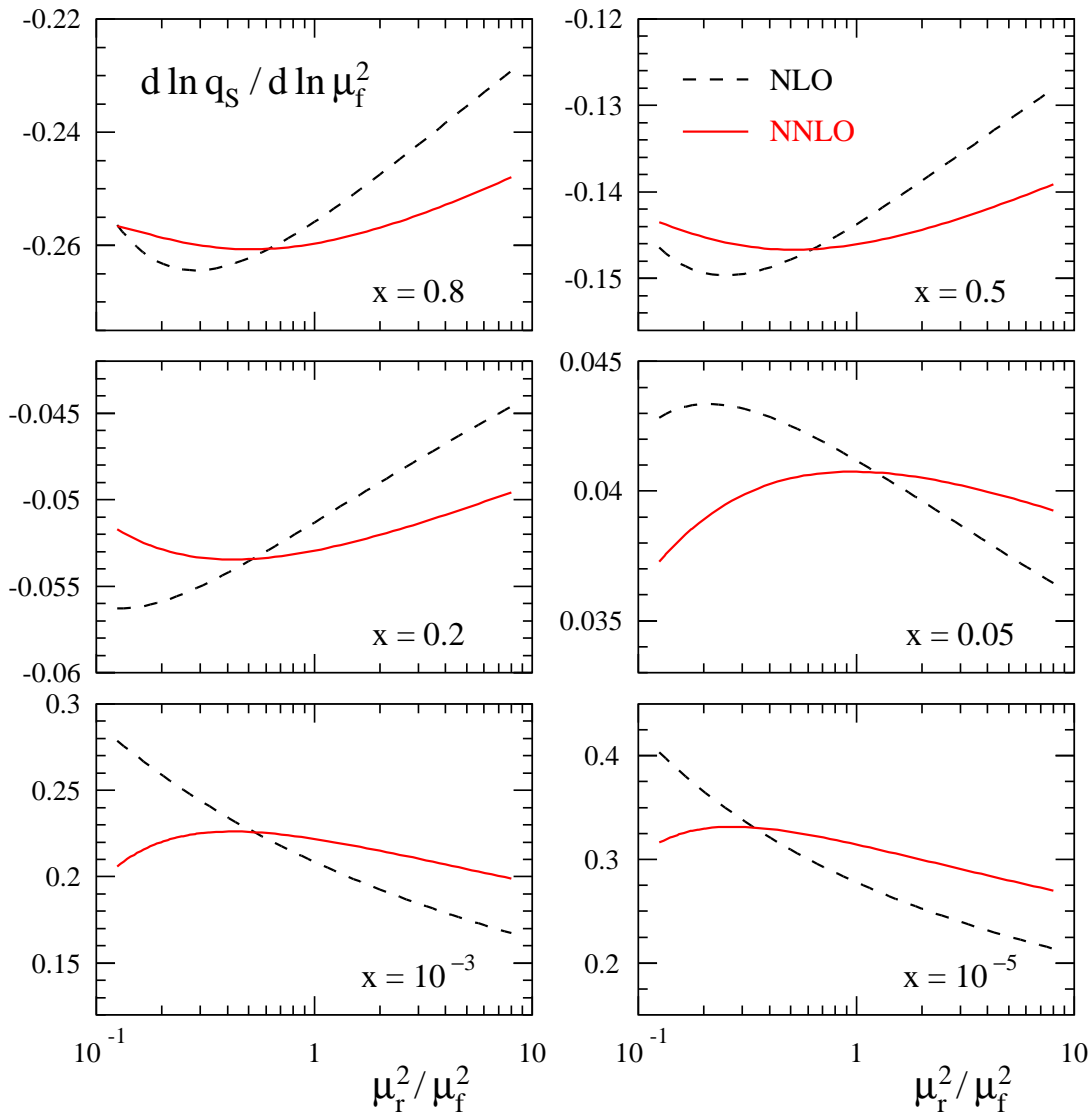,width=14.5cm,angle=0}}
\vspace{-2mm}
\caption{The dependence of the NLO and NNLO predictions for the
 derivative $\,d \ln q_{\rm s}/ d\ln\mu_f^{\,2}\,$ of the singlet-quark
 distribution on the renormalization scale $\mu_{\,r}$ for six typical 
 values of $x$.  The initial conditions are given in 
 Eqs.~(\ref{eq:shapes}) and (\ref{eq:asvalue}).}
\end{figure}

\begin{figure}[htb]
\label{pic:scales2}
\vspace*{-2mm}
\centerline{\epsfig{file=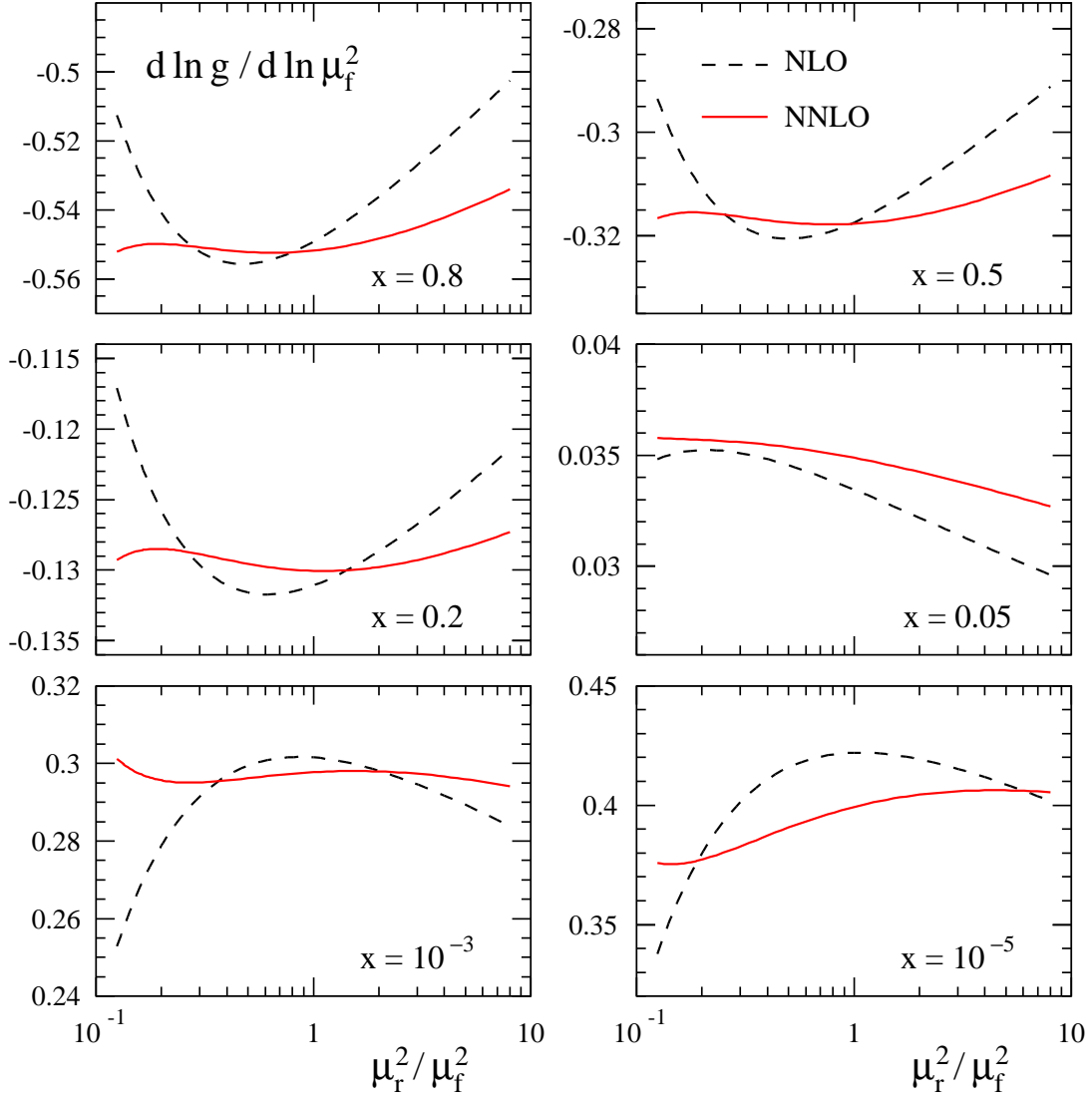,width=14.5cm,angle=0}}
\vspace{-2mm}
\caption{As Fig.~10, but for the derivative $\dot{g} \,\equiv\, d \ln g
 / d\ln\mu_f^{\,2}$ of the gluon distribution. Notice that the scales
 of the ordinates of the graphs differ within as well as between the
 two figures.}
\end{figure}

The relative scale uncertainties $\Delta \dot{q}_{\rm s}$ and
$\Delta \dot{g}$ of the average derivatives, estimated using the
conventional interval $\frac{1}{2}\,\mu_f\,\leq\,\mu_{\,r}\,\leq\,
2\mu_f$,
\beq
\label{eq:screl}
 \Delta \dot{f} \: \equiv \:
 \frac{\max\, [ \dot{f}(x,\mu_{\,r} = \frac{1}{2}
 \mu_f \ldots 2\mu_f)] - \min\, [\dot{f}(x,\mu_{\,r} = \frac{1}{2}\mu_f
 \ldots 2 \mu_f)] }
 { 2\, |\, {\rm average}\, [\dot{f}(x, \mu_{\,r} = \frac{1}{2}\mu_f
 \ldots 2 \mu_f)]\, | } \:\: ,
\eeq
are finally shown in Fig.~12.
For the singlet-quark (gluon) distribution, these uncertainty estimates 
amount to 2\% (1\%) or less at $x > 10^{-2}$ ($4\cdot 10^{-3\,}$), an
improvement by more than a factor of three with respect to the
corresponding NLO results. Taking into account also the apparent
convergence of the series in Figs.~6 and 7, it is not unreasonable
to expect that the effect of the higher-order singlet splitting 
functions will be about 1\% or less for $x \gsim 10^{-3}$. Larger 
corrections have to be expected at small $x$. One should also keep in 
mind that at fourth order also terms with the colour structure 
$d^{abc\,}d_{abc}/n_c$ --- which enter the non-singlet case already at 
three loops and have a large effect at $\,x<10^{-3\,}$ \cite
{Moch:2004pa} --- will contribute to the singlet splitting functions. 

\begin{figure}[thb]
\label{pic:scales3}
\vspace{-2mm}
\centerline{\epsfig{file=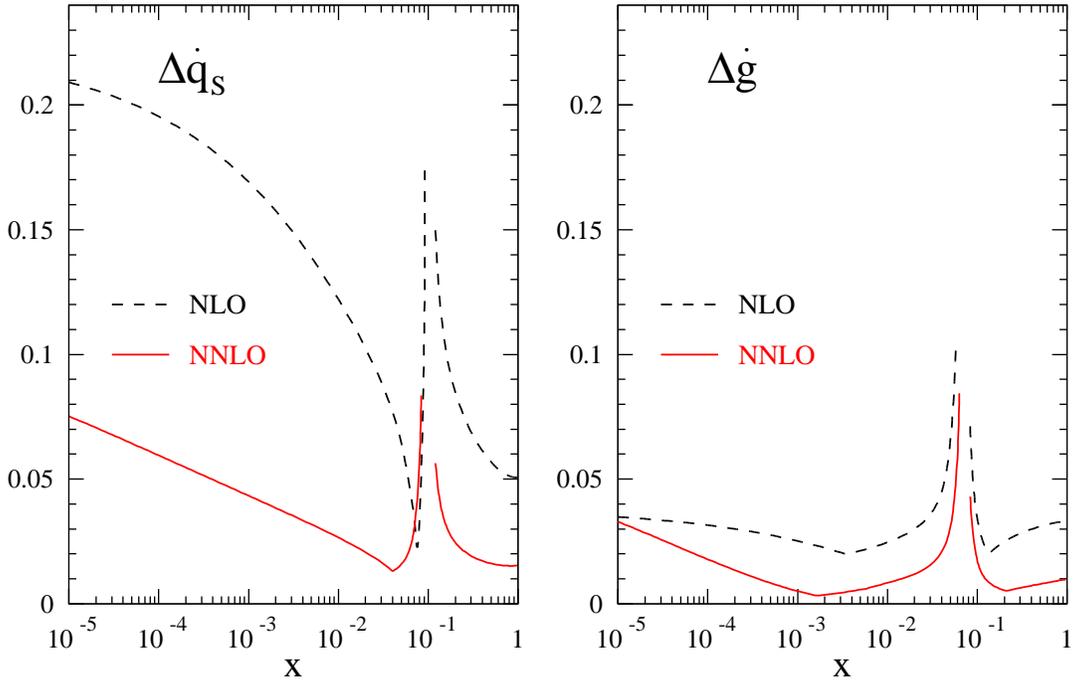,width=14.6cm,angle=0}}
\vspace{-2mm}
\caption{The renormalization scale uncertainty of the NLO and NNLO
 predictions for the scale derivatives of the singlet-quark density
 (right) and the gluon distribution (left) as estimated by the 
 respective quantities $\Delta \dot{q}_{\rm s}$ and $\Delta \dot{g}$ 
 defined in Eq.~(\ref{eq:screl}).}
\end{figure}

%
\section{Summary}
\label{sec:summary}
%
%
We have calculated the complete third-order contributions to the
splitting functions governing the evolution of unpolarized 
flavour-singlet parton distribution in perturbative QCD. 
Our calculation is performed in Mellin-$N$ space and follows the 
previous fixed-$N$ computations \cite{Larin:1997wd,Retey:2000nq} 
inasmuch as we compute the partonic structure functions in 
deep-inelastic scattering at even $N$ using the optical theorem 
and a dispersion relation as discussed in \cite{Larin:1997wd}. 
Our calculation, however, is not restricted to low fixed values of $N$ 
but provides the complete $N$-dependence from which the 
\mbox{$x$-space} splitting functions can be obtained by a (by now) 
standard Mellin inversion.
This progress has been made possible by an improved understanding of
the mathematics of harmonic sums, difference equations and harmonic
polylogarithms \cite{Vermaseren:1998uu,Remiddi:1999ew,Moch:1999eb}, and
the implementation of corresponding tools, together with other new
features \cite{Vermaseren:2002rp}, in the symbolic manipulation program
{\sc Form}~\cite{Vermaseren:2000nd} which we have employed to handle
the almost prohibitively large intermediate expressions.

Our results have been presented in both Mellin-$N$ and Bjorken-$x$
space, in the latter case we have also provided easy-to-use accurate
parametrizations. We agree with all partial results available
in the literature, in particular we reproduce the lowest six even-%
integer moments computed before \cite{Larin:1997wd,Retey:2000nq}. 
We also agree with the resummation predictions of Refs.~\cite
{Catani:1994sq,Fadin:1998py} for the leading small-$x$ logarithms 
$(\ln x)/x$ of the splitting functions $P_{\rm qq}$, $P_{\rm qg}$
and $P_{\rm gg}$, and with the large-$\nf$ result \cite{Bennett:1997ch}
for the simple $\ca\n2f$ part of $P_{\rm gg}$. Our results respect
the supersymmetric relation between all four splitting functions for
$\,\ca =\cf =\nf\, $ to the extend expected for the \MSb\ scheme. 
At large $x$ we verify the expected simple relation between the leading 
$1/(1-x)_+$ terms of $P_{\rm qq}$ and $P_{\rm gg}$. We find that also 
for the gluon-gluon splitting function the coefficients of the leading 
integrable term $\ln (1-x)$ at order $n=2,\,3$ are proportional to the 
coefficient of the $+$-distribution $1/(1-x)_+$ at order $n-1$, in 
complete analogy with our surprising findings in the non-singlet case 
\cite{Moch:2004pa}.

We have investigated the numerical impact of the three-loop (NNLO)
contributions on the evolution of the singlet-quark and gluon densities.
At $x \gsim 10^{-3}$ the perturbative expansion for the scale 
derivatives $\dot{f}\,\equiv\,d\ln f(x,\mu_f^{\,2})/d\ln\mu_f^{\,2\,}$, 
$f = q_{\rm s},\: g$ appears to be very well convergent and suggests a
residual higher-order uncertainty of about 1\% or less at $\as \lsim 
0.2$. Consequently the perturbative evolution can be safely extended to 
considerably larger values of $\as$, hence lower scales, in this range 
of $x$. 
The situation is much less clear at smaller $x$. For $\as = 0.2$ and 
realistic initial distributions with $xq_{\rm s},\, xg \sim x^{\,-0.3}$ 
at small $x$, the NNLO corrections for $\dot{q}_{\rm s}$ and $\dot{g}$ 
rise towards $x\ra 0$, respectively reaching 13\% and $-6\%$ at 
$x = 10^{-5}$. 
We stress that the results of the small-$x$ resummation alone cannot 
help here. For example, not even a qualitatively reliable prediction 
can be expected for the convolution $P_{\rm gg}\otimes g$, by which 
$P_{\rm gg}$ enters the evolution equations, even when all $1/x$ terms 
are included. 
Besides knowledge of as many of these terms as possible, further 
progress at small $x$ would require at least a four-loop generalization 
of the fixed-$N$ calculations \cite{Larin:1997wd,Retey:2000nq} and of
the $x$-space approximations~\cite{vanNeerven:2000wp} linking them to
the small-$x$ limits.

{\sc Form} files of our results, and {\sc Fortran} subroutines of our
exact and approximate splitting functions can be obtained from the 
preprint server \ {\tt http://arXiv.org} by downloading the source.
Furthermore they are available from the authors upon request.
\subsection*{Acknowledgments}
The preparations for this calculation have been started by J.V.
following a suggestion by S.A.~Larin. For stimulating discussions
during various stages of this project we would like to thank, in
chronological order, S.A.~Larin, F.~J.~Yndurain, E.~Remiddi, 
E.~Laenen, W.~L.~van Neerven, P.~Uwer, S.~Weinzierl and J.~Bl\"umlein. 
M.~Zhou has contributed some {\sc Form} routines during an early stage 
of the calculation.
The work of S.M. has been supported in part by Deutsche 
Forschungsgemeinschaft in Sonderforschungsbereich/Transregio 9.
The work of J.V. and A.V. has been part of the research program of the
Dutch Foundation for Fundamental Research of Matter (FOM).
 

\end{document}